\documentclass[12pt]{article}
\usepackage{amsmath}
\usepackage{graphicx}
\usepackage{enumerate}
\usepackage{url} 

\newcommand{\blind}{0}

\usepackage{amssymb,amsthm,enumitem,mathtools,microtype,natbib,xr-hyper}
\usepackage[dvipsnames]{xcolor}
\usepackage[colorlinks,allcolors=NavyBlue]{hyperref}

\newtheorem{proposition}{Proposition}

\begin{document}

\def\spacingset#1{\renewcommand{\baselinestretch}%
{#1}\small\normalsize} \spacingset{1}


\if0\blind
{
  \title{\bf Graph of Graphs: From Nodes to Supernodes in Graphical Models}

\author{
Maria De Iorio \\
    Yong Loo Lin School of Medicine,
    National University of Singapore \\
    Institute for Human Development and Potential, A*STAR, Singapore\\ 
   and \\
Willem van den Boom\thanks{
    Email: \href{mailto:willem@wvdboom.nl}{willem@wvdboom.nl}. This work was supported by the Singapore Ministry of Education Academic Research Fund Tier~2 under Grant MOE2019-T2-2-100.}\hspace{.2cm}\\
    Institute for Human Development and Potential, A*STAR, Singapore\\ 
    and \\
    Alexandros Beskos \\
    Department of Statistical Science, University College London\\
    and \\
    Ajay Jasra \\
    School of Data Science,
    Chinese University of Hong Kong, Shenzhen \\
    and \\ 
    Andrea Cremaschi \\ School of Science and Technology, IE University, Madrid
} \date{}
  \maketitle
} \fi

\if1\blind
{
  \bigskip
  \bigskip
  \bigskip
  \begin{center}
    {\LARGE\bf Graph of Graphs: \vspace{0.2cm}
    From Nodes to Supernodes in Graphical Models}
\end{center}
  \medskip
} \fi

\bigskip
\begin{abstract}  
High-dimensional data analysis typically focuses on low-dimensional structure, often to aid interpretation and computational efficiency. Graphical models provide a powerful methodology for learning the conditional independence structure in multivariate data by representing variables as nodes and dependencies as edges. Inference is often focused on individual edges in the latent graph. Nonetheless, there is increasing interest in determining more complex structures, such as communities of nodes, for multiple reasons, including more effective information retrieval and better interpretability. In this work, we propose a
hierarchical
graphical model where we first cluster nodes and then, at the higher 
level, investigate the relationships among groups of nodes. Specifically, nodes are partitioned into \emph{supernodes} with a data-coherent size-biased tessellation prior which combines ideas from Bayesian nonparametrics and Voronoi tessellations. This construct also allows accounting for the dependence of nodes within supernodes. At the higher
level, dependence structure among supernodes is modeled through a Gaussian graphical model, where the focus of inference is on \emph{superedges}. We provide theoretical justification for our modeling choices. We design tailored Markov chain Monte Carlo schemes, which also enable parallel computations. We demonstrate the effectiveness of our approach for large-scale structure learning in simulations and a transcriptomics application.
\end{abstract}


\if1\thepage{\vfill \newpage}\fi
\spacingset{1.9} 


\section{Introduction}
\label{sec:intro}

In applications with many variables, interest often lies in identifying large-scale structure.
For instance, groups of variables might carry meaning such as when dividing genes into co-expression modules \citep{Saelens2018}
or in item response theory \citep{Bock2021}, where latent traits are associated with sets of questionnaire items.
Furthermore, the relationship between variables and, more importantly at a larger scale, among the groups they belong to, can elucidate pre-eminent patterns in data.
We therefore introduce a
hierarchical
graphical model which clusters variables, and learns structure both within and among clusters.

Graphical models describe the dependencies in multivariate data by associating nodes of a graph with variables and the edges between them with conditional dependencies \citep{Lauritzen1996}. Recently, inferential focus has shifted from single edges to large-scale structure \citep{Fienberg2012,Barabasi2016}.
Such advances are driven by
the increasing amount of available data
and the complex patterns discovered in them.
Specifically, data exhibit mesoscopic patterns,
such as metabolic or signalling pathways,
that cannot be explained by models that use single edges as the main building block \citep{Iniguez2020}.
This new direction represents a change in perspective
from a reductionist viewpoint, with a shift from graph structures described through pairwise interaction between nodes, towards the use of large-scale structures \citep{Barabasi2012} for tackling the complexity present in empirical data.

Detecting substructures allows gaining better insight into the intricate patterns and dependencies within systems.
This is crucial in various fields such as bioinformatics (e.g.\ identifying functional motifs in biological networks) and social network analysis (e.g.\ detecting common structural patterns in social networks: for instance, subgraphs in criminal networks can reveal hidden patterns of criminal behavior).
A large-scale feature that has received specific attention is that of modularity or grouping of nodes \citep{Newman2012}
which for instance appears in genetics \citep{
Saelens2018
},
metabolomics \citep{Ravasz2002},
brain connectomes \citep{Sporns2016}
and protein-protein interactions \citep{Yook2004
}, shifting the focus from single edges to graph substructures. 

In the literature, there exist proposals on how to extend graphical models to learn groupings of nodes
 \citep[e.g.][]{Peixoto2019,vandenBoom2023}.
While these methods focus on larger structures, they are still based on inference of edges between individual variables, with the number of possible graphs growing superexponentially in the number of nodes. 
Also, in the context of Gaussian graphical models \citep[GGMs,][]{Dempster1972},
detection of individual edges reduces to testing for partial correlations which is particularly difficult \citep{Knudson2014}.
The effects on inference are exemplified
by the GGM simulation study in online Supplementary Material~\ref{ap:ROC_simul}
where increasingly many observations are required for reasonable recovery of edges with a larger number of nodes.

To overcome these challenges, we devise a 
hierarchical
construction which goes beyond edges between individual variables,  following a different strategy than
the existing literature. 
Specifically, we cluster nodes into groups and treat the groups of nodes as \emph{supernodes} (which represent macrostructure) and connect them using \emph{superedges} to form a \emph{supergraph}.
Within each supernode, the conditional independence structure is captured by a traditional GGM, more specifically a tree, with edges linking 
individual nodes. Note that edges between individual variables only appear within supernodes, but not across.
We refer to  this 
construction as \emph{graph of graphs}.

\begin{figure}
\centering
\includegraphics[width=\textwidth]{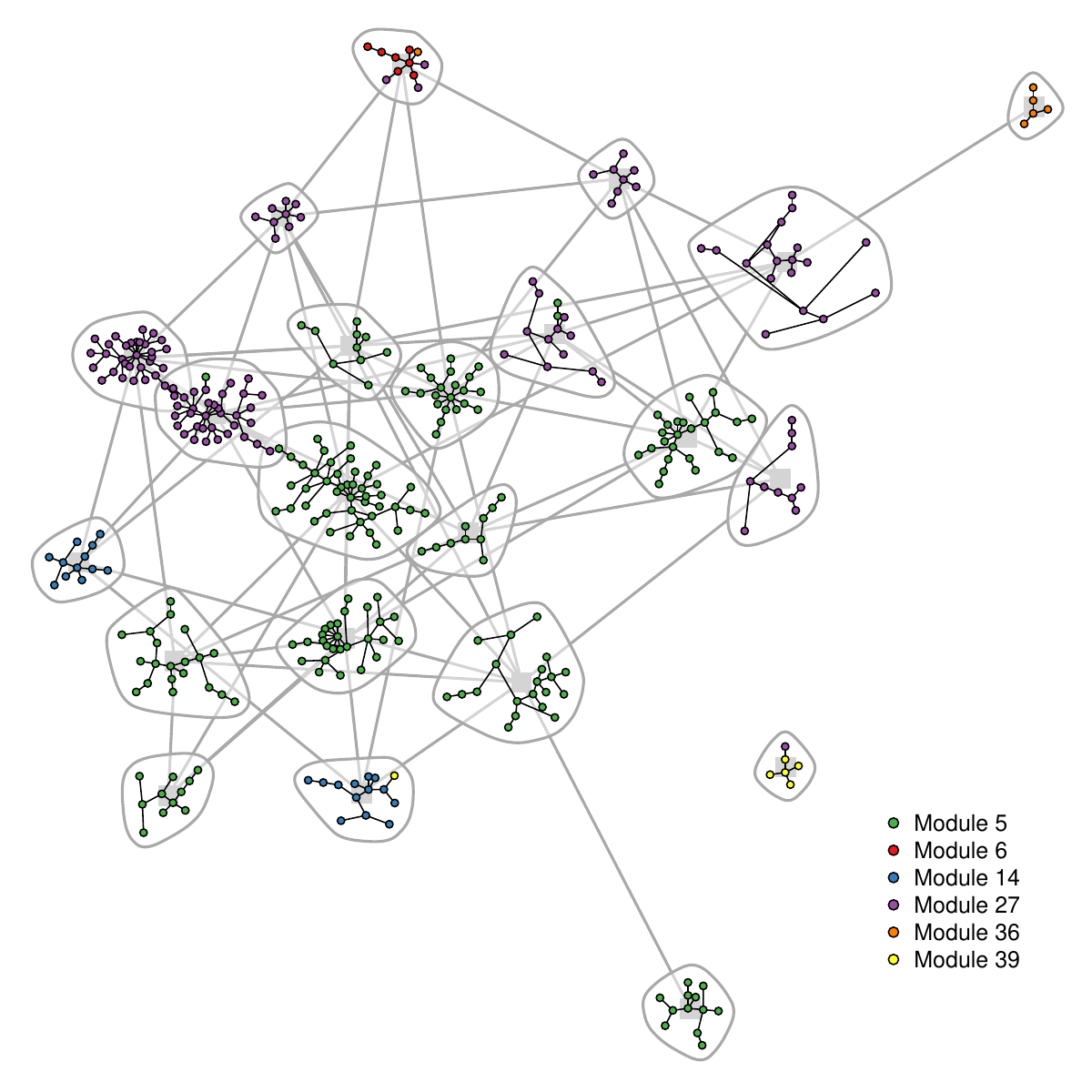}
\caption{
Gene expression data:
graph of graphs estimated from nested MCMC.
The circles represent nodes (i.e.\ genes) which are connected by the within-supernode graphs (trees) in black.
Trees are encircled in gray to mark the
supernodes.
Grey lines identify superedges between supernodes.
The nodes are colored according to the modules in \citet{Zhang2018}.
}
\label{fig:gene_sphere}
\end{figure}

To give an intuition of our modeling strategy,  Figure~\ref{fig:gene_sphere} shows an example of a graph of graphs inferred from gene expression data (see Section~\ref{sec:application} for details), alongside a modular structure found by \citet{Zhang2018} for the same data.
We can detect a rich structure among the genes which is notably more granular than the one found in \citet{Zhang2018}.  Figure~\ref{fig:gene_sphere} 
shows the partition of nodes into modules (supernodes) as well as the dependency structure among these (superedges), aiding interpretation and unveiling underlying biological mechanisms. This result needs to be contrasted with the single-level analysis in \citet{Zhang2018} and the analysis from a standard GGM model shown in Figure~\ref{fig:gene_glasso} in online Supplementary Material~\ref{ap:gene}.
We note that further inspection reveals that the finer granularity in the identified modules is supported by the literature.

Within the Bayesian framework, we construct a data-coherent prior on the clustering of nodes into supernodes which has two main components: (i)~a random tessellation \citep{Denison2002,Denison2002b} 
to enforce that highly correlated variables are grouped together; (ii)~a size-biased term to inform the size of the grouping \citep{Betancourt2022}.
Thus, the prior is highly informative and driven by the structure in the data.

Informative priors are
common in high-dimensional problems, like the horseshoe prior for sparse linear regression \citep{Bhadra2019}.
Such priors often do not reflect prior beliefs, but facilitate posterior inference, for instance asymptotically and relative to uninformative priors.
Although priors should represent subjective beliefs, 
there is in principle no reason against the use of data-dependent or data-coherent priors \citep{Martin2019
}.
Furthermore, they are sometimes preferred, as they lead to posterior distributions satisfying desirable frequentist properties.
For instance,
the variable selection priors in
\citet{Martin2017, Liu2021}
depend on data through centering at a maximum likelihood estimate.
We similarly use data information in our prior to obtain inference in line with our goals of dimensionality reduction and large-scale dependency discovery.

As the likelihood for the supergraph, we specify a GGM that links supernodes, specifically on the first principal components obtained from the variables within each supernode. We theoretically justify our modeling strategy.
The resulting likelihood does not correspond to a data generating process.
Such likelihoods are gaining popularity as it is increasingly difficult to specify a model that fully captures data complexity in high-dimensional problems.
In this context, 
our approach has connections with different methods:
(i)~indirect likelihood, which derives from an (auxiliary) model on a transformation of the data \citep{Drovandi2015};
(ii)~restricted likelihood, which is defined through an insufficient statistic of the data \citep[see][for an overview]{Lewis2021};
(iii)~the likelihood in
a Gibbs posterior, which is based on a loss function instead of a distribution on the data \citep[e.g.][]{Jiang2008
}.
Such likelihoods are used for various reasons, such as robustness to model misspecification.
Our motivation is more in line with
\citet{Pratt1965}
who considers restricted likelihoods (i.e.\ based on summary statistics) to focus inference on certain aspects of the data. We note that also Approximate Bayesian Computation methods follow a similar strategy.

In summary, the main contribution of our work, the graph of graphs, is a
hierarchical
graphical model able to detect macrostructures within a graph. Such macrostructures are described by supernodes and represent interpretable modules, capturing latent phenomena. Within each supernode, the microstructure is identified by a tree that provides a granular description of the dependence among the original variables.
The paper is structured as follows.
Section~\ref{sec:ggm} introduces graphical models.
Section~\ref{sec:model} details the model construction.
Posterior computations are described in Section~\ref{sec:inference}.
Section~\ref{sec:ggm_to_GoG} shows a simulated example.
Section~\ref{sec:application} presents
an application to gene expression data.
Section \ref{sec:discussion} concludes with a discussion.
An extensive overview of related work, simulation studies and a discussion of the methods are presented in the 
online Supplementary Material.

\subsection{Gaussian graphical models}
\label{sec:ggm}

Let a graph $G=(V,E)$ be defined by a set of edges $E \subset {\{(i,j)\mid 1\leq i<j\leq p \}}$
that represent links among the nodes in $V=\{1,\ldots,p\}$.
The nodes correspond to variables.
A graphical model \citep{Lauritzen1996} is a family of distributions which is Markov over $G$.
That is, the distribution is such that
the $i$\textsuperscript{th} and $j$\textsuperscript{th} variables are dependent conditionally on the other variables if and only if
$(i,j)\in E$.
In the special case of a GGM \citep{Dempster1972},
the distribution is the multivariate Gaussian
$\mathcal{N}(0_{p\times 1},\, \Psi^{-1})$
with precision matrix $\Psi$.
By properties of the multivariate Gaussian,
$\Psi_{ij} = 0$
implies that the $i$\textsuperscript{th} and $j$\textsuperscript{th} variables are independent
conditionally on the rest.
Thus, the conditional independence
structure specified by $G$
requires that
$\Psi_{ij} = 0$
if and only if nodes $i$ and $j$ are not connected.

A popular choice of prior for the precision matrix $\Psi$ conditional on $G$ is the $G$-Wishart distribution, as it induces conjugacy and allows working with non-decomposable graphs \citep{
Roverato2002}.
It is parameterized by the degrees of freedom $\delta > 0$
and a positive-definite rate matrix $D$.
Its density is not analytically available for general, non-decomposable $G$ due to an intractable normalizing constant.
For decomposable $G$,
the $G$-Wishart is tractable and coincides with the inverse Hyper Inverse Wishart
distribution \citep{rove:00}.

\section{Model description}
\label{sec:model}

\subsection{Graph of graphs}
\label{sec:graph_sphere}

The primary objective is to capture dependence structure at various complexity levels. For $p$ variables, applications often show high pairwise correlations among subsets of them due to common underlying phenomena or correlation with an unobserved variable. To capture this primary level of strong dependence, we divide the $p$ variables into $K$ groups (with $K$ being random), referred to as supernodes. Additionally, we aim to understand the dependence structure among these supernodes using a GGM. This hierarchical organization simplifies interpretation by capturing coarser dependence at the upper level among supernodes.

Let $X$ be an $n\times p$ matrix consisting of $n$ observations on $p$ variables.
We assume that data are standardized, so that $\sum_{i=1}^n X_{ij} = 0$
and $\|X_j\|^2 = n$ for all $j$,
where
$X_j$ is the $j$\textsuperscript{th} column of $X$ and
$\|\cdot\|$ is the Euclidean norm.
Denote the partition of the set of nodes $V=\{1,\dots,p\}$ into supernodes by ${\mathcal{T}} = \{S_k\}_{k=1}^K$
where the supernodes $S_k\subset V$ are such that $\bigcup_{k=1}^K S_k = V$
and $S_k\cap S_l = \emptyset$ for any $k\ne l$.
Then, the supergraph $G^\star = ({\mathcal{T}}, E^\star)$ 
has as vertices the set ${\mathcal{T}} = \{S_k\}$ of supernodes and as edges the set of superedges $E^\star\subset \{(S_k, S_l)\mid k< l\ \text{and}\ S_k, S_l \in {\mathcal{T}}\}$.

\begin{figure}[tb]
\centering
\includegraphics[width=.6\textwidth]{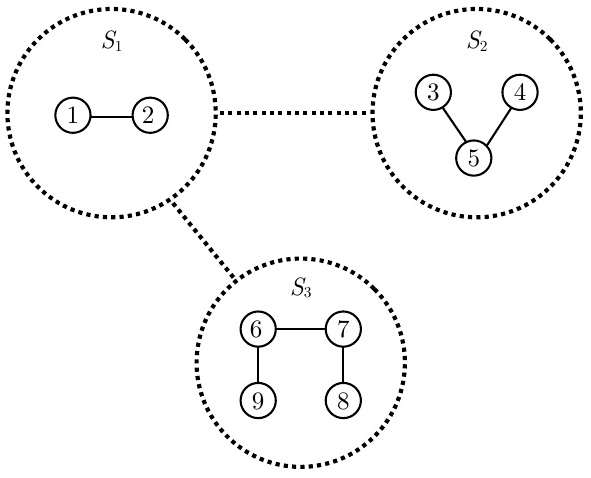}
\caption{
Visualization of a graph of graphs.
The dashed circles and lines represent supernodes and superedges, respectively, of a supergraph $G^\star$
consisting of the $K=3$ supernodes $S_1$, $S_2$ and $S_3$.
Within each supernode, the solid circles and solid lines show the tree among the original variables.
}
\label{fig:sphere}
\end{figure}

For the within-supernode structure,
given the set $S_k$, the nodes in $S_k$ correspond to vertices in the tree $T_k = (S_k, E_k)$, where $E_k\subset\{(i,j)\mid i<j\ \text{and}\ i,j\in S_k\}$ denotes the set of edges in $T_k$.
In summary,
the supergraph
$G^\star$ is a graph with vertices corresponding to supernodes $S_k$. Each supernode is a subset of the original variables $\{1, \ldots, p\}$ and the dependency structure among the variables within each supernode $S_k$ is described by a tree $T_k$.
The resulting hierarchical structure, with edges in $T_k$ connecting subsets of the original variables
and superedges in $G^\star$ connecting supernodes (i.e.\ trees),
motivates
the name \emph{graph of graphs}.
We visualize a graph of graphs in Figure~\ref{fig:sphere}.
The terminology `supernode', `supergraph' and `superedge' is borrowed from the literature on network compression 
\citep[e.g.][]{RodriguesJr2006}.

\subsection{Data-coherent size-biased tessellation prior}

The prior on $\mathcal{T}$
belongs to the class of data-dependent priors,
building on ideas from
Voronoi
tessellations \citep{Denison2002,Denison2002b},
exchangeable sequences of clusters \citep[ESC,][]{Betancourt2022} and
product partition models with covariates \citep[PPMx,][]{Muller2011}.

\subsubsection{Voronoi tessellation}
\label{sec:tessellation}

Our aim is that variables in a supernode refer to the same underlying phenomenon.
If that is the case, then
we expect variables in a supernode to be highly correlated due to the latent feature.
We impose such structure in ${\mathcal{T}}$ through a tessellation.
In a Voronoi tessellation,
elements of space are grouped together based on their distance to a set of centers.
In more detail, each center corresponds to a region. Then, the regions are defined by assigning the elements of the space to the center they are closest to in terms of some distance.

In our context, the space is the set of nodes $V$ and each region corresponds to a supernode.
For a set of centers $C\subset V$, each node is assigned to its closest center in terms of a distance based on correlation.
Denote the (sample) correlation between variables $X_i,X_j$ by
    $\hat{\rho}_{ij} =
    \frac{ X_i^\top X_j }{\|X_i\|\, \|X_j\|}$.
Then, node $i$ is assigned to the center $c\in C$ that minimizes the distance metric $\sqrt{2(1-|\hat{\rho}_{ic}|)}$ \citep{Chen2023}.
Thus, given the distance, $C$ identifies the tessellation ${\mathcal{T}}$ of nodes 
in a data-driven manner that encourages high correlation among nodes in a supernode.
Here,
we slightly abuse terminology by referring to the partition $\mathcal{T}$ of the discrete set $V$ as a tessellation,
while tessellations are usually defined over continuous spaces.
The relation between $C$ and $\mathcal{T}$ is deterministic and it depends on the choice of distance metric. In our case, we are interested
in capturing the correlation structure within $X$, but the choice of the distance is, in general, problem-specific.
Typically, the support of $\mathcal{T}$ is a small subset of the space of all possible partitions and such restriction alleviates posterior computation. \label{add:tessellation}
The approach can be generalized to probabilistic assignment of nodes to centers based on a function of distance, for instance, which would allow the exploration of a larger space of partitions.

We define the model on $\mathcal{T}$ hierarchically through the specification of a distribution on the set of centers $C$. Any node in $V$ can be a center, and any combination of nodes is possible. We assume that there is at least one center and we set $K=|C|$ as the number 
of regions/supernodes. Note that, given the discrete nature of $V$, different center combinations can give rise to the same tessellation $\mathcal{T}$ for a specific set of distances.
We build on ideas from exchangeable product partition models to specify a distribution for $C$. As a set of centers $C$ corresponds to only one partition, we use the terms centers and partition interchangeably. In the PPMx,
the probability assigned to a partition involves the product of a \emph{cohesion function} 
and a \emph{similarity function}. 
Typically, the cohesion function is derived from 
an exchangeable partition probability function
\citep[EPPF,][]{Hartigan1990},
such as the EPPF of the Dirichlet process or Gibbs-type priors \citep{DeBlasi2015}, which expresses a priori beliefs on the clustering structure. The similarity function usually exploits additional data information useful in the clustering process, biasing the prior probability of a partition towards clusters of subjects that share common ``relevant'' features.

\subsubsection{Tree activation function}
\label{sec:tree_activation}

For a given tessellation $\mathcal{T}$,
let $p_k = p_k(\mathcal{T}) = |S_k|$
denote the number of nodes in the $k$\textsuperscript{th} supernode
and
let $x_k=x_k(\mathcal{T},X)$ denote the $n\times p_k$ matrix consisting of the columns of $X$ assigned to the $k$\textsuperscript{th} supernode, i.e.\ $x_k = \{X_i\}_{i\in S_k}$. 
For the similarity $f_\textnormal{sim.}(x_k)$,
we choose a function that favors correlated variables to be grouped together, hence encouraging the supernodes to capture large-scale latent features.
We set $f_\textnormal{sim.}(x_k)=\widetilde{p}(x_k)$, where
$\widetilde{p}(x_k)$
is a probability distribution.
We refer to $\widetilde{p}(x_k)$ as \emph{tree activation function} and show that such distribution can effectively summarize the strength of pairwise correlations in $x_k$. 

As in the PPMx literature, we set the similarity function to coincide with a probability distribution for computational convenience (see Section~\ref{sec:inference}). Within each region of the tessellation $S_k$, we assume that the dependence structure among the variables $x_k$ is described by a tree. 
Then, $\widetilde{p}(x_k)$ is defined via a standard GGM, under the constraint of a tree structure for the graph.
That is, let $\Delta_k$ denote a precision matrix and $x_{ik}$ a row of $x_k$. We assume
\begin{align*}
    \widetilde{p}(x_k\mid \Delta_k ) &= \prod\nolimits_{i=1}^{n} \mathcal{N}(x_{ik}\mid 0_{p_k\times 1},\, \Delta_k^{-1}) \\
     \widetilde{p}(x_k) &= \sum\nolimits_{T_k} \widetilde{p}(T_k)  
     \int \widetilde{p}(x_k\mid \Delta_k )\, \widetilde{p}(\Delta_k\mid T_k)\, d\Delta_k
     = \sum\nolimits_{T_k} \widetilde{p}(T_k)\,  \widetilde{p}(x_k\mid T_k)
\end{align*}
The tree $T_k = (S_k,E_k)$
constrains $\Delta_k$
such that $(i,j)\in E_k$ if and only if the element of $\Delta_k$ corresponding to $i$\textsuperscript{th} and $j$\textsuperscript{th} variable in $x_k$ is nonzero. Since we assume a tree structure,
$\widetilde{p}(\Delta_k\mid T_k)$ is taken to be a 
Hyper-Wishart distribution w.r.t.\ the tree $T_k$,
with degrees of freedom $\delta > 0$ and positive-definite rate matrix $D$.
In this case, each edge is a (maximal) clique and each node is a separator.
While $\Delta_k$ is sparse, the tree constraint does not translate to sparsity in the covariance matrix $\Delta_k^{-1}$
which, with probability one, contains no zeros under $\widetilde{p}(\Delta_k\mid T_k)$.
As distribution on $T_k$,
we consider the uniform distribution over all trees.
That is $\widetilde{p}(T_k)=p_k^{2-p_k}$ since there are $p_k^{p_k-2}$ trees on $p_k$ nodes \citep{Cayley1889}.

The restriction to trees
has been found to be beneficial both empirically and theoretically \citep{Schwaller2019,Duan2023}, and here it
is desirable for two reasons.
First, an explicit evaluation of $\widetilde{p}(x_k)$ is feasible \citep{Meilă2006,Schwaller2019}. 
Second, we show that $\widetilde{p}(x_k)$ accurately captures the correlations in $x_k$ as $\widetilde{p}(x_k\mid T_k)$ turns out to be the product of edgewise terms which are a function of pairwise correlations.
Let $D_{\{i,j\}}$
be the $2\times 2$ submatrix of $D$ with rows and columns indexed by $\{i,j\}\subset S_k$, and
\[
     g_{ij}(\delta, D) = \frac{\Gamma\{(\delta+1)/2\}\,(D_{ii}\,D_{jj})^{\delta/2}}{\Gamma(\delta/2)\, |D_{\{i,j\}}|^{(\delta+1)/2}}
\]
Define the weights
$w_{ij} = g_{ij}(\delta^\star, D^\star) / g_{ij}(\delta,D)$
where $\delta^\star = \delta + n$ and $D^\star = D + x_k^\top x_k$.
Consider a
weighted graph with edge weight $w_{ij}$ between nodes $i,j\in S_k$.
Then, the Laplacian matrix corresponding to the weighted graph is the $p_k\times p_k$ matrix
$\Lambda$
defined by
$\Lambda_{ij}= -w_{ij}$
for $i\ne j$
and
$\Lambda_{ii}=\sum_{j\ne i} w_{ij}$.
Let $\Lambda^{\nu}$
denote the matrix obtained by removing the rows and columns indexed by $\nu\subset S_k$ from $\Lambda$.

The following result, which derives from \citet{Schwaller2019}, shows in (i–ii) how $\widetilde{p}(x_k)$
is a function of $w_{ij}$
and in (iii) how to efficiently compute $\widetilde{p}(x_k)$. 

\begin{proposition} \label{prop:mar_lik}
~
\begin{enumerate}[label=(\roman*), font=\upshape]
    \item
    The tree activation function is equal to
\[
    \widetilde{p}(x_k) = 
    \frac{p_k^{2-p_k}\, \Gamma(\delta^\star/2)^{p_k}\, (\prod_{i\in S_k} D_{ii})^{\delta/2}}{\pi^{n p_k/2}\, \Gamma(\delta/2)^{p_k}\, (\prod_{i\in S_k} D_{ii}^\star)^{\delta^\star/2}}
    \sum\nolimits_{E_k}\prod\nolimits_{(i,j)\in E_k} w_{ij}
\]
where
the sum is over all edge sets $E_k$ such that $T_k = (S_k, E_k)$ is a tree.
    \item
    $\widetilde{p}(x_k)$ is an increasing function of any weight $w_{ij}$.
    \item
    For any $u\in S_k$,
$\sum_{E_k}\prod_{(i,j)\in E_k} w_{ij} = |\Lambda^{\{u\}}|$ where $\Lambda$ is the Laplacian of the graph with edge weights $w_{ij}$.
\end{enumerate}
\end{proposition}

\begin{proof}
See online Supplementary Material~\ref{ap:proofs}.
\end{proof}

The weight $w_{ij}$ is a proxy for the correlation $\hat{\rho}_{ij}$. Consider the (improper) hyperparameter choice
$D=0_{p_k\times p_k}$. Then, 
\[
    w_{ij} \propto g_{ij}(\delta^\star, D^\star)
    \propto \frac{(D^\star_{ii}D^\star_{jj})^{\delta^\star/2}}{|D^\star_{\{i,j\}}|^{(\delta^\star+1)/2}}
    =
    \frac{(1 - \hat{\rho}_{ij}^2)^{-(\delta^\star+1)/2}}{\|X_i\|\, \|X_j\|}
\]
This suggests that
$w_{ij}$ and thus $\widetilde{p}(x_k)$
are increasing functions of the absolute correlation between $X_i$
and $X_j$.
Hence,
$\widetilde{p}(x_k)$ summarizes the strength of all correlations in the data.

To provide further insight into the role of $w_{ij}$ in the tree activation function,
we focus on the tree edge inclusion probabilities conditionally on $x_k$,
$\widetilde{\mathrm{Pr}}[{(i,j)\in E_k}\mid x_k]$.
We express $\widetilde{\mathrm{Pr}}[(i,j)\in E_k\mid x_k]$
in terms of $w_{ij}$, extending a result by \citet{Kirshner2007}.
Let $
r(i,j) \ =\  
\frac{|\Lambda^{\{i,j\}}|}{|\Lambda^{\{i\}}|}
$ be  the resistance distance
between nodes $i$ and $j$ in a graph with nodes $S_k$ and edge weights $w_{ij}$ \citep{
Bapat2004},
where
$\Lambda$
is the Laplacian corresponding to the weighted graph
\citep{Ali2020}. 

\begin{proposition} \label{prop:edge_prob}
We have that: (i)\, $\widetilde{\mathrm{Pr}}[(i,j)\in E_k\mid x_k] = w_{ij}\, r(i,j)$;\,\, (ii)\, $\widetilde{\mathrm{Pr}}[(i,j)\in E_k\mid x_k]$ is an increasing function of $w_{ij}$.
%
\end{proposition}

\begin{proof}
See online Supplementary Material~\ref{ap:proofs}. 
\end{proof}

Therefore, edge \emph{activation}, i.e.\ having an inclusion probability above a certain threshold,
depends on $w_{ij}$ being large enough.
This interpretation of $w_{ij}$ further highlights that the tree activation function $\widetilde{p}(x_k)$
captures the strength of the dependencies among the variables in the $k$\textsuperscript{th} supernode.
We conclude this section by noting that we could have used a Matrix 
$t$-distribution for $\widetilde{p}(x_k)$, which corresponds to a full graph, allowing us to capture global multicollinearity among the variables.
In Section~\ref{sec:simul_mvt} of online Supplementary Material, we find that using trees results in more accurate inference.
Moreover, we
stress that
we are interested in understanding the conditional independence structure within a supernode, therefore we opt for a more structured model.
Finally,  then the model can be extended to directed rooted trees, which we discuss in online Supplementary Material~\ref{ap:rooted}, where we obtain results analogous to Proposition~\ref{prop:mar_lik}.

\subsubsection{Size-biased cohesion function}

As cohesion function $f_\textnormal{coh.}(\cdot)$, we opt for an extension to the tessellation case of the ESC prior 
proposed by \citet{Betancourt2022} which provides additional prior control on cluster sizes. Specifically, 
we use a probability mass function on all positive integers
as in \citet{Betancourt2022}.
Here, the choice of $f_\textnormal{coh.}(\cdot)$ provides control over the size (and consequently the number) of the supernodes.  We point out that \citet{Betancourt2022} provide a constructive definition of their prior, while our extension to the 
graph of graphs
context is based on heuristics, leading to less effective control on cluster sizes (see online Supplementary Material~\ref{ap:size-biased}).

 Let  $\mathcal{C}(\mathcal{T})$ be the set of those combinations of centers $C$ that induce the same $\mathcal{T}$.
Moreover, we have $\binom{p}{K}$ ways of choosing $K$ centers among $p$ nodes, giving rise to the term $\binom{p}{|\mathcal{T}|}$ in \eqref{eq:tessellation_prior} below.
Then, the prior is defined by
\begin{equation} \label{eq:tessellation_prior}
    p(\mathcal{T}) \propto |\mathcal{C}(\mathcal{T})|\, \binom{p}{|\mathcal{T}|}^{-1} \prod_{k=1}^{|\mathcal{T}|} f_\textnormal{coh.}\{p_k(\mathcal{T})\}\, f_\textnormal{sim.}\{x_k(\mathcal{T},X)\}
\end{equation}
We refer to the prior $p(\mathcal{T})$ as the data-coherent size-biased tessellation prior.
Finally, 
we can formalize how \eqref{eq:tessellation_prior} biases the supernode sizes $p_k$ if $f_\textnormal{coh.}(\cdot)$ is a Geometric distribution with success probability $\pi$.

\begin{proposition} \label{prop:size-bias}
Let $f_\textnormal{sim.}(x_k)=1$ and $f_\textnormal{coh.}(p_k) = (1-\pi)^{p_k - 1}\, \pi$ in \eqref{eq:tessellation_prior}.
Then, a priori:
\begin{enumerate}[label=(\roman*), font=\upshape]
\item
$p(K)\propto \{\pi/(1-\pi)\}^K$
\item
Let $E$
denote the expectation w.r.t.\ the prior
and
$\overline{p} = \sum_{k=1}^K p_k / K$
 the mean supernode size. Then,
$E[\overline{p}]$
is a decreasing function of $\pi$
with $E[\overline{p}]\to p$ as $\pi\to 0$ and $E[\overline{p}]\to 1$ as $\pi\to 1$.
\end{enumerate}
\end{proposition}

\begin{proof}
See online Supplementary Material~\ref{ap:proofs}.
\end{proof}

As such, a cohesion function that assigns larger values $f_\textnormal{coh.}(p_k)$ to small supernode sizes, e.g.\ $p_k = 1$,
induces a prior on $\mathcal{T}$
that prefers more and smaller supernodes.

\subsection{Supergraph likelihood}
\label{sec:likelihood}

Given a tessellation ${\mathcal{T}}$,
we specify the supergraph likelihood involving
(i)~extraction of a latent feature from
each supernode and
(ii)~a GGM on these latent features.

\subsubsection{Latent feature extraction}

The data-coherent size-biased tessellation prior aims to group highly correlated variables into the same supernode.
To summarize the latent feature, captured by each supernode, 
we compute the first PC of $x_k$. 
Such use of principal component analysis (PCA) as a dimensionality reduction tool is supported by empirical and theoretical results \citep{Meyer1975,Malevergne2004,Stepanov2021,Whiteley2022}.

Let $\phi$ denote the proportion of variance in $x_k$ explained by the first PC of $x_k$. Let $\rho$ be
the average correlation, $s^2$ the average squared correlation in the $k$\textsuperscript{th} supernode 
and $\rho^i $ the average absolute correlation of variable $i$, with 
\[
\rho = \frac{2}{p_k(p_k-1)} \sum_{i>j\in S_k} \hat{\rho}_{ij};\qquad  
    s^2 = \frac{2}{p_k(p_k-1)} \sum_{i>j\in S_k} \hat{\rho}_{ij}^2;\qquad 
    \rho^i = \frac{1}{p_k-1} \sum_{j\in S_k: j\ne i} |\hat{\rho}_{ij}|
\]
Following \citet{Stepanov2021}, we can show the following proposition.

\begin{proposition} \label{prop:pca} 
Let $h_{p_k}(t) = \frac{1}{p_k} + ( 1 - \frac{1}{p_k}) t$. Define $h_\star(t) = 
\tfrac{1}{2}(1 + \sqrt{2t - 1})$ if $t\ge 1/2$ and
$h_\star(t) = t$ otherwise.
The proportion $\phi$ of variance explained by the first principal component satisfies:
\begin{enumerate}[label=(\roman*), font=\upshape]
    \item $\max[h_{p_k}(\rho),\, h_\star\{h_{p_k}(s^2)\} ] \leq \phi \leq \min\{ h_{p_k}(s),\, \max_i h_{p_k}(\rho^i)\}$ \label{point:bound}
    \item If all correlations $\hat{\rho}_{ij}$ are equal,
    then $\phi= h_{p_k}(\rho)$ if $\rho\ge 0$ and $\phi=\frac{1-\rho}{p_k}$ otherwise.
\end{enumerate}
\end{proposition}

\begin{proof}
See online Supplementary Material~\ref{ap:proofs}. 
\end{proof}

Thus, higher absolute correlations in $x_k$ imply that the first PC can better describe most of the variation in a supernode.
Moreover,
with perfect correlations ($s^2 = 1$),
the first PC captures all variation in $x_k$. See also 
Figure~\ref{fig:prop_var} 
in Supplementary Material.

\subsubsection{Model on the latent features}
\label{sec:supergraph_model}

We now specify a GGM that links the supernodes through their latent features.
What follows is conditional on a tessellation with $K$ regions and supernodes $\{S_k\}$.
Let the $n\times K$ matrix $Y^\star$
contain the first PCs corresponding to each supernode $x_k$.
The PCs are standardized, i.e.\ $\|Y^\star_k\|^2 = n$.
As model for $Y^\star$, we assume a  GGM, where each row of $Y^\star$ is normally distributed with mean zero and precision matrix $\Omega^\star$, conditionally on the supergraph $G^\star$.

\subsubsection{Augmented space}
\label{sec:augmented}

In the previous subsection,
we have defined the model for the supergraph conditional on the tessellation.
We note that the focus of inference is not only the supergraph, but also the tessellation of nodes.
This includes also inference on $K$ as well as each supernode composition $S_k$.
As such, when performing posterior inference using
Markov chain Monte Carlo (MCMC)
methods, this would require transdimensional moves and consequentially devising labor-intensive MCMC schemes since $Y^\star$
 changes dimension with $K$ and supernode membership changes with tessellation.
This issue is discussed more exhaustively in online Supplementary Material~\ref{ap:mcmc}.
Note that such changes in dimensions are different from those addressed by tools such as reversible jump MCMC \citep{Green1995} where the parameter instead of the data changes dimension.
Thus, to avoid the change in dimension, 
we resort to a data augmentation trick, which has been successfully exploited in other contexts \citep[e.g.][]{Royle2007,Walker2007}.
We define a GGM on an augmented space which has the same dimension $p$ as the original data $X$. We specify a GGM on all $p$ PCs across the supernodes instead of  just the $K$ first PCs. Note that if a supernode contains $p_k$ variables, then the number of PCs associated to $S_k$ is $p_k$ with $\sum_{k=1}^K p_k=p $.
Therefore,
let $Y$ denote the $n\times p$ matrix obtained by adding all lower ranked PCs to $Y^\star$, again standardized such that $\|Y_i\|^2 = n$ for every $i$.

We highlight that our main inferential focus is on \label{add:single_PC}
the links among supernodes, i.e.\ between first PCs, rather than on any weaker patterns involving lower rank PCs.
This strategy allows
for a reduction in complexity in terms of GGM inference
and improved
interpretation.
Let $G$ be a graph on $p$ nodes where edges (corresponding to superedges) can only exist between the $K$ nodes corresponding to first PCs such that the supergraph $G^\star$ uniquely determines $G$.
The other $(p - K)$ nodes are auxiliary to avoid changes in dimension.
Such use of an augmented parameter $G$ is similar in spirit to the use of pseudopriors by \citet{Carlin1995}, who exploit auxiliary parameters to avoid transdimensionality.

In more detail,
let the rows of $Y$ be independently distributed according to
$\mathcal{N}(0_{p\times 1},\, \Omega^{-1})$.
Conditionally on 
the graph $G$, the prior on the precision matrix $\Omega$
is the $G$-Wishart distribution
with degrees of freedom $\delta_G > 0$ and positive-definite rate matrix $D_G$. Note that nodes corresponding to lower ranked PCs are not connected among themselves or with any supernode. As such there will always be a zero element in the precision matrix in such entries. 
This gives rise to the marginal likelihood
\citep[e.g.][]{AtayKayis2005}
\begin{equation} \label{eq:likelihood}
    p(Y^\star\mid \mathcal{T},G^\star,X) \propto p(Y \mid \mathcal{T}, G, X) = \frac{I_{G}(\delta_G^\star, D_G^\star)}{(2\pi)^{np/2} I_{G}(\delta_G, D_G)}
\end{equation}
where
$\delta_G^\star = \delta_G + n$,
$D_G^\star = D_G + Y^\top Y$, and
$I_{G}(\delta_G, D_G)$ denotes the normalizing constant of the density of the $G$\nobreakdash-Wishart distribution with graph $G$,
degrees of freedom $\delta_G$ and rate matrix $D_G$.
Note that $p(Y \mid \mathcal{T},G,X)$ depends only on $Y^\star$ and not on the lower ranked PCs $(Y\setminus Y^\star)$
due to standardization.
Furthermore, the structure of the induced precision matrix $\Omega^\star$ of $Y^\star$ corresponds to $G^\star$.
In what follows, with abuse of terminology, we refer to $G$ as supergraph as $G^\star$ can be deterministically recovered from $G$.
Finally,
a choice of
prior
$p(G\mid \mathcal{T})$
on supergraphs,
which we discuss in Section~\ref{sec:application},
completes the model specification.

\section{Inference}
\label{sec:inference}

Main object of inference are the tessellation $\mathcal{T}$ and the supergraph $G$, from which we can recover $G^\star$. The target distribution is
\begin{equation} \label{eq:post}
p(\mathcal{T}, G, Y \mid X ) \propto p(\mathcal{T} )\, p(G\mid \mathcal{T})\, p(Y\mid \mathcal{T}, G, X)
\end{equation}
Sampling from the distribution in \eqref{eq:post} enables posterior inference on the graph of graphs $(\mathcal{T}, G^\star)$.
Many challenges need to be addressed to sample from this distribution.
First of all,
we have $n$ observations for each of the $p$ nodes. 
As a result,
the posterior 
on the tessellation
is highly concentrated unless $n$ is small:
see online Supplementary Material~\ref{ap:conc}
where we show that $p(\mathcal{T}\mid X)$ can be a point mass for moderate $n$. 
We remark that
the approaches for clustering of nodes by
\citet{Peixoto2019} and  \citet{vandenBoom2023} 
do not suffer from this collapsing of the posterior on the partition
as $n$ gets large.
That is because they cluster nodes based on edges in the graph,
with the graph representing a single (latent) observation.
In their setup,
a large $n$ results in concentration of the posterior on the graph, but typically not of the posterior on the partition.

Such concentration of $p(\mathcal{T}\mid X)$
inhibits MCMC convergence and mixing.
To still be able to use MCMC for inference,
we consider transformations of the posterior in \eqref{eq:post}
that are less concentrated.
Specifically, we propose two possible solutions: (i)~coarsening of the tree activation functions and of the likelihood; (ii)~nested MCMC with coarsening of the tree activation functions.
A coarsened likelihood is a likelihood raised to a power
as in \citet{Page2018}, with the goal of flattening it for better exploration of posterior space.

In what follows, we refer to the two proposed algorithms as coarsening of the likelihood and nested MCMC respectively. Both of them require coarsening of the tree activation function. 

\subsection{Coarsening of the tree activation functions}
\label{sec:prior_coarsening}

The tree activation function $\widetilde{p}(x_k)$,
defined as a probability distribution in Section~\ref{sec:tree_activation},
generally becomes more peaked as the sample size $n$ increases.
More specifically,
$\widetilde{p}(x_k)$ is defined through the conditional distribution $\widetilde{p}(x_k\mid T_k)$
with  tree $T_k=(S_k, E_k)$, where the $n$ rows of $x_k$ are independently distributed conditionally on the precision $\Delta_k$.
This scaling with $n$ results in the  prior $p(\mathcal{T})$
with the similarity function $f_\textnormal{sim.}(x_k) = \widetilde{p}(x_k)$
to be skewed too strongly by the similarity information for large sample sizes. Then, the size biasing from the cohesion function $f_\textnormal{coh.}(\cdot)$ in \eqref{eq:tessellation_prior} becomes negligible, and $p(\mathcal{T})$ becomes too peaked.

A similar phenomenon appears in PPMx if the number of covariates increases with the partition prior becoming very peaked \citep{Barcella2017}, and the posterior concentrating  
on either $K=1$ or $K=p$ clusters as the dimensionality increases \citep{Chandra2023}.
Relatedly,
the prior in PPMx with many covariates may dominate the posterior distribution, with the likelihood being much less peaked than the prior.
A variety of solutions has been explored to address this issue \citep{Page2018}
including
dimensionality reduction via covariate selection
and enrichment with clustering at two levels \citep{Wade2014}.

The issue of an overly concentrated prior is yet more pertinent in our context because we cluster variables/nodes, while PPMx considers clustering of observations. Thus, the tree activation functions dominate when the sample size is large.
Also, we treat the observations as exchangeable, such that approaches based on variable selection or enrichment are not sensible.
Instead, as in \citet{Page2018}, we coarsen the similarity functions.

We use a modified $\widetilde{p}(x_k)$ as similarity function $f_\textnormal{sim.}(x_k)$:
we replace the distribution $\widetilde{p}(x_k\mid T_k)$
by the coarsened version $\widetilde{p}(x_k\mid T_k)^{\zeta}$ for some power $\zeta\in(0,1]$.
We consider $\zeta\propto n^{-1}$
to reflect the $n$ i.i.d.\ observations.
The power balances how strongly the correlations in $X$ inform the tessellation.
Specifically,
instead of
$f_\textnormal{sim.}(x_k) = \widetilde{p}(x_k) = \sum_{T_k} \widetilde{p}(x_k\mid T_k)\, \widetilde{p}(T_k) $,
we use the coarsened similarity function
\begin{equation} \label{eq:coarsened_sim}
    f_\textnormal{sim.}^{(\zeta)}(x_k) = \sum_{T_k} \widetilde{p}(x_k\mid T_k)^{\zeta}\, \widetilde{p}(T_k)
\end{equation}
where the sum is over all possible  $T_k$ 
and  $\widetilde{p}(T_k) = p_k^{2-p_k}$ is the uniform distribution.
The
prior choice 
facilitates  the computation of the
similarity function, as
computation of $\widetilde{p}(x_k)$
via the determinant $|\Lambda^{\{u\}}|$ in
Proposition~\ref{prop:mar_lik}
is notoriously numerically unstable \citep{
Momal2021}.
This is due to
the relative sizes of the weights $w_{ij}$ diverging as $n$ and $p_k$ increase.
To alleviate the problem, we can replace $w_{ij}$ by $w_{ij}^{\zeta}$
in Proposition~\ref{prop:mar_lik}
to compute $f_\textnormal{sim.}^{(\zeta)}(x_k)$ instead of $\widetilde{p}(x_k)$. 

Concerning the choice of the power $\zeta$,
\citet{Page2018} use $\zeta = n^{-1}$, though coarsening with larger powers, still with $\zeta\to 0$ as $n\to\infty$, has been explored within PPMx \citep{Pedone2024} and in other contexts \citep{Miller2019}.
Alternatively,
a prior can be placed on $\zeta$ as it has been done in the context of \emph{power priors} \citep{Chen2000}. 
MCMC
with a prior on $\zeta$ is challenging as it leads to a doubly intractable posterior \citep{Carvalho2021}.
We further discuss our choice of $\zeta$ in Section~\ref{sec:application}.

\subsection{Coarsening of the likelihood and nested MCMC}

Like the tree activation function, the information provided by the supergraph likelihood $p(Y\mid \mathcal{T},G,X)$ in \eqref{eq:likelihood} scales with the sample size $n$.
Also,
this scaling causes posterior uncertainty for the tessellation $\mathcal{T}$ to vanish for large $n$. To counterbalance this phenomenon,
we consider two options: coarsening of the likelihood and nested MCMC.
In both cases, we need to devise tailored computational solutions, which, nevertheless, exploit the same techniques.
The resulting algorithms are detailed in online Supplementary Material~\ref{ap:mcmc}.

\subsubsection{Coarsening of the likelihood}

Coarsening or flattening the likelihood can undo its undesirable scaling with $n$.
However,
there is a need to
balance the information provided by the similarity function
$f_\textnormal{sim.}^{(\zeta)}(\cdot)$
and by the likelihood.
Therefore,
we use the same power $\zeta$ to coarsen both the tree activation functions and the likelihood.
Specifically,
we use a transformation
of the posterior in \eqref{eq:post}
as target distribution:
\begin{equation} \label{eq:joint_target}
\pi^{(\zeta)}(\mathcal{T},G,Y)\propto p^{(\zeta)}(\mathcal{T})\, p(G\mid \mathcal{T})\, p(Y\mid \mathcal{T},G,X)^\zeta
\end{equation}
where
$p(Y\mid \mathcal{T},G,X)^\zeta$ is the likelihood raised to the power $\zeta$
and
$p^{(\zeta)}(\mathcal{T})$ is the data-coherent size-biased tessellation prior in \eqref{eq:tessellation_prior} with the coarsened $f_\textnormal{sim.}^{(\zeta)}(\cdot)$ in \eqref{eq:coarsened_sim} as similarity function.

As in the context of model misspecification,
raising the likelihood to a power is done
to avoid undesired concentration of the posterior 
\citep[e.g.][]{Grunwald2017,Miller2019}.
Furthermore, such a power is standard in Gibbs posteriors where the likelihood derives from a loss function and the power balances the influence of the prior \citep[e.g.][]{Jiang2008
}.
Finally,
\citet{Martin2017} and \citet{Liu2021} coarsen the likelihood to avoid overconcentration of the posterior with data-coherent priors.

\subsubsection{Nested MCMC}

An alternative to coarsening of the likelihood to avoid overconcentration of the posterior
is to recast the problem  within \emph{cutting feedback} \citep{Plummer2015
},
cutting the feedback from $Y$ to $\mathcal{T}$
such that $Y$ does not inform $\mathcal{T}$.
The target becomes the cut distribution
\begin{equation} \label{eq:cf_target}
\pi_\textnormal{cut}(\mathcal{T},G,Y)\propto p^{(\zeta)}(\mathcal{T})\, p(G,Y\mid \mathcal{T},X)
\end{equation}
where
$p(G,Y\mid \mathcal{T},X)
\propto {p(G\mid \mathcal{T})}\, {p(Y\mid \mathcal{T},G,X)}$ is the posterior on the supergraph conditionally on $\mathcal{T}$.
Then,
the marginal distribution on $\mathcal{T}$
under the target
is $p^{(\zeta)}(\mathcal{T})$
which is sufficiently diffuse under enough coarsening,
i.e.\ small enough $\zeta$.
To sample $\pi_\textnormal{cut}(\mathcal{T},G,Y)$,
we use nested MCMC \citep{Plummer2015,Carmona2020} which allows for parallel computation. We note that the cut posterior in \eqref{eq:cf_target} is an approximation of the true posterior.
 
We employ cutting feedback to improve MCMC mixing
\citep[see also][]{Liu2009,Plummer2015}. Moreover, a random variable whose feedback is being cut might
provide information about a parameter that conflicts with other, more trusted parts of the model \citep{Plummer2015,Jacob2017}.
Such a conflict is present here:
the information in $p(Y\mid \mathcal{T},G,X) $ provides contrasting information  for the tessellation 
as compared to the data-coherent size-biased tessellation prior.

\section{Example}
\label{sec:ggm_to_GoG}

\begin{figure}[tb]
\centering
\includegraphics[width=\textwidth]{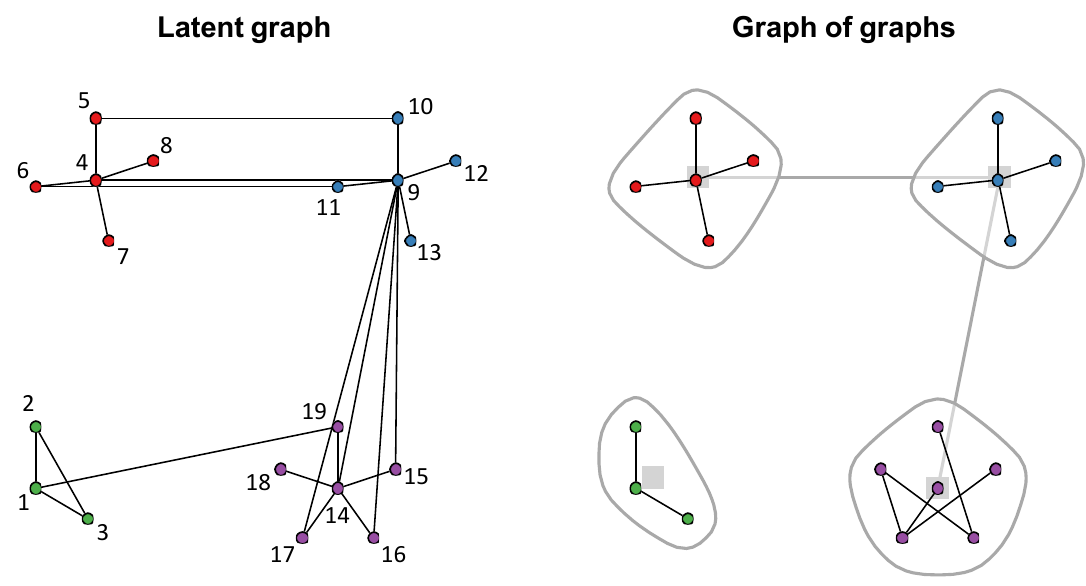}
\vspace{-1.7cm}
\caption{
Latent graph used to simulate data (left) and graph of graphs estimate (right).
The circles represent nodes which
are colored according to the block structure of the latent graph.
In the right plot,
nodes are connected by the within-supernode graphs (trees) in black.
Trees are encircled in gray to mark the
supernodes.
Grey lines identify superedges between supernodes
with a posterior inclusion probability greater than $0.5$.
\vspace{0.3cm}
}
\label{fig:ggm_to_GoG}
\end{figure}

To highlight how the graph of graphs differs from standard GGMs,
we simulate
the data matrix $X$ with $n=1000$ observations by sampling its rows independently from $\mathcal{N}(0_{p\times 1},\,\Psi^{-1})$
where the sparsity pattern in $\Psi$ corresponds to a latent graph (see Section~\ref{sec:ggm}).
For ease of exposition, we consider 
only $p=19$ nodes and we specify a graph where the nodes are subdivided into four densely connected blocks consisting of (i) 3, (ii) 5, (iii) 5 and (iv) 6 nodes.
The graph and its block structure are visualized in Figure~\ref{fig:ggm_to_GoG}.
We set $\Psi$ given the graph as follows. For the elements of $\Psi$ corresponding to edges within blocks, we have: $\Psi_{i,j} = \Psi_{j,i}
    = 0.1 / \sqrt{2}$, $1\leq i< j \leq 3$; $\Psi_{4,i} = \Psi_{i,4} = 0.1 / \sqrt{4}$,  $5\leq i\leq 8$ ;
    $\Psi_{9,i} = \Psi_{i,9} = 0.1 / \sqrt{4}$,  $10\leq i\leq 13$ ;
    $\Psi_{14,i} = \Psi_{i,14} = 0.1 / \sqrt{5}$,  $15\leq i\leq 19$.
For the elements corresponding to edges between blocks,
we specify: $
\Psi_{1,19} = \Psi_{19,1}
    = -0.2 ;
    \Psi_{i,5+i} = \Psi_{5+i,i}
    = -0.6,  4\leq i \leq 6 ;
\Psi_{9,i} = \Psi_{i,9}
    = -0.6,  14\leq i \leq 17.$
Finally, all diagonal elements are equal to one and all other elements are equal to zero.

For ease of visualization, we infer a graph of graphs while keeping the tessellation $\mathcal{T}$ fixed to the block structure of the latent graph used to simulate the data.
For the supergraph,
we draw from the posterior $p(G\mid \mathcal{T}, X)$
using the MCMC methodology for GGMs from \citet{vandenBoom2022a}.
We identify two superedges with a posterior inclusion probability greater than $0.5$.
The corresponding supergraph is visualized in Figure~\ref{fig:ggm_to_GoG}.
The trees $T_k$ shown are those that maximize $\widetilde{p}(T_k\mid x_k)$ in the tree activation functions, which we compute
using a maximum spanning tree algorithm \citep{Schwaller2019}.
The estimate of the graph of graphs summarizes the latent graph effectively:
superedges are assigned to the pairs of blocks in the latent graph that are connected by three or four edges, while pairs of blocks connected by no or one edge are not connected with a superedge.
The trees within the supernodes mostly match the within-block edges in the latent graph.

\section{Application to gene expression data}
\label{sec:application}

In online Supplementary Material~\ref{ap:simul},
we present simulation studies to investigate the performance of the graph of graphs model and show that it outperforms two-step approaches in terms of recovery of the partition of nodes.
Here, we apply our methodology to
gene expression levels,
the interactions between which are often represented as networks.
An important concept in the gene network literature is that of module, which is a densely connected subgraph of genes
with similar expression profiles \citep{Zhang2018}. Such genes are typically co-regulated and functionally related \citep{Saelens2018}.
Therefore,
a critical step in the analysis of large gene expression data sets is module detection to group genes into co-expression clusters.
The graph of graphs model treats each module as a supernode and has intrinsic advantages, when learning the supernode/module membership from data. On the other hand,
in the gene network literature,
typically, a two-step approach is adopted. First, the graph is estimated from the gene expressions, and then the modules are derived from the graph estimate \citep[see, e.g.,][]{Saelens2018,Zhang2018}.
Such an approach underestimates uncertainty, often leading to false positives, and does not capture the mesoscopic dependency structure between modules.

We analyze data on gene expressions
from $n=561$ ovarian cancer tissue samples
from
The Cancer Genome Atlas.
We focus on $p=373$ genes
identified in Table~2 of \citet{Zhang2018} as
spread across six estimated modules,
which are highly enriched in terms of Gene Ontology \citep[GO,][]{Ashburner2000} annotations.
The gene expressions are quantile-normalized to marginally follow a standard Gaussian distribution.
We apply the proposed graph of graphs methodology with the following prior specification.
As in \citet{Natarajan2024}, we choose a shifted Negative Binomial distribution started at $p_k=1$ for cohesion function $f_\textnormal{coh.}(p_k)$ in the tessellation prior.
The success probability is set equal to  $1/6$ and the size parameter to $2$, leading to an expected supernode size equal to 10 based on an initial exploratory data analysis.
Hyperparameters for the $G$-Wishart distribution
$\widetilde{p}(\Delta_k\mid T_k)$ in the tree activation function and the $G$-Wishart prior $p(\Omega\mid G)$ for the supergraph are set to the standard values $\delta=3$, $D=I_{p_k}$, $\delta_G=3$ and $D_G = I_p$.
We specify an Erd\"{o}s-R\'{e}nyi prior for the supergraph,
i.e.\
$p(G\mid \mathcal{T}) \propto \xi_{\text{se}}^{\mid E^\star \mid} \left( 1 - \xi_{\text{se}}\right)^{\binom{K}{2} - \mid E^\star \mid}$, with a priori superedge inclusion probability $\xi_{\text{se}} = 0.1$. This prior induces sparsity on the inferred graph $G$.
Finally, the coarsening parameter $\zeta$ is chosen small enough ($\zeta = 10 / n$) to allow for good MCMC mixing without flattening the target distribution excessively.
We run both  MCMC algorithms for $50000$ iterations, discarding the first $40000$ as burn-in and using a thinning of 10 on the remaining iterations, yielding 1000 samples to be used for posterior inference and in the inner part of the nested MCMC.

The inference on the tessellation (see online Supplementary Material~\ref{ap:gene} and Figure~\ref{fig:gene_sphere}) reveals a more refined grouping of genes into supernodes than the modules estimated
using a two-step approach\label{add:2step}
by \cite{Zhang2018}, but the partition of genes is otherwise similar, both with coarsening of the likelihood and with nested MCMC.
Here, we report the tessellation that minimizes the lower bound to the posterior expectation of the variation of information \citep{Wade2018}.
This results in $K=30$ and $22$ supernodes with coarsening of the likelihood and nested MCMC, respectively.
We note the difference in the number of supernodes obtained with the two algorithms. Recall that the nested MCMC does not allow for information transfer from the supergraph to the tessellation. The supergraph likelihood is obviously favoring more structure in the data.
Still, the two estimated partitions are very similar, yielding a rand index \citep{Rand1971} equal to 0.945.
We visualize the resulting graph of graphs in Figure~\ref{fig:gene_sphere}
(in Section~\ref{sec:intro})
only for nested MCMC,
as its smaller $K$
makes for easier exposition
than the larger $K$ with the coarsened likelihood.
The trees $T_k$ shown are
global maxima of \label{add:global_maxima}
$\widetilde{p}(T_k\mid x_k)$ in the tree activation functions
obtained using a maximum spanning tree algorithm \citep{Schwaller2019}.
For comparison,
consider Module~14 from \citet{Zhang2018},
which is the only module that we include for which \citet{Zhang2018} reports edge estimates.
All but one edge between genes from Module~14 in Figure~\ref{fig:gene_sphere}
are also inferred by \citet{Zhang2018}, which suggests that the inference on trees is appropriate.
For the supergraph,
we draw from the posterior $p(G\mid \mathcal{T}, X)$
with the tessellation $\mathcal{T}$ fixed to the point estimate
using the MCMC methodology for GGMs from \citet{vandenBoom2022a},
resulting in 84 superedges (approximately one third of the possible edges) with a posterior inclusion probability greater than $0.5$.
For visualization purposes, we include fewer superedges in Figure~\ref{fig:gene_sphere}, i.e. only those with a posterior probability greater than 0.99 (64 in total).
Again, there is consistency with the modules estimated by \citet{Zhang2018}:
each supernode corresponds to a single module as the vast majority of its nodes, if not all, come from the same module.
Note that
44.78\% of all pairs of supernodes corresponding to the same module
are connected by a superedge in Figure~\ref{fig:gene_sphere}.
This drops to 20.73\% for pairs corresponding to different modules.
In online Supplementary Material~\ref{ap:gene}, we describe how the presence of a superedge is associated with more interactions as derived from the STRING database \citep{Szklarczyk2021} between pairs of genes involved in the two linked supernodes, suggesting biological meaning of the superedge inference (see Supplementary Figure~\ref{fig:gene_interactions}).
Finally, such interactions are more prevalent within supernodes than between, which indicates that our more granular partition of nodes compared to \citet{Zhang2018} is justified.

To further inspect the tessellations,
we perform
GO overrepresentation analysis in online Supplementary Material~\ref{ap:gene}.
Such analysis detects GO terms that appear relatively more frequently among the genes in a supernode than among all 373 genes.
We summarize
the results in Figure~\ref{fig:GO} in Supplementary Material.
We find that
different supernodes are generally associated with distinct biological processes (i.e.\ GO terms),
suggesting that our model can capture underlying biological mechanisms.

\section{Discussion}
\label{sec:discussion}

In this work, we develop a hierarchical graphical model, the graph of graphs, clustering nodes into supernodes with superedges connecting them at a mesoscopic level. This structure improves statistical inference, scalability, and interpretability beyond individual node connections. We use a data-coherent size-biased tessellation prior for node grouping and a GGM on the supernodes to specify a likelihood over supergraphs. The model includes supernode-specific PCA and requires advanced nonstandard inference tools, such as coarsening likelihood terms and cutting feedback via nested MCMC. We provide theoretical justification for our modeling choices such as the use of supernode-specific PCA.
We demonstrate the model with gene expression data, yielding biologically relevant conclusions. Our model can be extended to alternative similarity functions, different clustering priors like the Dirichlet process, and even overlapping supernodes.
Such overlap
might, for instance, be desirable when inferring modules of genes involved in multiple pathways simultaneously \citep{Saelens2018}.
The proposed approach 
is applicable in various domains beyond genomics.

\bigskip
\begin{center}
{\large\bf SUPPLEMENTARY MATERIAL}
\end{center}

\begin{description}
\item[Supplement:] Simulation studies, overview of notation and related work, further material on the size-biased prior and the gene expression application, proofs of the propositions, and details of the MCMC algorithms. (.pdf file)
\item[Code:] The code to implement the model is available at\\
\if0\blind{\url{https://github.com/willemvandenboom/graph-of-graphs}}\fi
\if1\blind{\url{https://anonymous.4open.science/r/graph-of-graphs}}\fi
. (GitHub repository)
\end{description}

\bigskip
\pagebreak    
\begin{center}
{\large\bf ACKNOWLEDGEMENTS}
\end{center}

The data used in Section~\ref{sec:application}
are generated by The Cancer Genome Atlas Research Network: \url{https://www.cancer.gov/tcga}.

\bibliographystyle{apalike_modified.bst}
\bibliography{graph}

\begin{thebibliography}{}

\bibitem[Ali et~al., 2020]{Ali2020}
Ali, P., Atik, F., \& Bapat, R. (2020).
\newblock Identities for minors of the {L}aplacian, resistance and distance
  matrices of graphs with arbitrary weights.
\newblock {\em Linear and Multilinear Algebra}, 68(2):323--336.

\bibitem[Ashburner et~al., 2000]{Ashburner2000}
Ashburner, M., Ball, C., Blake, J., Botstein, D., Butler, H., Cherry, J.,
  et~al. (2000).
\newblock {G}ene {O}ntology: tool for the unification of biology.
\newblock {\em Nature Genetics}, 25(1):25--29.

\bibitem[Atay-Kayis \& Massam, 2005]{AtayKayis2005}
Atay-Kayis, A. \& Massam, H. (2005).
\newblock A {M}onte {C}arlo method for computing the marginal likelihood in
  nondecomposable {G}aussian graphical models.
\newblock {\em Biometrika}, 92(2):317--335.

\bibitem[Bapat, 2004]{Bapat2004}
Bapat, R. (2004).
\newblock Resistance matrix of a weighted graph.
\newblock {\em MATCH Communications in Mathematical and in Computer Chemistry},
  50:73--82.

\bibitem[Barab{\'a}si, 2012]{Barabasi2012}
Barab{\'a}si, A.-L. (2012).
\newblock The network takeover.
\newblock {\em Nature Physics}, 8(1):14--16.

\bibitem[Barabási, 2016]{Barabasi2016}
Barabási, A.-L. (2016).
\newblock {\em Network Science}.
\newblock Cambridge University Press, Cambridge, England.

\bibitem[Barcella et~al., 2017]{Barcella2017}
Barcella, W., De~Iorio, M., \& Baio, G. (2017).
\newblock A comparative review of variable selection techniques for covariate
  dependent {D}irichlet process mixture models.
\newblock {\em Canadian Journal of Statistics}, 45(3):254--273.

\bibitem[Betancourt et~al., 2022]{Betancourt2022}
Betancourt, B., Zanella, G., \& Steorts, R. (2022).
\newblock Random partition models for microclustering tasks.
\newblock {\em Journal of the American Statistical Association},
  117(539):1215--1227.

\bibitem[Bhadra et~al., 2019]{Bhadra2019}
Bhadra, A., Datta, J., Polson, N., \& Willard, B. (2019).
\newblock Lasso meets horseshoe: a survey.
\newblock {\em Statistical Science}, 34(3):405--427.

\bibitem[Bock \& Gibbons, 2021]{Bock2021}
Bock, D. \& Gibbons, R. (2021).
\newblock {\em Item Response Theory}.
\newblock Wiley, Hoboken, NJ.

\bibitem[Carlin \& Chib, 1995]{Carlin1995}
Carlin, B. \& Chib, S. (1995).
\newblock Bayesian model choice via {M}arkov chain {M}onte {C}arlo methods.
\newblock {\em Journal of the Royal Statistical Society: Series B
  (Methodological)}, 57(3):473--484.

\bibitem[Carmona \& Nicholls, 2020]{Carmona2020}
Carmona, C. \& Nicholls, G. (2020).
\newblock Semi-modular inference: enhanced learning in multi-modular models by
  tempering the influence of components.
\newblock {\em Proceedings of the Twenty Third International Conference on
  Artificial Intelligence and Statistics}, 4226--4235. PMLR.

\bibitem[Carvalho \& Ibrahim, 2021]{Carvalho2021}
Carvalho, L. \& Ibrahim, J. (2021).
\newblock On the normalized power prior.
\newblock {\em Statistics in Medicine}, 40(24):5251--5275.

\bibitem[Cayley, 1889]{Cayley1889}
Cayley, A. (1889).
\newblock A theorem on trees.
\newblock {\em The Quarterly Journal of Pure and Applied Mathematics},
  23:376--378.

\bibitem[Chandra et~al., 2023]{Chandra2023}
Chandra, N., Canale, A., \& Dunson, D. (2023).
\newblock Escaping the curse of dimensionality in {B}ayesian model-based
  clustering.
\newblock {\em Journal of Machine Learning Research}, 24(144).

\bibitem[Chen et~al., 2023]{Chen2023}
Chen, J., Ng, Y.~K., Lin, L., Zhang, X., \& Li, S. (2023).
\newblock On triangle inequalities of correlation-based distances for gene
  expression profiles.
\newblock {\em {BMC} Bioinformatics}, 24:40.

\bibitem[Chen \& Ibrahim, 2000]{Chen2000}
Chen, M.-H. \& Ibrahim, J. (2000).
\newblock Power prior distributions for regression models.
\newblock {\em Statistical Science}, 15(1):46--60.

\bibitem[{De Blasi} et~al., 2015]{DeBlasi2015}
{De Blasi}, P., Favaro, S., Lijoi, A., Mena, R., Prunster, I., \& Ruggiero, M.
  (2015).
\newblock Are {G}ibbs-type priors the most natural generalization of the
  {D}irichlet process?
\newblock {\em {IEEE} Transactions on Pattern Analysis and Machine
  Intelligence}, 37(2):212--229.

\bibitem[Dempster, 1972]{Dempster1972}
Dempster, A. (1972).
\newblock Covariance selection.
\newblock {\em Biometrics}, 28(1):157--175.

\bibitem[Denison et~al., 2002a]{Denison2002}
Denison, D., Adams, N., Holmes, C., \& Hand, D. (2002a).
\newblock Bayesian partition modelling.
\newblock {\em Computational Statistics \& Data Analysis}, 38(4):475--485.

\bibitem[Denison et~al., 2002b]{Denison2002b}
Denison, D., Holmes, C., Mallick, B., \& Smith, A. (2002b).
\newblock {\em Bayesian methods for nonlinear classification and regression}.
\newblock Wiley Series in Probability and Statistics. John Wiley \& Sons,
  Chichester, UK.

\bibitem[Drovandi et~al., 2015]{Drovandi2015}
Drovandi, C., Pettitt, A., \& Lee, A. (2015).
\newblock Bayesian indirect inference using a parametric auxiliary model.
\newblock {\em Statistical Science}, 30(1):72--95.

\bibitem[Duan \& Dunson, 2023]{Duan2023}
Duan, L. \& Dunson, D. (2023).
\newblock Bayesian spanning tree: Estimating the backbone of the dependence
  graph.
\newblock {\em Journal of Machine Learning Research}, 24:397.

\bibitem[Fienberg, 2012]{Fienberg2012}
Fienberg, S. (2012).
\newblock A brief history of statistical models for network analysis and open
  challenges.
\newblock {\em Journal of Computational and Graphical Statistics},
  21(4):825--839.

\bibitem[Green, 1995]{Green1995}
Green, P. (1995).
\newblock Reversible jump {M}arkov chain {M}onte {C}arlo computation and
  {B}ayesian model determination.
\newblock {\em Biometrika}, 82(4):711--732.

\bibitem[Gr{\"u}nwald \& van Ommen, 2017]{Grunwald2017}
Gr{\"u}nwald, P. \& van Ommen, T. (2017).
\newblock Inconsistency of {B}ayesian inference for misspecified linear models,
  and a proposal for repairing it.
\newblock {\em Bayesian Analysis}, 12(4):1069--1103.

\bibitem[Hartigan, 1990]{Hartigan1990}
Hartigan, J. (1990).
\newblock Partition models.
\newblock {\em Communications in Statistics - Theory and Methods},
  19(8):2745--2756.

\bibitem[I{\~{n}}iguez et~al., 2020]{Iniguez2020}
I{\~{n}}iguez, G., Battiston, F., \& Karsai, M. (2020).
\newblock Bridging the gap between graphs and networks.
\newblock {\em Communications Physics}, 3:88.

\bibitem[Jacob et~al., 2017]{Jacob2017}
Jacob, P., Murray, L., Holmes, C., \& Robert, C. (2017).
\newblock Better together? {S}tatistical learning in models made of modules.
\newblock arXiv:1708.08719v1.

\bibitem[Jiang \& Tanner, 2008]{Jiang2008}
Jiang, W. \& Tanner, M. (2008).
\newblock Gibbs posterior for variable selection in high-dimensional
  classification and data mining.
\newblock {\em The Annals of Statistics}, 36(5):2207--2231.

\bibitem[Kirshner, 2007]{Kirshner2007}
Kirshner, S. (2007).
\newblock Learning with tree-averaged densities and distributions.
\newblock {\em Advances in Neural Information Processing Systems}. Curran
  Associates, Inc.

\bibitem[Knudson \& Lindsey, 2014]{Knudson2014}
Knudson, D. \& Lindsey, C. (2014).
\newblock Type {I} and type {II} errors in correlations of various sample
  sizes.
\newblock {\em Comprehensive Psychology}, 3:1.

\bibitem[Lauritzen, 1996]{Lauritzen1996}
Lauritzen, S. (1996).
\newblock {\em Graphical Models}.
\newblock Oxford Statistical Science Series. The Clarendon Press, Oxford.

\bibitem[Lewis et~al., 2021]{Lewis2021}
Lewis, J., MacEachern, S., \& Lee, Y. (2021).
\newblock Bayesian restricted likelihood methods: conditioning on insufficient
  statistics in {B}ayesian regression (with discussion).
\newblock {\em Bayesian Analysis}, 16(4):1393--1462.

\bibitem[Liu et~al., 2021]{Liu2021}
Liu, C., Yang, Y., Bondell, H., \& Martin, R. (2021).
\newblock Bayesian inference in high-dimensional linear models using an
  empirical correlation-adaptive prior.
\newblock {\em Statistica Sinica}, 31:2051--2072.

\bibitem[Liu et~al., 2009]{Liu2009}
Liu, F., Bayarri, M., \& Berger, J. (2009).
\newblock Modularization in {B}ayesian analysis, with emphasis on analysis of
  computer models.
\newblock {\em Bayesian Analysis}, 4(1):119--150.

\bibitem[Malevergne \& Sornette, 2004]{Malevergne2004}
Malevergne, Y. \& Sornette, D. (2004).
\newblock Collective origin of the coexistence of apparent random matrix theory
  noise and of factors in large sample correlation matrices.
\newblock {\em Physica A: Statistical Mechanics and its Applications},
  331(3--4):660--668.

\bibitem[Martin et~al., 2017]{Martin2017}
Martin, R., Mess, R., \& Walker, S. (2017).
\newblock Empirical {B}ayes posterior concentration in sparse high-dimensional
  linear models.
\newblock {\em Bernoulli}, 23(3):1822--1847.

\bibitem[Martin \& Walker, 2019]{Martin2019}
Martin, R. \& Walker, S. (2019).
\newblock Data-driven priors and their posterior concentration rates.
\newblock {\em Electronic Journal of Statistics}, 13(2):3049--3081.

\bibitem[Meilă \& Jaakkola, 2006]{Meilă2006}
Meilă, M. \& Jaakkola, T. (2006).
\newblock Tractable {B}ayesian learning of tree belief networks.
\newblock {\em Statistics and Computing}, 16(1):77--92.

\bibitem[Meyer, 1975]{Meyer1975}
Meyer, E. (1975).
\newblock A measure of the average intercorrelation.
\newblock {\em Educational and Psychological Measurement}, 35(1):67--72.

\bibitem[Miller \& Dunson, 2019]{Miller2019}
Miller, J. \& Dunson, D. (2019).
\newblock Robust {B}ayesian inference via coarsening.
\newblock {\em Journal of the American Statistical Association},
  114(527):1113--1125.

\bibitem[Momal et~al., 2021]{Momal2021}
Momal, R., Robin, S., \& Ambroise, C. (2021).
\newblock Accounting for missing actors in interaction network inference from
  abundance data.
\newblock {\em Journal of the Royal Statistical Society Series C: Applied
  Statistics}, 70(5):1230--1258.

\bibitem[M{\"u}ller et~al., 2011]{Muller2011}
M{\"u}ller, P., Quintana, F., \& Rosner, G. (2011).
\newblock A product partition model with regression on covariates.
\newblock {\em Journal of Computational and Graphical Statistics},
  20(1):260--278.

\bibitem[Natarajan et~al., 2024]{Natarajan2024}
Natarajan, A., De~Iorio, M., Heinecke, A., Mayer, E., \& Glenn, S. (2024).
\newblock Cohesion and repulsion in {B}ayesian distance clustering.
\newblock {\em Journal of the American Statistical Association},
  119(546):1374--1384.

\bibitem[Newman, 2012]{Newman2012}
Newman, M. (2012).
\newblock Communities, modules and large-scale structure in networks.
\newblock {\em Nature Physics}, 8(1):25--31.

\bibitem[Page \& Quintana, 2018]{Page2018}
Page, G. \& Quintana, F. (2018).
\newblock Calibrating covariate informed product partition models.
\newblock {\em Statistics and Computing}, 28(5):1009--1031.

\bibitem[Pedone et~al., 2024]{Pedone2024}
Pedone, M., Argiento, R., \& Stingo, F. (2024).
\newblock Personalized treatment selection via product partition models with
  covariates.
\newblock {\em Biometrics}, 80(1):ujad003.

\bibitem[Peixoto, 2019]{Peixoto2019}
Peixoto, T. (2019).
\newblock Network reconstruction and community detection from dynamics.
\newblock {\em Physical Review Letters}, 123(12):128301.

\bibitem[Plummer, 2015]{Plummer2015}
Plummer, M. (2015).
\newblock Cuts in {B}ayesian graphical models.
\newblock {\em Statistics and Computing}, 25(1):37--43.

\bibitem[Pratt, 1965]{Pratt1965}
Pratt, J. (1965).
\newblock Bayesian interpretation of standard inference statements.
\newblock {\em Journal of the Royal Statistical Society: Series B
  (Methodological)}, 27(2):169--192.

\bibitem[Rand, 1971]{Rand1971}
Rand, W. (1971).
\newblock Objective criteria for the evaluation of clustering methods.
\newblock {\em Journal of the American Statistical Association},
  66(336):846--850.

\bibitem[Ravasz et~al., 2002]{Ravasz2002}
Ravasz, E., Somera, A., Mongru, D., Oltvai, Z., \& Barabási, A.-L. (2002).
\newblock Hierarchical organization of modularity in metabolic networks.
\newblock {\em Science}, 297:1551--1555.

\bibitem[{Rodrigues Jr.} et~al., 2006]{RodriguesJr2006}
{Rodrigues Jr.}, J., Traina, A., Faloutsos, C., \& {Traina Jr.}, C. (2006).
\newblock {SuperGraph} visualization.
\newblock {\em Eighth {IEEE} International Symposium on Multimedia
  ({ISM}{\textquotesingle}06)}. {IEEE}.

\bibitem[Roverato, 2000]{rove:00}
Roverato, A. (2000).
\newblock Cholesky decomposition of a hyper inverse {W}ishart matrix.
\newblock {\em Biometrika}, 87(1):99--112.

\bibitem[Roverato, 2002]{Roverato2002}
Roverato, A. (2002).
\newblock Hyper inverse {W}ishart distribution for non-decomposable graphs and
  its application to {B}ayesian inference for {G}aussian graphical models.
\newblock {\em Scandinavian Journal of Statistics}, 29(3):391--411.

\bibitem[Royle et~al., 2007]{Royle2007}
Royle, A., Dorazio, R., \& Link, W. (2007).
\newblock Analysis of multinomial models with unknown index using data
  augmentation.
\newblock {\em Journal of Computational and Graphical Statistics},
  16(1):67--85.

\bibitem[Saelens et~al., 2018]{Saelens2018}
Saelens, W., Cannoodt, R., \& Saeys, Y. (2018).
\newblock A comprehensive evaluation of module detection methods for gene
  expression data.
\newblock {\em Nature Communications}, 9:1090.

\bibitem[Schwaller et~al., 2019]{Schwaller2019}
Schwaller, L., Robin, S., \& Stumpf, M. (2019).
\newblock Closed-form {B}ayesian inference of graphical model structures by
  averaging over trees.
\newblock {\em Journal de la Société Française de Statistique},
  160(2):1--23.

\bibitem[Sporns \& Betzel, 2016]{Sporns2016}
Sporns, O. \& Betzel, R. (2016).
\newblock Modular brain networks.
\newblock {\em Annual Review of Psychology}, 67:613--640.

\bibitem[Stepanov et~al., 2021]{Stepanov2021}
Stepanov, Y., Herrmann, H., \& Guhr, T. (2021).
\newblock Generic features in the spectral decomposition of correlation
  matrices.
\newblock {\em Journal of Mathematical Physics}, 62(8):083505.

\bibitem[Szklarczyk et~al., 2021]{Szklarczyk2021}
Szklarczyk, D., Gable, A., Nastou, K., Lyon, D., Kirsch, R., Pyysalo, S.,
  et~al. (2021).
\newblock The {STRING} database in 2021: customizable protein–protein
  networks, and functional characterization of user-uploaded gene/measurement
  sets.
\newblock {\em Nucleic Acids Research}, 49(D1):D605--D612.

\bibitem[van~den Boom et~al., 2022]{vandenBoom2022a}
van~den Boom, W., Beskos, A., \& {De Iorio}, M. (2022).
\newblock The \mbox{$G$-Wishart} weighted proposal algorithm: efficient
  posterior computation for {G}aussian graphical models.
\newblock {\em Journal of Computational and Graphical Statistics},
  31(4):1215--1224.

\bibitem[van~den Boom et~al., 2023]{vandenBoom2023}
van~den Boom, W., De~Iorio, M., \& Beskos, A. (2023).
\newblock Bayesian learning of graph substructures.
\newblock {\em Bayesian Analysis}, 18(4):1311--1339.

\bibitem[Wade et~al., 2014]{Wade2014}
Wade, S., Dunson, D., Petrone, S., \& Trippa, L. (2014).
\newblock Improving prediction from {D}irichlet process mixtures via
  enrichment.
\newblock {\em Journal of Machine Learning Research}, 15(30):1041--1071.

\bibitem[Wade \& Ghahramani, 2018]{Wade2018}
Wade, S. \& Ghahramani, Z. (2018).
\newblock Bayesian cluster analysis: point estimation and credible balls (with
  discussion).
\newblock {\em Bayesian Analysis}, 13(2):559--626.

\bibitem[Walker, 2007]{Walker2007}
Walker, S. (2007).
\newblock Sampling the {D}irichlet mixture model with slices.
\newblock {\em Communications in Statistics - Simulation and Computation},
  36(1):45--54.

\bibitem[Whiteley et~al., 2022]{Whiteley2022}
Whiteley, N., Gray, A., \& Rubin-Delanchy, P. (2022).
\newblock Discovering latent topology and geometry in data: a law of large
  dimension.
\newblock arXiv:2208.11665v2.

\bibitem[Yook et~al., 2004]{Yook2004}
Yook, S.-H., Oltvai, Z., \& Barab{\'{a}}si, A.-L. (2004).
\newblock Functional and topological characterization of protein interaction
  networks.
\newblock {\em Proteomics}, 4(4):928--942.

\bibitem[Zhang, 2018]{Zhang2018}
Zhang, S. (2018).
\newblock Comparisons of gene coexpression network modules in breast cancer and
  ovarian cancer.
\newblock {\em {BMC} Systems Biology}, 12(S1):57--87.

\end{thebibliography}


\begin{thebibliography}{}

\bibitem[Amini et~al., 2024]{Amini2024}
Amini, A., Paez, M., \& Lin, L. (2024).
\newblock Hierarchical stochastic block model for community detection in
  multiplex networks.
\newblock {\em Bayesian Analysis}, 19(1):319--345.

\bibitem[Arora et~al., 2012]{Arora2012}
Arora, S., Ge, R., \& Moitra, A. (2012).
\newblock Learning topic models -- going beyond {SVD}.
\newblock {\em 2012 {IEEE} 53rd Annual Symposium on Foundations of Computer
  Science}.

\bibitem[Ashburner et~al., 2000]{Ashburner2000}
Ashburner, M., Ball, C., Blake, J., Botstein, D., Butler, H., Cherry, J.,
  et~al. (2000).
\newblock {G}ene {O}ntology: tool for the unification of biology.
\newblock {\em Nature Genetics}, 25(1):25--29.

\bibitem[Betancourt et~al., 2022]{Betancourt2022}
Betancourt, B., Zanella, G., \& Steorts, R. (2022).
\newblock Random partition models for microclustering tasks.
\newblock {\em Journal of the American Statistical Association},
  117(539):1215--1227.

\bibitem[Bhattacharyya et~al., 2024]{Bhattacharyya2024}
Bhattacharyya, A., Gayen, S., John, P., Sen, S., \& Vinodchandran, N. (2024).
\newblock Distribution learning meets graph structure sampling.
\newblock arXiv:2405.07914v1.

\bibitem[Bing et~al., 2020]{Bing2020}
Bing, X., Bunea, F., Ning, Y., \& Wegkamp, M. (2020).
\newblock Adaptive estimation in structured factor models with applications to
  overlapping clustering.
\newblock {\em The Annals of Statistics}, 48(4):2055--2081.

\bibitem[Carmona \& Nicholls, 2020]{Carmona2020}
Carmona, C. \& Nicholls, G. (2020).
\newblock Semi-modular inference: enhanced learning in multi-modular models by
  tempering the influence of components.
\newblock {\em Proceedings of the Twenty Third International Conference on
  Artificial Intelligence and Statistics}, 4226--4235. PMLR.

\bibitem[Castelletti \& Mascaro, 2022]{Castelletti2022}
Castelletti, F. \& Mascaro, A. (2022).
\newblock {BCDAG}: an {R} package for {B}ayesian structure and causal learning
  of {G}aussian {DAG}s.
\newblock arXiv:2201.12003v1.

\bibitem[Chandra et~al., 2022]{Chandra2022}
Chandra, N., Müller, P., \& Sarkar, A. (2022).
\newblock Bayesian scalable precision factor analysis for massive sparse
  {G}aussian graphical models.
\newblock {arXiv:2107.11316v4}.

\bibitem[Chen et~al., 2022]{Chen2022}
Chen, J., Saad, Y., \& Zhang, Z. (2022).
\newblock Graph coarsening: from scientific computing to machine learning.
\newblock {\em {SeMA} Journal}, 79(1):187--223.

\bibitem[Cheng et~al., 2017]{Cheng2017}
Cheng, L., Shan, L., \& Kim, I. (2017).
\newblock Multilevel {G}aussian graphical model for multilevel networks.
\newblock {\em Journal of Statistical Planning and Inference}, 190:1--14.

\bibitem[Colombi et~al., 2024]{Colombi2024}
Colombi, A., Argiento, R., Paci, L., \& Pini, A. (2024).
\newblock Learning block structured graphs in {G}aussian graphical models.
\newblock {\em Journal of Computational and Graphical Statistics},
  33(1):152--165.

\bibitem[Cremaschi et~al., 2023]{Cremaschi2023}
Cremaschi, A., Argiento, R., De~Iorio, M., Cai, S., Chong, Y., Meaney, M., \&
  Kee, M. (2023).
\newblock Seemingly unrelated multi-state processes: a {B}ayesian
  semiparametric approach.
\newblock {\em Bayesian Analysis}, 18(3):753--775.

\bibitem[Csárdi et~al., 2023]{Csardi2023}
Csárdi, G., Nepusz, T., Traag, V., Horvát, S., Zanini, F., Noom, D., \&
  Müller, K. (2023).
\newblock {\em {igraph}: Network Analysis and Visualization in R}.
\newblock R package version 1.6.0.

\bibitem[D{\textquotesingle}Agostino \& Scala, 2014]{DAgostino2014}
D{\textquotesingle}Agostino, G. \& Scala, A., editors (2014).
\newblock {\em Networks of Networks: The Last Frontier of Complexity}.
\newblock Understanding Complex Systems. Springer Cham, Heidelberg.

\bibitem[De~Leenheer, 2020]{DeLeenheer2020}
De~Leenheer, P. (2020).
\newblock An elementary proof of a matrix tree theorem for directed graphs.
\newblock {\em {SIAM} Review}, 62(3):716--726.

\bibitem[Deo, 1974]{Deo1974}
Deo, N. (1974).
\newblock {\em Graph Theory with Applications to Engineering and Computer
  Science}.
\newblock Prentice-Hall Series in Automatic Computation. Prentice-Hall,
  Englewood Cliffs, NJ.

\bibitem[Dickey, 1967]{Dickey1967}
Dickey, J. (1967).
\newblock Matricvariate generalizations of the multivariate $t$ distribution
  and the inverted multivariate $t$ distribution.
\newblock {\em The Annals of Mathematical Statistics}, 38(2):511--518.

\bibitem[Duan \& Dunson, 2023]{Duan2023}
Duan, L. \& Dunson, D. (2023).
\newblock Bayesian spanning tree: Estimating the backbone of the dependence
  graph.
\newblock {\em Journal of Machine Learning Research}, 24:397.

\bibitem[Ferreira \& Lee, 2007]{Ferreira2007}
Ferreira, M. \& Lee, H. (2007).
\newblock {\em Multiscale Modeling}.
\newblock Springer, New York.

\bibitem[Fosdick et~al., 2019]{Fosdick2019}
Fosdick, B., McCormick, T., Murphy, T., Ng, T., \& Westling, T. (2019).
\newblock Multiresolution network models.
\newblock {\em Journal of Computational and Graphical Statistics},
  28(1):185--196.

\bibitem[Friedman et~al., 2007]{Friedman2007}
Friedman, J., Hastie, T., \& Tibshirani, R. (2007).
\newblock Sparse inverse covariance estimation with the graphical lasso.
\newblock {\em Biostatistics}, 9(3):432--441.

\bibitem[Galloway, 2018]{Galloway2018}
Galloway, M. (2018).
\newblock {\em {CVglasso}: Lasso Penalized Precision Matrix Estimation}.
\newblock R package version 1.0.
  \url{https://CRAN.R-project.org/package=CVglasso}.

\bibitem[Jin et~al., 2021]{Ha2021}
Jin, M., Stingo, F., \& Baladandayuthapani, V. (2021).
\newblock Bayesian structure learning in multilayered genomic networks.
\newblock {\em Journal of the American Statistical Association},
  116(534):605--618.

\bibitem[Jin et~al., 2022]{Jin2022}
Jin, W., Zhao, L., Zhang, S., Liu, Y., Tang, J., \& Shah, N. (2022).
\newblock Graph condensation for graph neural networks.
\newblock {\em The Tenth International Conference on Learning Representations}.

\bibitem[Josephs et~al., 2023]{Josephs2023}
Josephs, N., Amini, A., Paez, M., \& Lin, L. (2023).
\newblock Nested stochastic block model for simultaneously clustering networks
  and nodes.
\newblock {arXiv:2307.09210v1}.

\bibitem[Kim \& Kim, 2020]{Kim2020}
Kim, G. \& Kim, S. (2020).
\newblock Multi-level {G}aussian graphical models conditional on covariates.
\newblock {\em Proceedings of the Twenty Third International Conference on
  Artificial Intelligence and Statistics}, 4216--4225. PMLR.

\bibitem[Kim et~al., 2023]{Kim2023}
Kim, H., Ghosh, S., \& Hector, E. (2023).
\newblock Bayesian estimation of clustered dependence structures in functional
  neuroconnectivity.
\newblock arXiv:2305.18044v1.

\bibitem[Korte \& Vygen, 2002]{Korte2002}
Korte, B. \& Vygen, J. (2002).
\newblock {\em Combinatorial Optimization. Theory and Algorithms}.
\newblock Algorithms and Combinatorics. Springer, Berlin, 2nd ed.

\bibitem[Kuipers et~al., 2014]{Kuipers2014}
Kuipers, J., Moffa, G., \& Heckerman, D. (2014).
\newblock Addendum on the scoring of {G}aussian directed acyclic graphical
  models.
\newblock {\em The Annals of Statistics}, 42(4):1689--1691.

\bibitem[Kumar et~al., 2023]{Kumar2023}
Kumar, M., Sharma, A., \& Kumar, S. (2023).
\newblock A unified framework for optimization-based graph coarsening.
\newblock {\em Journal of Machine Learning Research}, 24(118).

\bibitem[Legramanti et~al., 2022]{Legramanti2022}
Legramanti, S., Rigon, T., Durante, D., \& Dunson, D. (2022).
\newblock Extended stochastic block models with application to criminal
  networks.
\newblock {\em The Annals of Applied Statistics}, 16(4):2369--2395.

\bibitem[Lenkoski \& Dobra, 2011]{Lenkoski2011}
Lenkoski, A. \& Dobra, A. (2011).
\newblock Computational aspects related to inference in {G}aussian graphical
  models with the {G}-{W}ishart prior.
\newblock {\em Journal of Computational and Graphical Statistics},
  20(1):140--157.

\bibitem[Li et~al., 2022]{Li2022}
Li, T., Lei, L., Bhattacharyya, S., Van~den Berge, K., Sarkar, P., Bickel, P.,
  \& Levina, E. (2022).
\newblock Hierarchical community detection by recursive partitioning.
\newblock {\em Journal of the American Statistical Association},
  117(538):951--968.

\bibitem[Lin et~al., 2016]{Lin2016}
Lin, J., Basu, S., Banerjee, M., \& Michailidis, G. (2016).
\newblock Penalized maximum likelihood estimation of multi-layered {G}aussian
  graphical models.
\newblock {\em Journal of Machine Learning Research}, 17(146).

\bibitem[Lyzinski et~al., 2017]{Lyzinski2017}
Lyzinski, V., Tang, M., Athreya, A., Park, Y., \& Priebe, C. (2017).
\newblock Community detection and classification in hierarchical stochastic
  blockmodels.
\newblock {\em {IEEE} Transactions on Network Science and Engineering},
  4(1):13--26.

\bibitem[Majumdar \& Michailidis, 2022]{Majumdar2022}
Majumdar, S. \& Michailidis, G. (2022).
\newblock Joint estimation and inference for data integration problems based on
  multiple multi-layered {G}aussian graphical models.
\newblock {\em Journal of Machine Learning Research}, 23(1):1--53.

\bibitem[Meilă \& Jaakkola, 2006]{Meilă2006}
Meilă, M. \& Jaakkola, T. (2006).
\newblock Tractable {B}ayesian learning of tree belief networks.
\newblock {\em Statistics and Computing}, 16(1):77--92.

\bibitem[Meyer, 1975]{Meyer1975}
Meyer, E. (1975).
\newblock A measure of the average intercorrelation.
\newblock {\em Educational and Psychological Measurement}, 35(1):67--72.

\bibitem[Moghaddam et~al., 2009]{Moghaddam2009}
Moghaddam, B., Khan, E., Murphy, K., \& Marlin, B. (2009).
\newblock Accelerating {B}ayesian structural inference for non-decomposable
  {G}aussian graphical models.
\newblock {\em Advances in Neural Information Processing Systems}. Curran
  Associates, Inc.

\bibitem[Moran et~al., 2022]{Moran2022}
Moran, G., Sridhar, D., Wang, Y., \& Blei, D. (2022).
\newblock Identifiable deep generative models via sparse decoding.
\newblock {\em Transactions on Machine Learning Research}.

\bibitem[Morrison, 2005]{Morrison2005}
Morrison, D. (2005).
\newblock {\em Multivariate Statistical Methods}.
\newblock Duxbury Advanced Series. Brooks/Cole Thomson Learning, Belmont, CA,
  4th ed.

\bibitem[Newman, 2006]{Newman2006}
Newman, M. (2006).
\newblock Finding community structure in networks using the eigenvectors of
  matrices.
\newblock {\em Physical Review E}, 74(3):036104.

\bibitem[Newman \& Girvan, 2004]{Newman2004}
Newman, M. \& Girvan, M. (2004).
\newblock Finding and evaluating community structure in networks.
\newblock {\em Physical Review E}, 69(2):026113.

\bibitem[Ni et~al., 2015]{Ni2015}
Ni, J., Tong, H., Fan, W., \& Zhang, X. (2015).
\newblock Flexible and robust multi-network clustering.
\newblock {\em Proceedings of the 21th {ACM} {SIGKDD} International Conference
  on Knowledge Discovery and Data Mining}, 835--844. {ACM}.

\bibitem[Peixoto, 2014]{Peixoto2014}
Peixoto, T. (2014).
\newblock Hierarchical block structures and high-resolution model selection in
  large networks.
\newblock {\em Physical Review X}, 4(1):011047.

\bibitem[Peixoto, 2019]{Peixoto2019}
Peixoto, T. (2019).
\newblock Network reconstruction and community detection from dynamics.
\newblock {\em Physical Review Letters}, 123(12):128301.

\bibitem[Peluso \& Consonni, 2020]{Peluso2020}
Peluso, S. \& Consonni, G. (2020).
\newblock Compatible priors for model selection of high-dimensional {G}aussian
  {DAGs}.
\newblock {\em Electronic Journal of Statistics}, 14(2):4110--4132.

\bibitem[Plummer, 2015]{Plummer2015}
Plummer, M. (2015).
\newblock Cuts in {B}ayesian graphical models.
\newblock {\em Statistics and Computing}, 25(1):37--43.

\bibitem[Rand, 1971]{Rand1971}
Rand, W. (1971).
\newblock Objective criteria for the evaluation of clustering methods.
\newblock {\em Journal of the American Statistical Association},
  66(336):846--850.

\bibitem[Roverato, 2002]{Roverato2002}
Roverato, A. (2002).
\newblock Hyper inverse {W}ishart distribution for non-decomposable graphs and
  its application to {B}ayesian inference for {G}aussian graphical models.
\newblock {\em Scandinavian Journal of Statistics}, 29(3):391--411.

\bibitem[Schwaller et~al., 2019]{Schwaller2019}
Schwaller, L., Robin, S., \& Stumpf, M. (2019).
\newblock Closed-form {B}ayesian inference of graphical model structures by
  averaging over trees.
\newblock {\em Journal de la Société Française de Statistique},
  160(2):1--23.

\bibitem[Shan et~al., 2020]{Shan2020}
Shan, L., Qiao, Z., Cheng, L., \& Kim, I. (2020).
\newblock Joint estimation of the two-level {G}aussian graphical models across
  multiple classes.
\newblock {\em Journal of Computational and Graphical Statistics},
  29(3):562--579.

\bibitem[Shen et~al., 2024]{Shen2024}
Shen, L., Amini, A., Josephs, N., \& Lin, L. (2024).
\newblock Bayesian community detection for networks with covariates.
\newblock {\em Bayesian Analysis}.
\newblock Advance online publication.

\bibitem[Stepanov et~al., 2021]{Stepanov2021}
Stepanov, Y., Herrmann, H., \& Guhr, T. (2021).
\newblock Generic features in the spectral decomposition of correlation
  matrices.
\newblock {\em Journal of Mathematical Physics}, 62(8):083505.

\bibitem[Szklarczyk et~al., 2021]{Szklarczyk2021}
Szklarczyk, D., Gable, A., Nastou, K., Lyon, D., Kirsch, R., Pyysalo, S.,
  et~al. (2021).
\newblock The {STRING} database in 2021: customizable protein–protein
  networks, and functional characterization of user-uploaded gene/measurement
  sets.
\newblock {\em Nucleic Acids Research}, 49(D1):D605--D612.

\bibitem[{The Gene Ontology Consortium} et~al., 2023]{GO_2023}
{The Gene Ontology Consortium}, Aleksander, S., Balhoff, J., Carbon, S.,
  Cherry, J., Drabkin, H., et~al. (2023).
\newblock {The Gene Ontology knowledgebase in 2023}.
\newblock {\em Genetics}, 224(1):iyad031.

\bibitem[Tr{\"a}uble et~al., 2021]{Trauble2021}
Tr{\"a}uble, F., Creager, E., Kilbertus, N., Locatello, F., Dittadi, A., Goyal,
  A., et~al. (2021).
\newblock On disentangled representations learned from correlated data.
\newblock {\em Proceedings of the 38th International Conference on Machine
  Learning}, 10401--10412. PMLR.

\bibitem[van~den Boom et~al., 2022]{vandenBoom2022a}
van~den Boom, W., Beskos, A., \& {De Iorio}, M. (2022).
\newblock The \mbox{$G$-Wishart} weighted proposal algorithm: efficient
  posterior computation for {G}aussian graphical models.
\newblock {\em Journal of Computational and Graphical Statistics},
  31(4):1215--1224.

\bibitem[van~den Boom et~al., 2023]{vandenBoom2023}
van~den Boom, W., De~Iorio, M., \& Beskos, A. (2023).
\newblock Bayesian learning of graph substructures.
\newblock {\em Bayesian Analysis}, 18(4):1311--1339.

\bibitem[Wade \& Ghahramani, 2018]{Wade2018}
Wade, S. \& Ghahramani, Z. (2018).
\newblock Bayesian cluster analysis: point estimation and credible balls (with
  discussion).
\newblock {\em Bayesian Analysis}, 13(2):559--626.

\bibitem[Williamson, 1985]{Williamson1985}
Williamson, G. (1985).
\newblock {\em Combinatorics for Computer Science}.
\newblock Computers and Math Series. Computer Science Press, Rockville, MD.

\bibitem[Wu et~al., 2021]{Wu2021}
Wu, T., Hu, E., Xu, S., Chen, M., Guo, P., Dai, Z., et~al. (2021).
\newblock {clusterProfiler} 4.0: a universal enrichment tool for interpreting
  omics data.
\newblock {\em The Innovation}, 2(3):100141.

\bibitem[Yang et~al., 2023]{Yang2023}
Yang, B., Wang, K., Sun, Q., Ji, C., Fu, X., Tang, H., et~al. (2023).
\newblock Does graph distillation see like vision dataset counterpart?
\newblock {\em Advances in Neural Information Processing Systems 36}. Curran
  Associates, Inc.

\bibitem[Yekutieli \& Benjamini, 1999]{Yekutieli1999}
Yekutieli, D. \& Benjamini, Y. (1999).
\newblock Resampling-based false discovery rate controlling multiple test
  procedures for correlated test statistics.
\newblock {\em Journal of Statistical Planning and Inference},
  82(1-2):171--196.

\bibitem[Yoshida \& West, 2010]{Yoshida2010}
Yoshida, R. \& West, M. (2010).
\newblock Bayesian learning in sparse graphical factor models via variational
  mean-field annealing.
\newblock {\em Journal of Machine Learning Research}, 11(59):1771--1798.

\bibitem[Zhang, 2018]{Zhang2018}
Zhang, S. (2018).
\newblock Comparisons of gene coexpression network modules in breast cancer and
  ovarian cancer.
\newblock {\em {BMC} Systems Biology}, 12(S1):57--87.

\end{thebibliography}

\end{document}


\title{\bf Supplement to ``Graph of Graphs: From Nodes to Supernodes in Graphical Models''}
\author{\if0\blind Maria De Iorio, Willem van den Boom, Alexandros Beskos,\\ Ajay Jasra and Andrea Cremaschi\fi}
\date{}
\maketitle

\appendix
\numberwithin{equation}{section}
\renewcommand{\theequation}{\thesection\arabic{equation}}

\section{Simulation study on inferring individual edges}
\label{ap:ROC_simul}

We investigate the ability of Gaussian graphical models (GGMs) to recover single edges.
Specifically,
we consider the Bayesian GGM described in Section~\ref{sec:ggm} of the main manuscript with the default hyperparameter choices $\delta = 3$ and $D=I_p$ for the $G$-Wishart prior on the precision matrix, and a uniform prior on graphs, i.e.\ $p(G) = 2^{-p(p-1)/2}$.
We simulate data for $p=20,40,100$, $n=50,100,500,1000$ and a graph density of 25\%, 50\% or 75\% as follows:
\begin{enumerate}
    \item
    We select a graph $G$ uniformly at random from all graphs on $p$ nodes with the specified graph density, i.e.~proportion of edges present.
    \item
    Given $G$, we sample a precision matrix $\Psi$ from the $G$-Wishart prior.
    \item
    Finally, we sample $n$ observations independently from $\mathcal{N}(0_{p\times 1},\,\Psi^{-1})$.
\end{enumerate}
We generate 10 replicate data sets for each scenario.
Then, we estimate the posterior edge inclusion probabilities by running the algorithm  proposed in \citet{vandenBoom2022a}
for $100000$ iterations and discarding the first $10000$ iterations as burn-in.

\begin{figure}
\centering
\includegraphics[width=0.9\textwidth]{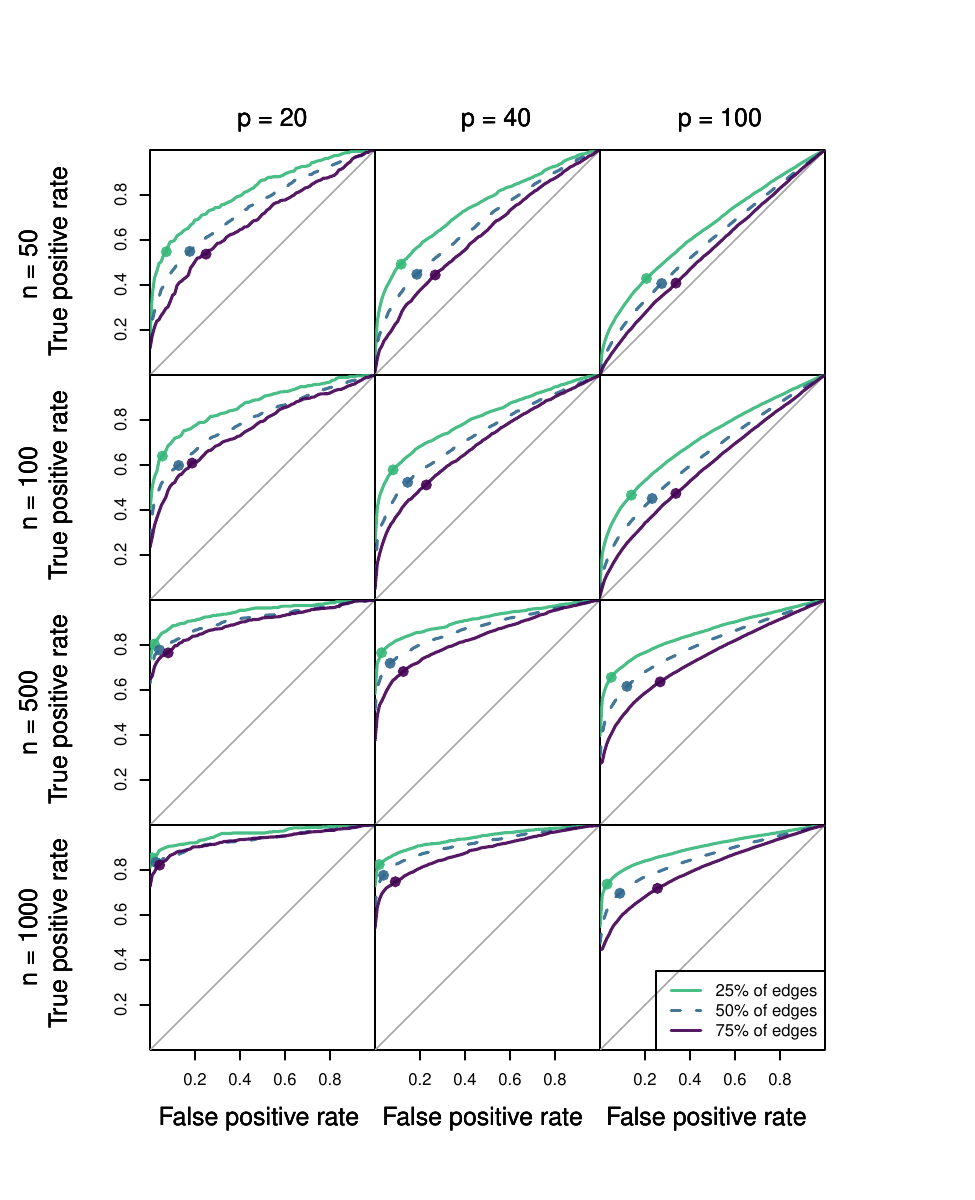}
\caption{
Receiver operating characteristic (ROC) curves
for edge detection in a GGM for varying
number of nodes $p$,
number of observations $n$
and
the proportion
of edges present in the true underlying graph (out of the maximum number possible).
The dots mark the performance when selecting edges with a posterior inclusion probability greater than $0.5$.
}
\label{fig:ROC_simul}
\end{figure}

A typical way to compute a point estimate of the graph in a Bayesian GGM is by selecting those edges whose posterior inclusion probability is above a certain threshold.
We compare the resulting graph with the true underlying graph $G$ from Step~1 above for different threshold values.
Furthermore, we compute the corresponding true positive and false positive rates of edge detection.
We do so for each scenario, and we aggregate the results across the 10 replicates.
Figure~\ref{fig:ROC_simul} summarizes the results: recovery of individual edges is increasingly challenging and requires substantially larger sample sizes
for more complex graphs, i.e.\ those with more edges or nodes.

\pagebreak

\section{Overview of notation}
\label{ap:overview}

We provide an overview of the notation used in Table~\ref{tab:overview_notation}.
\begin{table}[h]
\centering
\caption{Overview of notation used.}
\begin{tabular}{r|l}
     \textbf{Symbol} & \textbf{Description} \\
     \hline
     $n$ & Number of observations \\
     $p$ & Number of nodes \\
     $p_k$ & Number of nodes in the $k$\textsuperscript{th} supernode \\
     $K$ & Number of supernodes \\
     $X$ & $n\times p$ data matrix \\
     $X_j$ & $j$\textsuperscript{th} column of $X$ \\
     $\hat{\rho}_{ij}$ & Sample correlation between variables $X_i$ and $X_j$ \\
     $V$ & Set $\{1,\dots,p\}$ of all nodes \\
     $C$ & Subset of nodes that are centers of supernodes \\
     $S_k$ & Set of nodes in the $k$\textsuperscript{th} supernode \\
     $\mathcal{T}$ & Tessellation $\{S_k\}_{k=1}^K$ of $V$ \\
     $T_k$ & Tree $(S_k, E_k)$ with $S_k$ as nodes and edge set $E_k$ \\
     $G^\star$ & Supergraph $(\mathcal{T}, E^\star)$ with $K$ supernodes \\
     $G$ & Augmented version of supergraph $G^\star$ with $p$ nodes \\
     $x_k$ & $n\times p_k$ matrix of variables in supernode $S_k$ \\
     $Y^\star$ & $n\times K$ matrix of first principal components of each $x_k$ \\
     $Y$ & $n\times p$ matrix with all principal components of each $x_k$ \\
     $f_\textnormal{coh.}(p_k)$ & Cohesion function \\
     $f_\textnormal{sim.}(x_k)$ & Similarity function \\
     $\widetilde{p}(x_k)$ & Tree activation function \\
     $\zeta$ & Power used for coarsening \\
     $f_\textnormal{sim.}^{(\zeta)}(x_k)$ & Coarsened similarity function \\
     $p^{(\zeta)}(\mathcal{T})$ & Tessellation prior with $f_\textnormal{sim.}^{(\zeta)}(x_k)$ as similarity function \\
     $\pi^{(\zeta)}(\mathcal{T},G,Y)$ & Target distribution when coarsening the likelihood \\
     $\pi_\textnormal{cut}(\mathcal{T},G,Y)$ & Target distribution of nested MCMC \\
     $\delta$, $\delta_G$ & Degrees of freedom of the $G$-Wishart priors \\
     $D$, $D_G$ & Rate matrices of the $G$-Wishart priors
\end{tabular}
\label{tab:overview_notation}
\end{table}

\section{Size-biased  prior}
\label{ap:size-biased}

Here,
we discuss how the size biasing in the tessellation prior is different from the one 
proposed for 
exchangeable sequences of clusters \citep[ESC,][]{Betancourt2022}.
Let $\Pi$ denote a partition of the set $\{1,\dots, p\}$.
For number of clusters $K = |\Pi|$,
let
$p_1,\dots,p_K$
denote the cluster sizes in $\Pi$.

\subsection{ESC priors}

\citet{Betancourt2022}
specify a prior distribution $p(\Pi)$
such that each possible
ordered sequence $p_1,\dots,p_K$
appears with probability
proportional to
$\prod_{k=1}^K f(p_k)$
for some probability distribution $f(\cdot)$.
Specifically,
they construct the distribution on the number of clusters and their sizes in $p(\Pi)$ by (i)~drawing an infinite number of $p_k$ as i.i.d.\ random variables from $f(\cdot)$ and by (ii) conditioning on the event that $\sum_{k=1}^K p_k = p$. Then,
\begin{equation} \label{eq:goal}
    p(p_1,\dots,p_K) \propto \prod_{k=1}^K f(p_k)
\end{equation}
Furthermore,
by Proposition~2 of \citet{Betancourt2022},
\[
    p(\Pi)
    \propto
    \frac{\prod_{k=1}^K f(p_k)}{\binom{p}{p_1\,\dots\, p_K} / K!}
\]

\subsection{Size-biased prior for tessellations}

Assume now a set of distances among the elements of the set $\{1,\dots,p\}$ and the following tessellation.
For a set of centers
$C\subset\{1,\dots,p\}$,
assign each element in $\{1,\dots,p\}$
to the center that they are closest to.
Then, $C$ parameterizes a partition $\Pi$.
However,
not necessarily all combinations $\{p_1,\dots,p_K\}$
of cluster sizes with $\sum_{k=1}^K p_k = p$
can result from this tessellation construction.
Denote the set of those $\{p_1,\dots,p_K\}$ that can be obtained through a tessellation by $\mathcal{S}$.
The set $\mathcal{S}$ is determined by the distances.

The goal is 
to specify a prior distribution $p(C)$ in the same spirit as
\eqref{eq:goal}, though now truncated to
$\{p_1,\dots,p_K\}\in\mathcal{S}$.
If we let the distribution on $C$ be uniform conditionally on
$\{p_1,\dots,p_K\}$,
then we would specify
\begin{equation} \label{eq:goal_S}
    p(C) \propto \frac{\prod_{k=1}^K f(p_k)}{|\{C': \{p_1,\dots,p_K\} \}|}
\end{equation}
Here,
$|\{C': \{p_1,\dots,p_K\} \}|$
denotes the number of sets of centers $C'$ that result in the cluster sizes
$\{p_1,\dots,p_K\}$.
This can be computed by enumerating all $\binom{p}{K}$ possible $C$ with $|C|=K$.
Without a more efficient algorithm
to compute $|\{C': \{p_1,\dots,p_K\} \}|$,
this construction for size biasing of tessellations results in a prior that is computationally too expensive to evaluate.

\subsection{A heuristic alternative}

To avoid the computation of $|\{C': \{p_1,\dots,p_K\} \}|$,
we instead consider a prior that replaces this term by
$|\{C': |C'|=K \}| = \binom{p}{K}$.
That is,
\[
    p(C) \propto \frac{\prod_{k=1}^K f(p_k)}{|\{C': |C'|=K \}|}
    = \frac{\prod_{k=1}^K f(p_k)}{\binom{p}{K}}
\]
The rationale behind this modification is that, while it does not satisfy \eqref{eq:goal_S},
it leads to a similar marginal distribution on the number of clusters $K$.
Under the last definition of $p(C)$ above,
we have
\[
    p(K) = \sum_{C\, :\, |C|=K} p(C)
    \propto \sum_{\{p_1,\dots,p_K\}\, :\, |C|=K} \frac{|\{C': \{p_1,\dots,p_K\} \}|}{|\{C': |C'|=K \}|} \prod_{k=1}^K f(p_k)
\]
Instead,
under the definition in \eqref{eq:goal_S},
we have
\begin{align*}
    p(K) &= \sum_{C\, :\, |C|=K} p(C)
    \propto \sum_{\{p_1,\dots,p_K\}\, :\, |C|=K}\   |\{C': \{p_1,\dots,p_K\} \}| 
 \frac{\prod_{k=1}^K f(p_k)}{|\{C': \{p_1,\dots,p_K\} \}|} \\
   & = \sum_{\{p_1,\dots,p_K\}\, :\, |C|=K}\ \prod_{k=1}^K f(p_k)
\end{align*}
To see how the two distributions are similar,
note that
\[
    \sum_{\{p_1,\dots,p_K\}\, :\, |C|=K} \frac{|\{C': \{p_1,\dots,p_K\} \}|}{|\{C': |C'|=K \}|} = 1
\]

\section{Proofs of propositions}
\label{ap:proofs}

The proofs of Propositions~\ref{prop:mar_lik} and \ref{prop:edge_prob}
involve the likelihoods of (pairs of) columns of $x_k = \{X_i\}_{i\in S_k}$
under the tree-based GGM
in Section~\ref{sec:tree_activation}
conditionally on the edge set $E_k$.
Derivations of these (in a slightly more elaborate GGM setting with a nonzero mean for the multivariate Gaussian)
are presented by \citet{Kuipers2014}.
We therefore omit their derivations and state them directly.
For $i\in S_k$,
\begin{equation} \label{eq:edge_lik1}
    \widetilde{p}(X_i) = \widetilde{p}(X_i\mid E_k) = \frac{\Gamma(\delta^\star/2)\, D_{ii}^{\delta/2}}{\pi^{n/2}\, \Gamma(\delta/2)\, (D_{ii}^\star)^{\delta^\star/2}}
\end{equation}
where $\delta^\star$ and $D^\star$
are as in Proposition~\ref{prop:mar_lik}.
Furthermore,
\begin{equation} \label{eq:edge_lik2}
    \widetilde{p}(X_i,X_j\mid (i,j)\in E_k) = \frac{\Gamma_2\{(\delta^\star+1)/2\}\, |D_{\{i,j\}}|^{(\delta+1)/2}}{\pi^n\, \Gamma_2\{(\delta+1)/2\}\, |D_{\{i,j\}}^\star|^{(\delta^\star+1)/2}}
\end{equation}
where $\Gamma_2(\cdot)$ is
the multivariate gamma function of dimension two: $\Gamma_2(t) = \pi^{1/2}\, \Gamma(t)\, \Gamma(t-1/2)$.

\begin{proof}[Proof of Proposition~\ref{prop:mar_lik}]
\label{proof:mar_lik}
Firstly, if $(i,j)\in E_k$,
\begin{equation} \label{eq:w_def}
    w_{ij} = \frac{\widetilde{p}(X_i,X_j\mid (i,j)\in E_k)}{\widetilde{p}(X_i)\, \widetilde{p}(X_j)}
\end{equation}
by
\eqref{eq:edge_lik1}, \eqref{eq:edge_lik2} and the definition of $w_{ij}$ in Proposition~\ref{prop:mar_lik}.
Note that the setup of this proposition fits Section~3 of \citet{Schwaller2019}.
Their Equation~(4)
gives:
\[
\begin{aligned}
\widetilde{p}(x_k\mid E_k)
&= \prod\nolimits_{i\in S_k} \widetilde{p}(X_i) \prod\nolimits_{(i,j)\in E_k} \frac{\widetilde{p}(X_i,X_j\mid (i,j)\in E_k)}{\widetilde{p}(X_i)\, \widetilde{p}(X_j)} \\
&= \frac{\Gamma(\delta^\star/2)^{p_k}\, (\prod_{i\in S_k} D_{ii})^{\delta/2}}{\pi^{n p_k/2}\, \Gamma(\delta/2)^{p_k}\, (\prod_{i\in S_k} D_{ii}^\star)^{\delta^\star/2}} \prod\nolimits_{(i,j)\in E_k} w_{ij}
\end{aligned}
\]
where the last equality follows from \eqref{eq:edge_lik1} and \eqref{eq:w_def}.
Then,
part~(i) of the required result follows from
\[
    \widetilde{p}(x_k) = \sum_{T_k} \widetilde{p}(x_k\mid T_k)\, \widetilde{p}(T_k)
\]
and the prior definition
$\widetilde{p}(T_k) = p_k^{2 - p_k}$.
Also,
$\widetilde{p}(x_k) \propto \sum_{E_k} \prod_{(i,j)\in E_k} w_{ij}$
provides part~(ii), that $\widetilde{p}(x_k)$ is an increasing function of any weight $w_{ij}$.
Finally,
part~(iii) is
Theorem~1 of \citet{Schwaller2019}.
\end{proof}

\begin{proof}[Proof of Proposition~\ref{prop:edge_prob}]
\label{proof:edge_prob}
Consider the mapping of our setup to \citet{Schwaller2019}
from the proof of Proposition~\ref{prop:mar_lik}.
\citet{Schwaller2019} write the prior on trees as
$\widetilde{p}(T_k) = \prod_{(i,j)\in E_k} \beta_{ij}$.
Our uniform prior corresponds to
a constant $\beta_{ij}= b = p_k^{(2-p_k)/(p_k - 1)}$.
Then,
the weights in
Equation~(7) of \citet{Schwaller2019} reduce to
\begin{equation} \label{eq:omega}
    \omega_{ij} = \beta_{ij}\, \frac{\widetilde{p}(X_i,X_j\mid (i,j)\in E_k)}{\widetilde{p}(X_i)\, \widetilde{p}(X_j)}
    = b\, w_{ij}
\end{equation}
where the last equality follows from \eqref{eq:w_def} and $\beta_{ij}=b$.
Note that the Laplacian matrix of the graph with weights $\omega_{ij}$ is $b\Lambda$
where $\Lambda$ is defined in Section~\ref{sec:tree_activation}.
Then,
Theorem~3 of \citet{Schwaller2019}
states
\begin{equation} \label{eq:thm3}
    \widetilde{\mathrm{Pr}}[(i,j)\in E_k\mid x_k] = \omega_{ij} M_{ij}
\end{equation}
where $M_{ij} = Q_{ii} + Q_{jj} - 2Q_{ij}$
with, for some node $u\in S_k$,
\[
    Q_{ij} = \begin{cases}
        \{(b\Lambda^{\{u\}})^{-1}\}_{ij}\quad &i,j\ne u \\
        0,\quad &\text{otherwise}
    \end{cases}
\]
Consider $u=i$.
Then, $Q_{ii}=0$ and $Q_{ij}=0$ such that
$M_{ij} = Q_{jj} = \{(b\Lambda^{\{i\}})^{-1}\}_{ij}$.
Inserting this expression for $M_{ij}$ and \eqref{eq:omega} into \eqref{eq:thm3} yields
\[
    \widetilde{\mathrm{Pr}}[(i,j)\in E_k\mid x_k] = w_{ij} \{(\Lambda^{\{i\}})^{-1}\}_{ij}
\]
Expressing the inverse $(\Lambda^{\{i\}})^{-1}$ in terms of the cofactors of $\Lambda^{\{i\}}$ gives
\[
    \{(\Lambda^{\{i\}})^{-1}\}_{ij} = \frac{|\Lambda^{\{i,j\}}|}{|\Lambda^{\{i\}}|}=r(i,j)
\]
Combining the last two displays
provides
part~(i) of the required result.

To see that $\widetilde{\mathrm{Pr}}[(i,j)\in E_k\mid x_k]$ is increasing in $w_{ij}$,
apply
part~(iii) of Proposition~\ref{prop:mar_lik}
to obtain
\[
    |\Lambda^{\{i\}}| = \sum\nolimits_{E_k}\prod\nolimits_{(l,m)\in E_k} w_{lm} 
    = w_{ij} A + B
\]
where
\begin{align*}
    A &= \sum\nolimits_{E_k:(i,j)\in E_k}\prod\nolimits_{(l,m)\in E_k\setminus\{(i,j)\}} w_{kl} \\
    B &= \sum\nolimits_{E_k:(i,j)\notin E_k}\prod\nolimits_{(l,m)\in E_k} w_{lm}
\end{align*}
do not involve $w_{ij}$.
Also
$|\Lambda^{\{i,j\}}|$
does not involve $w_{ij}$ by the definition of the Laplacian $\Lambda$.
Thus,
\[
    \widetilde{\mathrm{Pr}}[(i,j)\in E_k\mid x_k] = w_{ij} \frac{|\Lambda^{\{i,j\}}|}{|\Lambda^{\{i\}}|}
    = \frac{|\Lambda^{\{i,j\}}|}{A + B/w_{ij}}
\]
which is an increasing function of $w_{ij}$
\end{proof}

\begin{proof}[Proof of Proposition~\ref{prop:size-bias}]
\label{proof:size-bias}
The choice $p(C) = p(\mathcal{T}) / |\mathcal{C}(\mathcal{T})|$
with \eqref{eq:tessellation_prior}
and $K=|\mathcal{T}|=|C|$
implies
\begin{equation} \label{eq:prior_C}
    p(C) \propto \binom{p}{K}^{-1} \prod_{k=1}^{K} f_\textnormal{coh.}(p_k)\, f_\textnormal{sim.}(x_k)
\end{equation}
Assuming $f_\textnormal{sim.}(x_k)=1$ yields
\[
    p(K)
    =  \sum_{C: |C| = K} p(C)
    \propto \binom{p}{K}^{-1} \sum_{C: |C| = K} \prod_{k=1}^K f_\textnormal{coh.}(p_k)
\]
where
\[
    \prod_{k=1}^K f_\textnormal{coh.}(p_k) = (1-\pi)^{p-K} \pi^K
\]
as  $f_\textnormal{coh.}(p_k) = (1-\pi)^{p_k-1} \pi$.
Since there are $\binom{p}{K}$ sets $C$ with $|C|=K$,
\[
    p(K)\propto (1-\pi)^{p-K} \pi^K
\]
from which
part~(i)
follows.

Note that $\overline{p}=p/K$.
The distribution $p(K)$ concentrates on $K=1$ (respectively, $K=p$)
as $\pi\to 0$ ($\pi\to 1$),
from which the required limit $E[\overline{p}]\to p$ ($E[\overline{p}]\to 1$) follows.

To see that $E[\overline{p}]=E[p/K]$ is a decreasing function of $\pi$,
it suffices to show that $E[1/K]$ is a decreasing function of $\alpha=\pi/(1-\pi)$.
Note that
\[
    E[1/K] = \frac{\sum_{K=1}^p \alpha^K/K}{\sum_{K=1}^p \alpha^K}
\]
Therefore,
\[
    \frac{dE[1/K]}{d\alpha}
    = \frac{1 - E[K]\, E[1/K]}{\alpha}
\]
Now, Jensen's inequality provides $E[1/K] > 1/E[K]$
from which $\frac{dE[1/K]}{d\alpha} < 0$ follows as required for part~(ii).
\end{proof}

\begin{proof}[Proof of Proposition~\ref{prop:pca}]
\label{proof:pca}
Principal component analysis of $x_k$
corresponds to the eigenvalue decomposition of $x_k^\top x_k$.
Since $X$ is standardized such that $\|X_i\|^2=n$ for each variable $i$,
we have
$x_k^\top x_k = n R$
where $R_{ij}=\hat{\rho}_{ij}$ for $i\ne j$ and $R_{ii}=1$.
Thus,
$\phi$
can be computed using the eigenvalues $\lambda_1\geq \dots\geq \lambda_{p_k}$ of $R$
as \citep[page~268]{Morrison2005}
\[
    \phi = \frac{\lambda_1}{\sum_{i=1}^{p_k} \lambda_i}
    = \frac{\lambda_1}{\mathrm{tr}(R)}
    = \frac{\lambda_1}{p_k}
\]
Now,
Equation~(52) of
\citet{Stepanov2021} provides the lower bound in part~\ref{point:bound} of the required result.

For the upper bound in \ref{point:bound}, we use
\[
    \lambda_1\leq \max_i \sum_j |R_{ij}|
\]
from page~285 in \citet{Morrison2005}.
Note that
\[
    \sum_j |R_{ij}| = 1 + (p_k-1)\, \rho^i = p_k\, h_{p_k}(\rho^i)
\]
Combining the last three displays provides $\phi\leq \max_i h_{p_k}(\rho^i)$.
The other part of the upper bound follows from Equation~(13) of \citet{Meyer1975} which states
\[
    \lambda_1\leq 1 + (p_k-1) s
    = p_k\, h_{p_k}(s)
\]

Consider the case with constant correlation
$\hat{\rho}_{ij}=\rho$ next.
\citet[page~283]{Morrison2005}
provides
$\phi=h_{p_k}(\rho)$
for $\rho>0$,
which also follows from part~\ref{point:bound} of this proposition and the fact that $\rho = \rho^i$ in this case.
For $\rho< 0$,
note that
\[
    R = \rho\, 1_{p_k\times p_k} + (1-\rho) I_{p_k}
\]
Solving for the eigenvalue $\lambda$ in $Rv=\lambda v$
with the eigenvectors $v = 1_{p_k\times 1}$
and $v = e_1 - e_i$ ($i=2,\dots,p_k$),
where $e_i$ is the vector of all zeros except for its $i$\textsuperscript{th} element being equal to one,
yields
\[
    \lambda = p_k\,\rho + 1 - \rho
    \quad\text{and}\quad
    \lambda = 1-\rho
\]
respectively.
Thus,
for $\rho<0$,
the largest eigenvalue is $1-\rho$,
resulting in $\phi = (1-\rho)/p_k$.
\end{proof}

\begin{figure}
\centering
\includegraphics[width=0.8\textwidth]{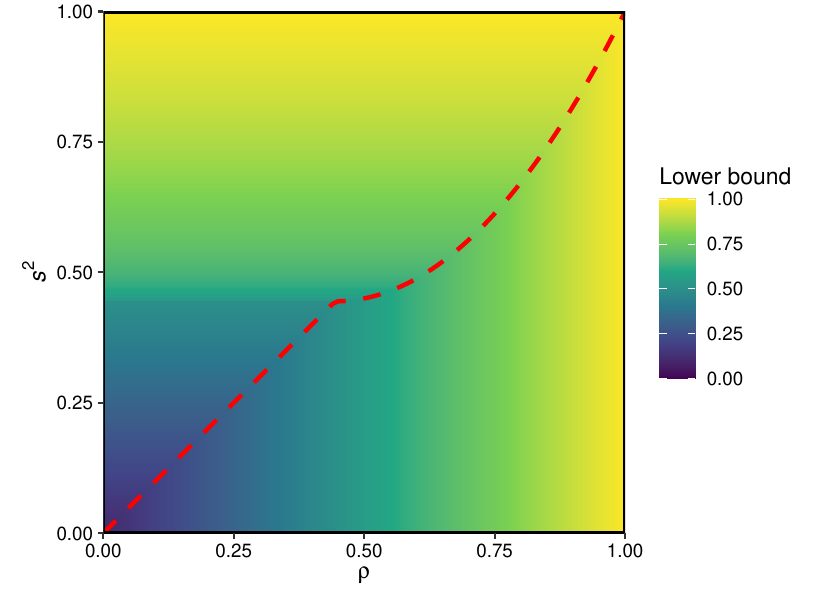}
\caption{ 
Visualization of the lower bound $\max[h_{p_k}(\rho),\, h_\star\{h_{p_k}(s^2)\} ]$ on $\phi$
from Proposition~\ref{prop:pca}\ref{point:bound} for $p_k = 10$.
The region below the dashed line marks where
$h_{p_k}(\rho) > h_\star\{h_{p_k}(s^2)\}$.
}
\label{fig:prop_var}
\end{figure}

Figure~\ref{fig:prop_var} visualizes the lower bound of Proposition~\ref{prop:pca}\ref{point:bound}.

\section{Tree activation function for directed rooted trees}
\label{ap:rooted}

In Section~\ref{sec:tree_activation} of the main manuscript,
the tree activation function $\widetilde{p}(x_k)$
is constructed
via a GGM with undirected trees $T_k$
to describe the dependence structure among the variables $x_k$.
The model can be extended to
rooted
trees, in the case of directed edges.
Also in this case
the resulting probability model $\overline{p}(x_k)$ can be computed efficiently and analytically.
For undirected trees,
$\widetilde{p}(x_k)$  is derived from 
part~(iii) of Proposition~\ref{prop:mar_lik},
which is the result of Kirchhoff's matrix tree theorem \citep{Schwaller2019}.
Here, we establish a similar result for $\overline{p}(x_k)$
using Tutte's theorem \citep{DeLeenheer2020}
which extends Kirchhoff's matrix tree theorem for undirected graphs to directed graphs.
In the remainder of this appendix,
we 
(i)~introduce the directed trees;
(ii)~modify the tree-based GGM to take the direction of edges into account; (iii)~specify a corresponding tree activation function $\overline{p}(x_k)$; (iv)~show that $\overline{p}(x_k)$ can be evaluated in a computationally efficient manner.

A rooted tree is a
(typically undirected)
tree where one node has been designated
as root \citep{Deo1974}, to which other nodes are connected to either directly or indirectly. In our context, the concept of root node, which corresponds to a variable in a supernode,
could be of interest and can be incorporated,
for instance by designating the center of a supernode as root. 

The edges of a rooted tree can
be assigned a natural orientation, either away from or towards  the root, in which
case the structure becomes a directed rooted tree.
Designate a node in supernode $S_k$ as root.
Then, a tree $T_k=(S_k, E_k)$ can be transformed into a corresponding directed tree called an
\emph{arborescence}\footnote{Other names for arborescence are \emph{out-arborescence} \citep{Bhattacharyya2024} and 
\emph{directed rooted tree} \citep{Williamson1985}.}
$\overline{T}_k = (S_k,\overline{E}_k)$ by directing edges outward
from the root \citep{Deo1974,Meilă2006}.
Specifically, $(i,j)\in\overline{E}_k$ if and only if there is a directed edge from node~$i$ to node~$j$,
and all edge directions follow from the requirement that there is a directed path from the root to any other node in $S_k$.
Also,
the tree $T_k$
and the choice of root
uniquely determine $\overline{T}_k$
since the root is the only node without a parent in $\overline{T}_k$, i.e.\ without an incoming edge.

Alternatively,
a directed tree could be constructed by directing edges inward to the root, a construction that has been referred to as 
\emph{anti-arborescence} \citep{Korte2002}
or an \emph{in-arborescence} \citep{Bhattacharyya2024}.
In principle,
such a directed tree can also be used to specify a Bayesian network.
However, we do not explore the scenario further because such networks are not common in applications since, if edges are directed inward to the root,
then variables corresponding to leaf nodes are marginally independent in the Bayesian network.
On the other hand, all variables are dependent in a Bayesian network corresponding to an arborescence.

\subsection{Arborescence}

In the case that the
directed acyclic graphs (DAG)
in a Bayesian network
is an arborescence $\overline{T}_k = (S_k,\overline{E}_k)$ rooted at $r\in S_k$,
the probability distribution of  $\{X_i\}_{i\in S_k}$
 factorizes
across the columns of $x_k$
as
\citep{Castelletti2022}
\begin{equation} \label{eq:DAG-model}
    \overline{p}(x_k\mid \overline{E}_k) = \overline{p}(X_r\mid \text{$r$ is root})\,
    \prod\nolimits_{(i,j)\in\overline{E}_k} \overline{w}_{ij},
    \qquad
    \overline{w}_{ij} = \overline{p}(X_j\mid X_i, (i,j)\in\overline{E}_k)
\end{equation}
Additionally, conditionally on a precision matrix $\Delta_k$, we assume the same Gaussian distribution on $x_k$ as in Section~\ref{sec:tree_activation}:
\[
    \overline{p}(x_k\mid \Delta_k ) = \prod_{i=1}^n \mathcal{N}(x_{ik}\mid 0_{p_k\times 1},\, \Delta_k^{-1})
\]
For the distribution to satisfy \eqref{eq:DAG-model},
$\mathcal{N}(0_{p_k\times 1},\, \Delta_k^{-1})$ needs to be Markov over $\overline{T}_k$,
i.e.\ it needs to satisfy the conditional independencies implied by the factorization in \eqref{eq:DAG-model} \citep[see][for details]{Peluso2020}.
To this end,
the prior $\overline{p}(\Delta_k\mid\overline{E}_k)$
is taken
to be a Compatible DAG-Wishart distribution \citep{Peluso2020,Castelletti2022}
with degrees of freedom $\overline{\delta}> p_k - 1$
and positive-definite rate matrix $\overline{D}$.\footnote{Also the $G$-Wishart prior from Section~\ref{sec:tree_activation} satisfies \eqref{eq:DAG-model} \citep{Meilă2006}. However, that choice would ignore direction of edges and result in $\overline{p}(x_k)=\widetilde{p}(x_k)$.}

With this prior specification,
we have that
\[\overline{p}(x_k\mid \overline{E}_k) = \int\overline{p}(x_k\mid \Delta_k)\, \overline{p}(\Delta_k\mid\overline{E}_k)\, d\Delta_k\]
satisfies \eqref{eq:DAG-model} by Equation~(8) in \citet{Castelletti2022}.
Moreover,
by Equation~(9) in \citet{Castelletti2022},
\begin{align*}
    \overline{p}(X_r\mid \text{$r$ is root}) &=
    \frac{
    \overline{g}(\overline{\delta}^\star,\, \overline{D}^\star_{rr})
    }{
    \pi^{n/2}\,
    \overline{g}(\overline{\delta},\, \overline{D}_{rr})
    } \\[10pt]
    \overline{w}_{ij} &=
    \frac{
    \overline{D}_{ii}^{1/2}\, 
    \overline{g}\{\overline{\delta}^\star + 1,\, \overline{D}^\star_{jj} - (\overline{D}^\star_{ij})^2 / \overline{D}^\star_{ii}\}
    }{
    \pi^{n/2}\,
    (\overline{D}^\star_{ii})^{1/2}\,
    \overline{g}(\overline{\delta} + 1,\, \overline{D}_{jj} - \overline{D}_{ij}^2 / \overline{D}_{ii})
    }
\end{align*}
where $\overline{\delta}^\star = \overline{\delta} + n$, $\overline{D}^\star = \overline{D} + x_k^\top x_k$
and
\[
    \overline{g}(\nu,d) = d^{-(\nu-p_k + 1) / 2}\, \Gamma\!\left(\frac{\nu - p_k + 1}{2}\right)
\]

We consider the uniform distribution over all 
arborescences with node set $S_k$, i.e.\ $\overline{p}(\overline{E}_k) = p_k^{1 - p_k}$ since
there are $p_k^{p_k - 2}$ undirected trees
and $p_k$ possible roots.
Now, the corresponding tree activation function is equal to
\[
    \overline{p}(x_k) = \sum\nolimits_{\overline{E}_k} \overline{p}(\overline{E}_k)\, \overline{p}(x_k\mid \overline{E}_k)
    = p_k^{1 - p_k}
    \sum\nolimits_{\overline{E}_k} \overline{p}(x_k\mid \overline{E}_k)
\]
where the sum is over all edge sets $\overline{E}_k$ such that $\overline{T}_k=(S_k,\overline{E}_k)$ is an arborescence.
Splitting the sum over arborescences by root and inserting \eqref{eq:DAG-model},
we can write
\[
\begin{aligned}
    \overline{p}(x_k) &= p_k^{1 - p_k} \sum_{r\in S_k}
    \sum\nolimits_{\overline{E}_k\, :\, \text{$r$ is root}}
    \overline{p}(X_r\mid \text{$r$ is root})
    \prod\nolimits_{(i,j)\in\overline{E}_k} \overline{w}_{ij} \\
    &= \frac{p_k^{1 - p_k}}{\pi^{n/2}}
    \sum_{r\in S_k} \frac{
    \overline{g}(\overline{\delta}^\star,\, \overline{D}^\star_{rr})
    }{
    \overline{g}(\overline{\delta},\, \overline{D}_{rr})
    }
    \sum\nolimits_{\overline{E}_k\, :\, \text{$r$ is root}}
    \prod\nolimits_{(i,j)\in\overline{E}_k} \overline{w}_{ij}
\end{aligned}
\]
The last expression for $\overline{p}(x_k)$ reduces to 
the tree activation function
$\widetilde{p}(x_k)$ in part~(i) of Proposition~\ref{prop:mar_lik} when setting $p_k = 1$,
$\overline{\delta}=\delta$ and $\overline{D}=D$.

We now express the sum
$ \sum\nolimits_{\overline{E}_k\, :\, \text{$r$ is root}}
    \prod\nolimits_{(i,j)\in\overline{E}_k} \overline{w}_{ij}$
in the last display as the determinant
of a $(p_k-1)\times (p_k - 1)$ matrix,
providing a result that is analogous to
part~(iii) of Proposition~\ref{prop:mar_lik}.
Consider a weighted complete directed graph over the node set $S_k$ with weight $\overline{w}_{ij}$ for the edge from node~$i$ to $j$.
Then, the in-degree
Laplacian matrix corresponding to the graph is the $p_k\times p_k$ matrix $\overline{\Lambda}$ defined by $\overline{\Lambda}_{ij}=-\overline{w}_{ij}$ for $i\ne j$
and $\overline{\Lambda}_{jj} = \sum_{i\ne j} \overline{w}_{ij}$.
Let $\overline{\Lambda}^{\nu}$
denote the matrix obtained by removing the rows and columns indexed by $\nu\subset S_k$ from $\overline{\Lambda}$.
Then,
\[
    \sum\nolimits_{\overline{E}_k\, :\, \text{$r$ is root}}
    \prod\nolimits_{(i,j)\in\overline{E}_k} \overline{w}_{ij}
     = |\overline{\Lambda}^{\{r\}}|
\]
by Theorem~3 of \citet{DeLeenheer2020} and Corollary~C.7 of \citet{Bhattacharyya2024}.

In the above,
we implicitly specify a uniform distribution over the possible roots $r\in S_k$ by choosing a uniform distribution over all arborescences.
However, the development readily generalizes to any other distribution over roots.
For instance, the root could be fixed at the center $c\in C$ that corresponds to the supernode in the Voronoi tessellation to impose additional structure in the tree activation function.

As discussed in Section~2.3 of \citet{Duan2023},
the posterior $\overline{p}(\overline{E}_k\mid x_k)$
satisfies \emph{root exchangeability}
if the weights are symmetric,
i.e.\
$\overline{w}_{ij}=\overline{w}_{ji}$.
Such symmetry holds for $\overline{D}=I_{p_k}$ if $X$ is standardized as then both $\overline{D}$ and $\overline{D}^\star$ are symmetric with constant diagonal.
Root exchangeability means that
the posterior over arborescences conditional on the choice of root is the same regardless of root (up to the direction of edges).
The same also holds for the marginal likelihood $\overline{p}(x_k\mid \text{$r$ is root})$
if both $\overline{D}$ and $\overline{D}^\star$ are symmetric with constant diagonal.

\section{Related work}
\label{ap:related}

\subsection{Graphical models}
\label{sec:graphical_models}

The graph of graphs consists of two levels and thus leads to a multilevel graphical model.
Such models have been considered without inference on the clustering of nodes.
\citet{Cheng2017} and \citet{Shan2020}
consider a GGM where
they factorize the elements of the overall precision matrix
into the product of a low-level (i.e.\ edge-specific) and a high-level (i.e.\ superedge-specific) term.
The low-level terms are also present for pairs of nodes from different supernodes,
such that edges across supernodes can still exist.
\citet{Kim2020}
assume equally sized supernodes.
Then, they set the precision matrix
equal to the sum of a superedge-specific and an edge-specific term
where the edge-specific term is nonzero only for edges between nodes in the same supernode.
\citet{Cremaschi2023} and \citet{Colombi2024}
define the presence (absence) of a superedge as the presence (absence) of all possible edges between the corresponding supernodes.
Another line of work \citep{Lin2016,Ha2021,Majumdar2022} considers multilayer graphical models based on chain graphs.
There, each layer corresponds to one graph
and
directed edges link nodes across graphs.

The works mentioned so far assume a prespecified partition of nodes into supernodes.
Inference on the partition of nodes
has been considered in single-level graphical models \citep[see][for an overview]{vandenBoom2023}.
Most recently,
\citet{Peixoto2019} and \citet{vandenBoom2023}
use stochastic blockmodels (SBMs) as prior distribution on the graph 
to partition the nodes.
In the context of network data, SBMs have been extended to employ a product partition models with covariates (PPMx) prior to cluster nodes based on both their connectivity pattern and the homogeneity of their attributes \citep{Legramanti2022,Shen2024}.
Furthermore, \citet{Josephs2023} specify an SBM for multiple networks using a nested Dirichlet process. \citet{Amini2024} extend the SBM to multilayer networks using a hierarchical Dirichlet process.
With a similar goal as our tessellation, to group correlated variables,
\citet{Kim2023} use a Dirichlet process to learn the blocks of a block-diagonal covariance matrix.

\subsection{Factor models}
\label{sec:factor_model}

In the graph of graphs, each supernode represents a latent feature.
From this point of view, our approach has connections with sparse latent factor models,
since nodes (i.e.\ variables) can only be associated with one supernode (i.e.\ factor).
Variables which, like our nodes, are associated with only one factor,
are known as pure variables \citep{Bing2020} or anchor features \citep{Arora2012,Moran2022}.
Factor analysis typically assumes independence of factors a priori.
This contrasts with how superedges can capture dependence among latent features.
Such dependence is often warranted
in real-world data \citep{Trauble2021}.

Another example of the link between graphical models and factor analysis is
\citet{Yoshida2010}.
They
construct a sparse factor model such that the precision matrix is sparse and the model thus can be interpreted as a GGM.
\citet{Chandra2022} decompose the precision matrix, instead of the covariance as in standard factor analysis, into a low-rank and a diagonal matrix.
Then, they induce sparsity in the low-rank decomposition resulting in a sparse precision matrix and thus a GGM.

\subsection{Multiscale Markov random fields}

The multilevel nature of the graph of graphs is reminiscent of multiscale Markov random fields \citep{Ferreira2007}.
They are also undirected multilevel graphical models
though with both the graphs and the hierarchical grouping of the nodes fixed to model certain dynamic processes within and across resolutions.
Unlike our construction,
where the within-supernode structure has no direct relation to the supergraph,
these models aim for consistency across scales.
However, they
typically
specify distributions that are not probabilistically coherent
across scales.
They therefore call for nonstandard Bayesian inference.
Specifically, Jeffrey's rule of conditioning, a generalization of Bayesian updating, is employed in this context. Note that our supergraph construction does not correspond to a data generating process on the data $X$ observed at the within-supernode level and focus is on macrostructures.

\subsection{\texorpdfstring{Multilevel graph constructions}{Multilevel graph constructions}}
\label{sec:multilayer}

\citet{Ni2015} refer to a two-level graph construction similar to our graph of graphs as a network of networks.
They analyze an observed network of networks to cluster both nodes and supernodes.
We remark that, in contrast to our construction, 
the term `network of networks'
usually refers to multiple networks with additional edges that connect individual nodes across networks
instead of connecting supernodes \citep[e.g.][]{DAgostino2014}.

Other multilevel network models similarly lack the notion of a superedge:
\citet{Fosdick2019} develop a multiresolution generalization of the stochastic blockmodel
where edges within communities are modeled at a finer resolution than edges between communities.
Additionally,
hierarchical, nested blockmodels where blocks are repeatedly split into yet lower level blocks
have been developed
without a concept of edges that connect blocks instead of individual nodes \citep[e.g.][]{Peixoto2014,Lyzinski2017,Li2022}.

The notion of a superedge does exist in graph coarsening \citep{Chen2022}
where a graph is coarsened to obtain a supergraph.
In this context, there is typically no notion of node-specific data $X$ like the graph of graphs considers.
Exceptions that consider $X$
are the works by \citet{Jin2022} and \citet{Yang2023} on learning supergraphs in graph neural networks,
and \citet{Kumar2023}.
We review the method from \citet{Kumar2023}
to provide a detailed comparison with graph of graphs.

\subsection{Connections with \texorpdfstring{\citet{Kumar2023}}{Kumar et al. (2023)}}
\label{sec:graph_coarsening}

In graph coarsening,
a graph $H$ with $p$ nodes is coarsened to obtain a supergraph $H^\star$ with $K$ supernodes.
\citet{Kumar2023} consider such coarsening while also taking into account
node-specific data or features
represented by an $n\times p$ matrix $X$.
In more detail,
they learn the supergraph through a minimization scheme.
Let $L$ and $L^\star$ denote the Laplacian matrices of the graph $H$ and the supergraph $H^\star$, respectively.
Here, the edges of $H$ and superedges of $H^\star$ are assumed to be weighted.
Furthermore, \citet{Kumar2023} consider a $p\times K$ \emph{loading matrix} $C$
which links the uncoarsened data $X$ and an $n\times K$ coarsened data matrix $X^\star$ such that $X = X^\star\, C^\top$.
Then, a supergraph is inferred by minimizing the objective
\citep[Equation~(11)]{Kumar2023}
\begin{equation} \label{eq:coarsening_objective}
    -\xi_1 \log\left(\left|L^\star + \tfrac{1}{p} 1_{p\times p}\right|\right)
    + \mathrm{tr}(X^\star L^\star X^{\star\top})
    + \xi_2\, h(L^\star) + \tfrac{\xi_3}{2} g(C)
\end{equation}
with respect to $L^\star$, $Y^\star$ and $C$,
subject to the constraints $L^\star = C^\top L C$ and
$X = X^\star\, C^\top$
plus some standard constraints on $C$
for some tuning parameters $\xi_1$, $\xi_2$ and $\xi_3$,
and regularization functions $h(\cdot)$ and $g(\cdot)$.

We now discuss connections of graph of graphs with the objective function in \eqref{eq:coarsening_objective}.
In our context, let us consider the following augmented target:
\[
    p(\mathcal{T},G,\Omega,Y\mid X)
    \propto p(\mathcal{T})\,
    p(G\mid\mathcal{T})\,
    p(\Omega\mid \mathcal{T},G)\,
    p(Y\mid \mathcal{T},G,\Omega,X)
\]
where
\[
    p(\Omega\mid \mathcal{T},G)
    = \frac{1}{I_G(\delta_G, D_G)}
    |\Omega|^{\delta_G/2 - 1}
    \exp\left\{ -\frac{1}{2} \mathrm{tr}(\Omega D_G) \right\}
\]
and
\[
    p(Y\mid \mathcal{T},G,\Omega,X) = (2\pi)^{-np/2}
    |\Omega|^{n/2}\exp\left\{ -\frac{1}{2} \mathrm{tr}(Y\Omega Y^\top) \right\}
\]
where $Y$ is the matrix of all principal components, $G$ is the augmented graph derived from $G^\star$ and $\Omega$ is the corresponding precision matrix.
Maximizing $p(\mathcal{T},G,\Omega,Y\mid X)$
with respect to $\mathcal{T}$, $G$ and $\Omega$
is equivalent to minimizing
$-2\log\{p(\mathcal{T},G,\Omega,Y\mid X)\}$
with respect to $\mathcal{T}$ and $\Omega$.
(Here, we do not minimize also with respect to the supergraph $G$ since the sparsity pattern of the precision matrix $\Omega$ encodes $G$.)
Dropping terms that do not depend on $\mathcal{T}$ and $\Omega$,
the quantity to be minimized can be written as
\begin{equation} \label{eq:objective}
    -(\delta_G + n - 2)\log(|\Omega|)
    + \mathrm{tr}(Y\Omega Y^\top)
    + h(\Omega,\mathcal{T}) + g(\mathcal{T})
\end{equation}
where
$h(\Omega,\mathcal{T}) = \mathrm{tr}(\Omega D_G) - 2\log\{p(G\mid\mathcal{T})\}$
and
$g(\mathcal{T}) = -2\log\{p(\mathcal{T})\}$.

\begin{table}
\centering
\caption{Matching of terms in the minimization objective~\eqref{eq:objective} of graph of graphs and \eqref{eq:coarsening_objective} of \citet{Kumar2023}. Here, `supergraph' refers also to the precision matrix in the context of graph of graphs. \label{tab:graph_coarsening}}
\begin{tabular}{r|ll}
     \textbf{Role of the term} & \textbf{Equation~\eqref{eq:objective}} & \textbf{Equation~\eqref{eq:coarsening_objective}} \\
     \hline
     Regularize the supergraph & $\log(|\Omega|)$ & $\log\left(\left|L^\star + \tfrac{1}{p} 1_{p\times p}\right|\right)$ \\
     Match the supergraph to data & $\mathrm{tr}(Y \Omega Y^\top)$ & $\mathrm{tr}(X^\star L^\star X^{\star\top})$ \\
     Regularize the supergraph & $h(\Omega,\mathcal{T})$ & $h(L^\star)$ \\
     Regularize the mapping of nodes to supernodes & $g(\mathcal{T})$ & $g(C)$ \\
\end{tabular}
\end{table}

The objective functions in \eqref{eq:objective} and
\eqref{eq:coarsening_objective} are similar:
a matching of the terms in the objectives and their roles is provided in Table~\ref{tab:graph_coarsening}.
There are also notable differences.
Firstly,
the supergraph is encoded by the precision matrix $\Omega$ in \eqref{eq:objective}
and by the Laplacian matrix $L^\star$ in \eqref{eq:coarsening_objective}.
Furthermore,
instead of a (discrete) tessellation $\mathcal{T}$,
\citet{Kumar2023} consider the continuous loading matrix $C$ to map nodes to supernodes.
Finally,
\eqref{eq:objective} becomes more similar to \eqref{eq:coarsening_objective}
if we use the $K\times K$ matrix $\Omega^\star$ and the $n\times K$ matrix $Y^\star$ instead of the $p\times p$ matrix $\Omega$
and the $n\times p$ matrix $Y$
(see Sections~\ref{sec:supergraph_model} and \ref{sec:augmented} for the definitions of
$\Omega^\star$, $Y^\star$, $\Omega$ and $Y$):
in graph of graphs, $\Omega$ and $Y$ are used
instead of $\Omega^\star$ and $Y^\star$
to avoid issues with transdimensional moves.
Such moves do not appear in \citet{Kumar2023} since they fix the number of supernodes $K$.

In addition to differences in the objectives themselves, \citet{Kumar2023}
consider the constraint $L^\star = C^\top L C$ that represents a match between the known graph $H$ and the unknown supergraph $H^\star$.
In graph of graphs, there is no known uncoarsened graph $H$ to which the supergraph $G^\star$ is constrained.

\section{Concentration of the untransformed posterior}
\label{ap:conc}

We empirically show how the posterior $p(\mathcal{T},G,Y\mid X)$ in \eqref{eq:post} of the main manuscript can be very concentrated for moderate sample size $n$.
To do so, we consider $p=6$ nodes such that $p(\mathcal{T},G,Y\mid X)$ can be computed by exhaustive enumeration of all possible graphs of graphs $(\mathcal{T},G^\star)$.
We simulate $n=60$ observations by sampling independently from
$\mathcal{N}(0_{6\times 1}, \Psi^{-1})$
where $\Psi$ is a block-diagonal precision matrix with tridiagonal blocks:
$\Psi_{ii} = 1$ for $1\leq i \leq 6$, and
$\Psi_{12} = \Psi_{23} = \Psi_{45} = \Psi_{56} = 0.4$ for the nonzero superdiagonal elements
and the same for the corresponding subdiagonal elements.
Its other elements are equal to zero.
Then,
we compute $p(\mathcal{T},G,Y\mid X)$, including sequentially as rows of $X$ the first $n=10,20,30,40,50,60$ simulated observations and assuming a uniform prior on $G$.

\begin{figure}
\centering
\includegraphics[width=\textwidth]{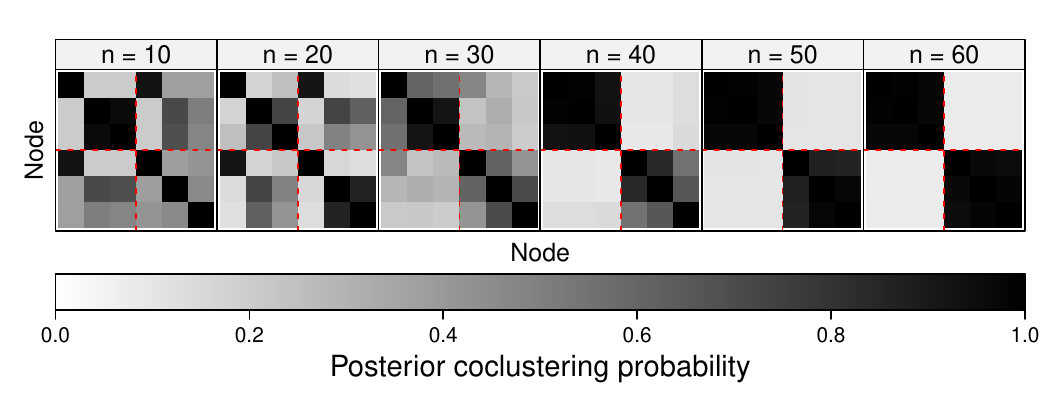}
\caption{
Simulation on the posterior concentration:
posterior co-clustering probabilities.
The panels visualize the posterior probability that a pair of nodes is allocated to the same supernode.
The dashed red lines demarcate the block structure of the matrix $\Psi$ used to simulate the data.
}
\label{fig:conc}
\end{figure}

Figure~\ref{fig:conc} summarizes the resulting posteriors on the tessellation $\mathcal{T}$.
Uncertainty in $\mathcal{T}$ decreases rapidly with $n$:
already at $n=50$, there is virtually no posterior uncertainty about the tessellation.

\section{MCMC algorithms}
\label{ap:mcmc}

We describe the Markov chain Monte Carlo (MCMC) algorithms for the two target distributions, i.e.\ $\pi^{(\zeta)}(\mathcal{T},G,Y)$ in \eqref{eq:joint_target} with the coarsened likelihood
and $\pi_\textnormal{cut}(\mathcal{T},G,Y)$
in \eqref{eq:cf_target} with nested MCMC, separately.
In both cases, we need to devise tailored computational solutions, which, nevertheless, exploit the same techniques.
Firstly, recall from Section~\ref{sec:tessellation} that the tessellation $\mathcal{T}$ is a deterministic function of the set of centers $C$.
For convenience, we choose to work directly with $C$ instead of $\mathcal{T}$ in the MCMC.

Working with $C$, we employ both a birth-death and a move Metropolis-Hastings steps.
The birth-death step uses as proposal
the addition or removal of an element from $C$,
which results in a restructuring of the supernodes configuration. These changes in $C$ are accompanied by suitable Metropolis-Hastings proposals for $G$.
Additionally, keeping the number of centers fixed, we propose to move a center from one node to another, which also implies a change in $G$, since the supernode membership might change.

\subsection{MCMC with coarsening of the likelihood}
\label{sec:joint_mcmc}

Recall that the tessellation $\mathcal{T}$ is a deterministic function of the set of centers $C$.
To enable convenient addition and removal of supernodes in $\mathcal{T}$ through changes to the allocation of centers, we consider the target distribution on $(C,G,Y)$ instead of on $(\mathcal{T},G,Y)$.
That is, the MCMC has as stationary distribution
\begin{equation} \label{eq:joint_target_C}
    \pi^{(\zeta)}(C,G,Y)\propto p^{(\zeta)}(C)\, p(G\mid C)\, p(Y\mid C,G,X)^\zeta
\end{equation}
where, analogously to \eqref{eq:prior_C},
\begin{equation} \label{eq:prior_C_coarse}
    p^{(\zeta)}(C) \propto \binom{p}{K}^{-1} \prod_{k=1}^{K} f_\textnormal{coh.}(p_k)\, f_\textnormal{sim.}^{(\zeta)}(x_k)
\end{equation}
Furthermore, $p(G\mid C) = p(G\mid \mathcal{T})$,
and
$p(Y\mid C,G,X) = p(Y\mid \mathcal{T},G,X)$ is the likelihood in \eqref{eq:likelihood} in the main manuscript.

\subsubsection{Evaluation of the target distribution}
\label{sec:eval_target}

When evaluating $\pi^{(\zeta)}(C,G,Y)$ in \eqref{eq:joint_target_C}, the expressions for $p(G\mid C)$ and $p(Y\mid C,G,X)$
include normalizing constants
while the normalizing constant is not available for
$p^{(\zeta)}(C)$.
The latter is not an issue for MCMC targeting $\pi^{(\zeta)}(C,G,Y)$ since the normalizing constant does not depend on $(C,G,Y)$
such that we can still evaluate $\pi^{(\zeta)}(C,G,Y)$ up to proportionality.

To evaluate
the likelihood $p(Y\mid C,G,X)$, we consider a factorization.
Let $G_C$ denote the subgraph of $G$ induced by the subset of nodes $C$, i.e.\ those corresponding to first principal components (PCs).
By construction, $G_C$ contains all edges of the augmented supergraph $G$ such that
\begin{multline} \label{eq:fac_lik}
    p(Y \mid C, G, X) = \frac{I_{G}(\delta_G^\star, D_G^\star)}{(2\pi)^{np/2} I_{G}(\delta_G, D_G)}
    =
    p(Y^\star \mid C, G_C, X) \prod_{i\ne\text{1st PC}} p(Y_i \mid C, X) \\
    =
    \frac{I_{G_C}(\delta^\star, D_C^\star)}{(2\pi)^{nK/2} I_{G_C}(\delta, D_C)}
    \prod_{i\ne\text{1st PC}} \frac{I_{(\{1\},\emptyset)}(\delta_G^\star,D^\star_{ii})}{(2\pi)^{n/2} I_{(\{1\},\emptyset)}(\delta_G,D_{ii})}
\end{multline}
where
$D_C$ and $D_C^\star$
are the $K\times K$ submatrices of $D$ and $D^\star$, respectively,
corresponding to the first PCs,
and $I_{(\{1\},\emptyset)}(\delta_G,D^\star_{ii})$
denotes the normalizing constant of the $G$-Wishart distribution with the graph $(\{1\},\,\emptyset)$ consisting of a single node and no edges.
Since the $G$-Wishart normalizing constant are intractable in general,
we evaluate them using
a Laplace approximation \citep{Lenkoski2011}.
Specifically,
we use the \emph{diagonal-Laplace} method from \citet{Moghaddam2009}
which uses a diagonal rather than a full Hessian matrix for speed.

We remark that $I_{(\{1\},\emptyset)}(\delta_G,D_{ii})$ and $I_{(\{1\},\emptyset)}(\delta_G^\star,D^\star_{ii})$ in \eqref{eq:fac_lik}
are available in closed form, e.g.\
$I_{(\{1\},\emptyset)}(\delta_G,D_{ii})
    = \Gamma(\delta_G/2)\, (D_{ii} / 2)^{-\delta_G / 2}$.
Nonetheless,
we use the Laplace approximation
$I_{(\{1\},\emptyset)}(\delta_G,D_{ii})
    \approx (2 \pi)^{1/2}\, e^{1 - \delta_G/2}\, (\delta_G - 2)^{(\delta_G - 1) / 2}\, D_{ii}^{-\delta_G / 2}$.
The approximation avoids inconsistency of how various parts of the likelihood are evaluated as the number of supernodes $K$ changes in the MCMC chain.

\subsubsection{Metropolis-Hastings algorithm}
\label{sec:coar_lik_MH}

We use Metropolis-Hastings to sample from $\pi^{(\zeta)}(C,G,Y)$ in \eqref{eq:joint_target_C}.
For the set of centers $C$, we alternate between two types of Metropolis-Hasting proposals: (i) a birth-death
step that considers the addition/deletion of a node in $C$ and (ii) a move step.
The birth-death step gives rise to the birth or death of the corresponding supernode and thus requires a corresponding proposed change in the supergraph $G$.
This is also true in the move step, where a randomly selected center $i\in C$ is moved to a node in $(V\setminus C)$ ($V=\{1,\dots,p\}$). This is because the center membership might change with a move step, and so should the superedges. We account for these changes by identifying those supernodes for which the variables assignments remain unchanged after the birth-death or move proposal, i.e. those groups of variables that are still clustered together after the proposed step. We indicate this set of centers as $A_{\text{old}}$. Similarly, we define the set $A_{\text{new}}$ as containing those supernodes for which the variables have been reassigned. Therefore, we write the proposal as follows:
\begin{gather*}
q_\textnormal{bdm}(C',G'; C,G) = q_G(G' \mid C'; C,G)\,q_C(C' \mid C, G) = \\
q_{\text{old}}(G'_{\text{old}} \mid C'; C,G)\,q_{\text{new}}(G'_{\text{new}} \mid C'; C,G)\,q_C(C' \mid C, G)
\end{gather*}
where $q_C$ indicates the proposal obtained by adding/removing/moving a center in the partition uniformly at random among the possible centers in the current configuration. The part of the proposal indicated by $q_G$, referring to the supergraph, is instead split into two parts corresponding to the sets $A_{\text{old}}$ and $A_{\text{new}}$ introduced above. The proposal $q_G$ is obtained by starting with a graph equal to $G$ and where the edges connecting the supernodes in $A_{\text{old}}$ are unchanged, while all other possible superedges are resampled with proposal probability equal to $\xi_q$. Therefore, changes to the graph structure only affect the set of edges where at least one node is in the set $A_{\text{new}}$.

We also update $G$ by itself with a Metropolis-Hastings proposal that adds or removes one superedge at a time, after each move or birth/death step.
Algorithm~\ref{alg:joint_mcmc} summarizes the resulting MCMC.

We consider in Sections~\ref{sec:application} and Appendix~\ref{ap:simul} the Erd\"{o}s-R\'{e}ny prior for $G$: $p(G\mid \mathcal{T}) \propto \xi_{\text{se}}^{\mid E^\star \mid} \left(1 - \xi_{\text{se}}\right)^{\binom{K}{2} - \mid E^\star \mid}$ with superedge inclusion probability $\xi_{\text{se}}$.

\begin{algorithm}
\caption{MCMC step with $\pi^{(\zeta)}(C,G,Y)$ in \eqref{eq:joint_target_C} as invariant distribution \label{alg:joint_mcmc}}
\begin{enumerate}
    \item
    \emph{Birth-death step:}
    Propose a birth or a death with equal probability, or as dictated by $K=|C|=1$ (birth) or $K=p$ (death):
    \begin{enumerate}
        \item
        \emph{Birth:}
        Generate a proposal $C'$ by adding a uniformly sampled element from $V\setminus C$ to $C$. Recompute the tessellation. Identify the sets $A_{\text{old}}$ and $A_{\text{new}}$.
        Generate a corresponding supergraph proposal $G'$ by creating edges with at least one supernode in the set $A_{\text{new}}$, with probability $\xi_q$.
        \item
        \emph{Death:}
        Generate a proposal $C'$ by removing a uniformly sampled element from $C$. Recompute the tessellation. 
        Identify the sets $A_{\text{old}}$ and $A_{\text{new}}$.
        Generate a corresponding supergraph proposal $G'$ by creating edges with at least one supernode in the set $A_{\text{new}}$, with probability $\xi_q$.
        \item
        Compute $Y'$ corresponding to the proposal $(C',G')$.
        Accept $(C',G',Y')$, i.e.\ set $C = C'$, $G=G'$ and $Y=Y'$, with probability
    \[
        \min\left\{1,\, \frac{\pi^{(\zeta)}(C',G',Y')\, q_\textnormal{bdm}(C',G'; C,G)}{\pi^{(\zeta)}(C,G,Y)\, q_\textnormal{bdm}(C,G; C',G')}\right\}
    \]
    \end{enumerate}
    \item
    \emph{Move step:} If $K<p$,
    sample an element $i$ uniformly at random from $C$. Then propose to ``move'' the associated supernode to a new center by sampling a new center from $(V\setminus C)$ uniformly at random. Recompute the tessellation. Identify the sets $A_{\text{old}}$ and $A_{\text{new}}$. Generate a corresponding supergraph proposal $G'$ by creating edges with at least one node in the set $A_{\text{new}}$, with probability $\xi_q$.
    Accept this proposal $(C',G',Y')$ with probability
    \[
        \min\left\{1,\, \frac{\pi^{(\zeta)}(C',G',Y')\, q_\textnormal{bdm}(C',G'; C,G)}{\pi^{(\zeta)}(C,G,Y)\, q_\textnormal{bdm}(C,G; C',G')}\right\}
    \]
    \item \label{step:supergraph}
    \emph{Supergraph move}:  after each birth/death and after each move step,
    sample a pair of supernodes uniformly at random.
    Then,
    generate a proposal $G'$ by changing
    whether this pair is connected by a superedge.
    Accept, i.e.\ set $G=G'$, with probability
    \[
        \min\left\{1,\, \frac{\pi^{(\zeta)}(C,G',Y)}{\pi^{(\zeta)}(C,G,Y)}\right\}
    \]
\end{enumerate}
\end{algorithm}

\clearpage

\subsection{Nested MCMC}

Recall that the cut distribution in \eqref{eq:cf_target}
factorizes as
$\pi_\textnormal{cut}(\mathcal{T},G,Y)\propto p^{(\zeta)}(\mathcal{T})\, p(G,Y\mid \mathcal{T},X)$.
We can therefore use
nested MCMC \citep{Plummer2015,Carmona2020}
for posterior computation,
with an \emph{inner MCMC} nested inside iterations of an \emph{outer MCMC}:
the outer MCMC first samples from
the coarsened data-coherent size-biased tessellation prior
$p^{(\zeta)}(\mathcal{T})$.
Then, the inner MCMC samples supergraphs from the conditional posterior $p(G,Y\mid \mathcal{T},X)$
using the tessellations $\mathcal{T}$ from the outer MCMC.

For the outer MCMC,
we consider birth-death and move Metropolis-Hastings steps similar to Algorithm~\ref{alg:joint_mcmc}.
As in Section~\ref{sec:joint_mcmc},
it is more convenient to work with the set of centers $C$ instead of $\mathcal{T}$.
Thus, we consider $p^{(\zeta)}(C)$ in \eqref{eq:prior_C_coarse} as target distribution.

We summarize the outer MCMC in Algorithm~\ref{alg:outer_mcmc} where
$q_\textnormal{bd}(C; C')$ denotes  the birth-death or move Metropolis-Hastings proposal.

\begin{algorithm}
\caption{Outer MCMC step with $p^{(\zeta)}(C)$ in \eqref{eq:prior_C_coarse} as invariant distribution \label{alg:outer_mcmc}}
\begin{enumerate}
    \item
    \emph{Birth-death step:}
    Propose a birth or death with equal probability, or as dictated by $K=|C|=1$ (birth) or $K=p$ (death):
    \begin{enumerate}
        \item
        \emph{Birth:}
        Generate a proposal $C'$ by adding a uniformly sampled element from $V\setminus C$ to $C$.
        \item
        \emph{Death:}
        Generate a proposal $C'$ by removing a uniformly sampled element from $C$.
        \item
        Accept the proposal $C'$, i.e.\ set $C = C'$, with probability
    \[
        \min\left\{1,\, \frac{p^{(\zeta)}(C')\, q_\textnormal{bd}(C'; C)}{p^{(\zeta)}(C)\, q_\textnormal{bd}(C; C')}\right\}
    \]
    \end{enumerate}
    \item
    \emph{Move step:} If $K<p$,
    sample an element $i$ uniformly at random from $C$. Then propose to ``move'' the associated supernode to a new center by sampling a new center from $(V\setminus C)$ uniformly at random.
    Accept this proposal $C'$ with probability
    \[
        \min\left\{1,\, \frac{p^{(\zeta)}(C') }{p^{(\zeta)}(C) }\right\}
    \]
\end{enumerate}
\end{algorithm}

For the inner MCMC,
consider
\begin{multline} \label{eq:inner_mcmc}
    p(G,Y\mid \mathcal{T},X) = p(G,Y\mid C,X)
    \propto p(G\mid C)\, p(Y\mid C,G,X) \\
    \propto p(G\mid C)\, p(Y^\star \mid C, G_C, X)
    \propto p(G\mid C)\, 
    \frac{I_{G_C}(\delta^\star, D_C^\star)}{(2\pi)^{nK/2} I_{G_C}(\delta, D_C)}
\end{multline}
where the second line follows from \eqref{eq:fac_lik}.
Recall that, conditionally on $C$,
there is a one-to-one relation between $G_C$ and $G$.
Also,
the last expression in \eqref{eq:inner_mcmc}
corresponds to the posterior of a Bayesian GGM with graph $G_C$ and data matrix $Y^\star$.
Thus, we can sample $G_C$ and therefore $G$
from $p(G\mid \mathcal{T},X)$
using an MCMC algorithm for GGMs from \citet{vandenBoom2022a}.

To combine the outer and inner MCMC,
we follow
\citet{Carmona2020}
and run the inner MCMC only after discarding of burn-in iterations and thinning in the outer MCMC.
This reduces computation time substantially by limiting the number of times the inner MCMC chain needs to be run without notably affecting the quality of inference.
Algorithm~\ref{alg:nested_mcmc} details the resulting nested MCMC.
For the inner MCMC,
the number of iterations $N_\textnormal{inner}$ should be large enough for $G^{(s)}$
to be approximately distributed according to $p(G,Y^{(s)}\mid \mathcal{T}^{(s)},X)$.
That is,
$N_\textnormal{inner}$ can be interpreted as the number of burn-in iterations used with the MCMC from \citet{vandenBoom2022a}.
See Section~\ref{sec:trace} for MCMC diagnostics on whether $N_\textnormal{inner}$ is large enough in the gene expression application.
Note that Step~\ref{step:inner} of Algorithm~\ref{alg:nested_mcmc} is embarrassingly parallel as the MCMC can be run independently for each $s$.

\begin{algorithm}
\caption{Nested MCMC to generate a sample $\{\mathcal{T}^{(s)},G^{(s)}\}_{s=1}^{N_\textnormal{outer}}$ from
$\pi_\textnormal{cut}(\mathcal{T},G,Y)$ in \eqref{eq:cf_target} \label{alg:nested_mcmc}}
\begin{enumerate}
    \item
    \emph{Outer MCMC:}
    Generate
    $N_\textnormal{outer}$
    MCMC samples
    $\{C^{(s)}\}_{s=1}^{N_\textnormal{outer}}$
    from $p^{(\zeta)}(C)$
    by iterating Algorithm~\ref{alg:outer_mcmc},
    discarding burn-in iterations and thinning.
    Record the corresponding tessellations $\{\mathcal{T}^{(s)}\}_{s=1}^{N_\textnormal{outer}}$.
    \item \label{step:inner}
    \emph{Inner MCMC:}
    For $s=1,\dots,N_\textnormal{outer}$:
    \begin{enumerate}
        \item
        Compute $Y^{(s)}$ corresponding to $\mathcal{T}^{(s)}$.
        \item
        Run MCMC from \citet{vandenBoom2022a}
        with respect to $p(G,Y^{(s)}\mid \mathcal{T}^{(s)},X)$
        for $N_\textnormal{inner}$ iterations.
        \item
        Record the last sample as
    $G^{(s)}$.
    \end{enumerate}
\end{enumerate}
\end{algorithm}

\section{Simulation studies for the graph of graphs model}
\label{ap:simul}

In this section, we apply the MCMC methods described in Appendix~\ref{ap:mcmc} to simulated data.
We generate data 
by using latent factors shared across members of supernodes in Section~\ref{sec:simul_lat_fac}.
We use the same model specification and MCMC settings as for the application in Section~\ref{sec:application}, with the following differences.
Since we simulate data with larger average supernode size than in Section~\ref{sec:application}, we
use a lower success probability of $1/p$ in the Negative Binomial distribution for the cohesion function $f_\textnormal{coh.}(p_k)$ in the tessellation prior. For the joint MCMC update, we find that the same coarsening used  in Section~\ref{sec:application} equal to $10/n$ yields satisfactory results. For the nested MCMC, we opt for a coarsening of $1/n$ selected through sensitivity analysis (results not shown).
As for posterior computation, we run the MCMC algorithm for $20000$ iterations, discarding the first $15000$ as burn-in and employing a thinning of 5 for the inner part of the nested MCMC. 

Moreover, in Section~\ref{sec:simul_2_step}, we compare our results with two-step approaches.

\subsection{Simulation based on latent factors}
\label{sec:simul_lat_fac}

We generate data $X$ using latent factors.
Specifically, we simulate an $n\times 3$ matrix
$Z$ corresponding to three latent factors. The columns of $Z$ represent supernodes. The rows of $Z$ are sampled independently from a three-dimensional Gaussian distribution with mean zero and the precision matrix corresponding to either not having any superedges (i.e., the identity matrix) or to having only one superedge. Specifically, in the latter case, we set the diagonal elements of the precision matrix equal to 1 and the element between the first and third variable equal to 0.9, thus introducing a superedge connecting blocks one and three. The three blocks are of sizes 5, 10 and 20, respectively, yielding a total of $p = 35$ variables. We simulate $n = 1000$ data points.
For $i = 1, \dots, n$ and $j = 1, \dots, p$, we simulate the data matrix $X$ as follows:
\begin{align*}
&X_{ij} = Z_{ib_j} + \epsilon_{ij} \\   
&\epsilon_{ij} \overset{\textnormal{i.i.d.}}{\sim}\mathcal{N}\left(0, \sigma^2_{\epsilon}\right)
\end{align*}
where $b_j \in \left\{1, 2, 3\right\}$ indicates the block to which the $j$\textsuperscript{th} variable belongs to, and $\sigma^2_{\epsilon} \in \left\{0.01, 0.05\right\}$.
Finally, we standardize the columns of $X$ to have unit standard deviation.

\begin{figure}
\centering
\includegraphics[width=0.55\textwidth]{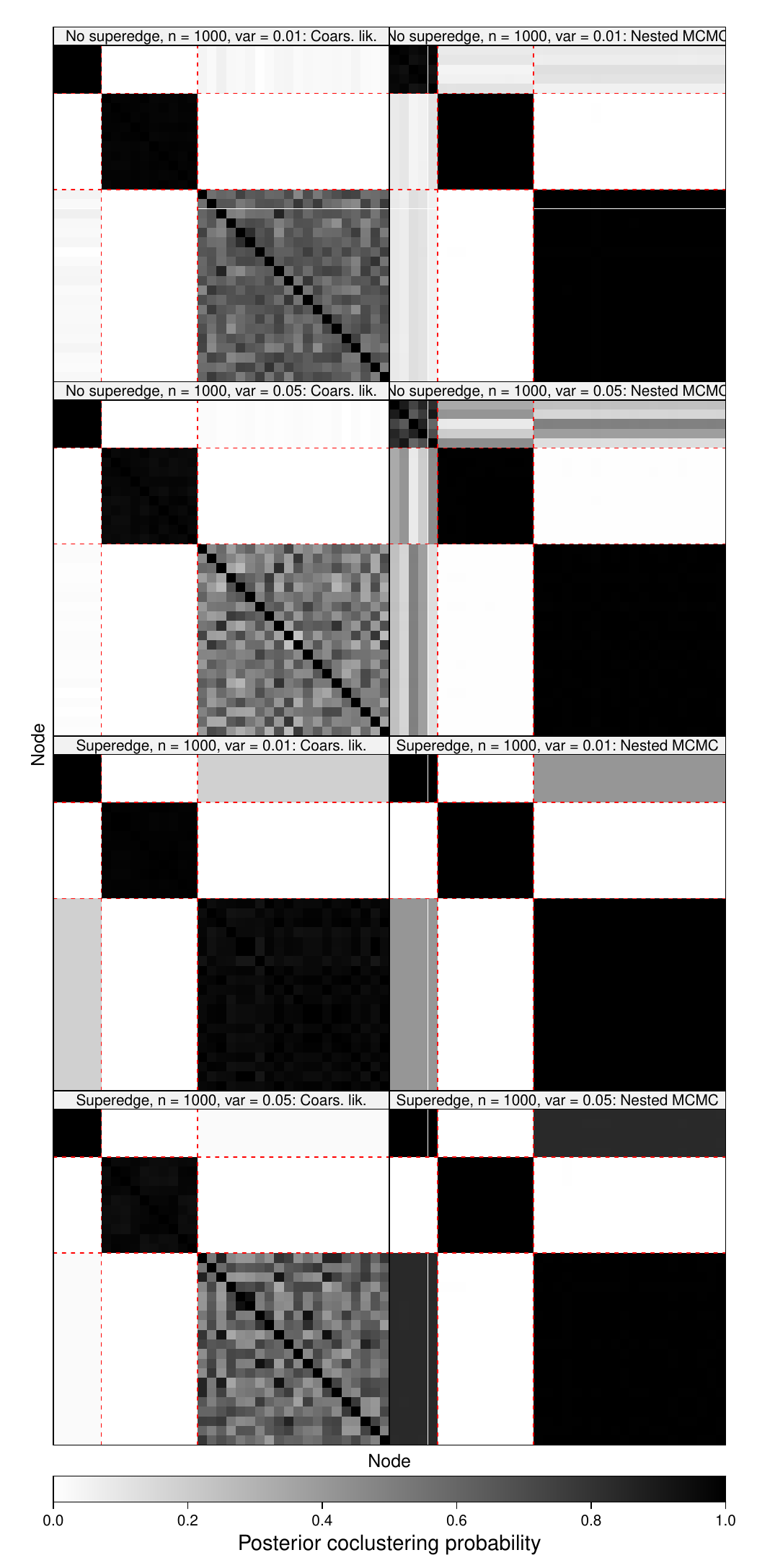}
\caption{
Simulation based on latent factors:
posterior co-clustering probabilities.
The panels display the posterior probability that a pair of nodes is allocated to the same supernode.
The dashed red lines indicate the latent factor structure used to simulate the data.
}
\label{fig:simul_lat_fac_post_sim}
\end{figure}

\begin{figure}
\centering
\includegraphics[width=0.55\textwidth]{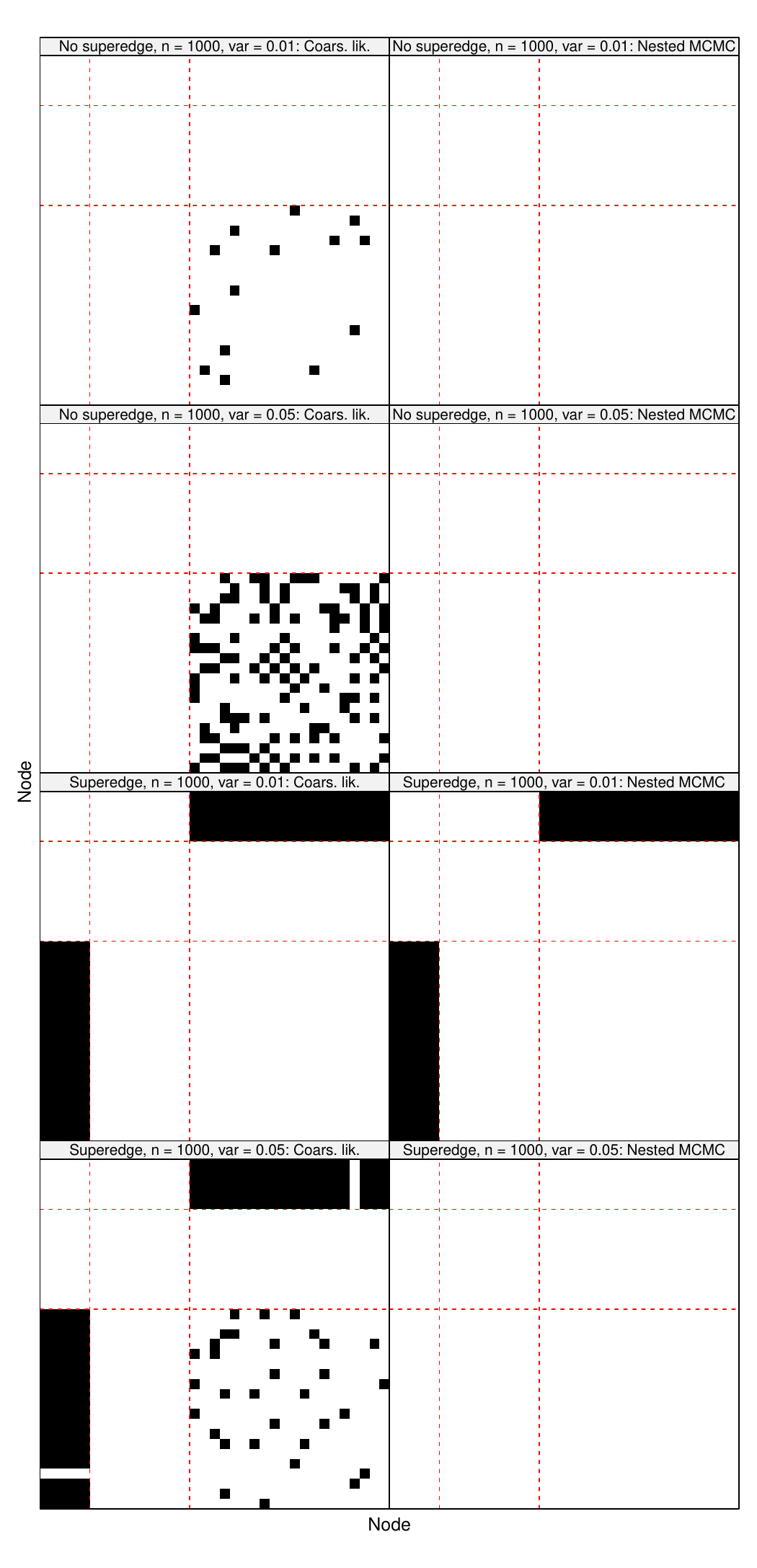}
\caption{
Simulation based on latent factors:
posterior estimate of the supergraph.
The panels display the superedges in the supergraph $G^\star$ with posterior inclusion probability greater than $0.5$. The posterior probabilities considered are those between nodes belonging to pairs of supernodes connected in the supergraph.
The dashed red lines indicate the latent factor structure used to simulate the data.
}
\label{fig:simul_lat_fac_postedge}
\end{figure}

We summarize posterior inference in Figures~\ref{fig:simul_lat_fac_post_sim} and \ref{fig:simul_lat_fac_postedge}
where the joint MCMC with coarsening of the likelihood is denoted by ``Coars.\ lik.''.
The group structure is accurately recovered in all settings, with better results obtained with the nested MCMC, as shown in Figure~\ref{fig:simul_lat_fac_post_sim}.
Regarding supergraph inference, Figure~\ref{fig:simul_lat_fac_postedge}
shows the identification of the correct supergraph, with the exception of one of the nested MCMC cases, and with some uncertainty in the joint MCMC update part. The results improve for smaller values of the variance parameter $\sigma^2_{\epsilon}$.

\clearpage

\subsection{Comparison with two-step approaches}
\label{sec:simul_2_step}

We apply two-step approaches to the simulated data from
Sections~\ref{sec:simul_lat_fac}.
Specifically, 
we perform the following two steps:
\begin{enumerate}
    \item
    Estimate a graph on $p$ nodes by fitting the graphical lasso on $X$ with
    the regularization parameter set using five-fold cross-validation, which is the default in the R package \texttt{CVglasso} \citep{Galloway2018}.
    \item
    Cluster the $p$ nodes based on the estimated graph.
\end{enumerate}
For clustering of nodes, we consider two approaches: one based on \emph{edge betweenness} \citep{Newman2004} and the other based on the \emph{leading eigenvector} of the modularity matrix of the graph \citep{Newman2006}.

The clustering method based on edge betweenness performs hierarchical clustering by repeatedly removing edges and keeping track of the connected components of the resulting graph. Then, the connected components correspond to clusters of nodes. The removal of edges is based on edge betweenness, i.e.\ the number of shortest paths between any pair of nodes that passes through the edge. At each iteration, the edge with the highest betweenness is removed.
From the resulting sequence of partitions of nodes,
the partition which maximizes the
\emph{modularity} is reported \citep{Newman2004}.

The second approach considers the leading eigenvector of the modularity matrix \citep{Newman2006}. 
Then, nodes are partitioned into two communities based on the signs of the elements in the eigenvector using the fact that each element corresponds to a node.
Such splits into two communities are repeated until the modularity of the node partition no longer increases \citep[see][for details]{Newman2006}.
The clustering methods are implemented in the R package \texttt{igraph} \citep{Csardi2023}.

\begin{figure}
\centering
\includegraphics[width=0.82\textwidth]{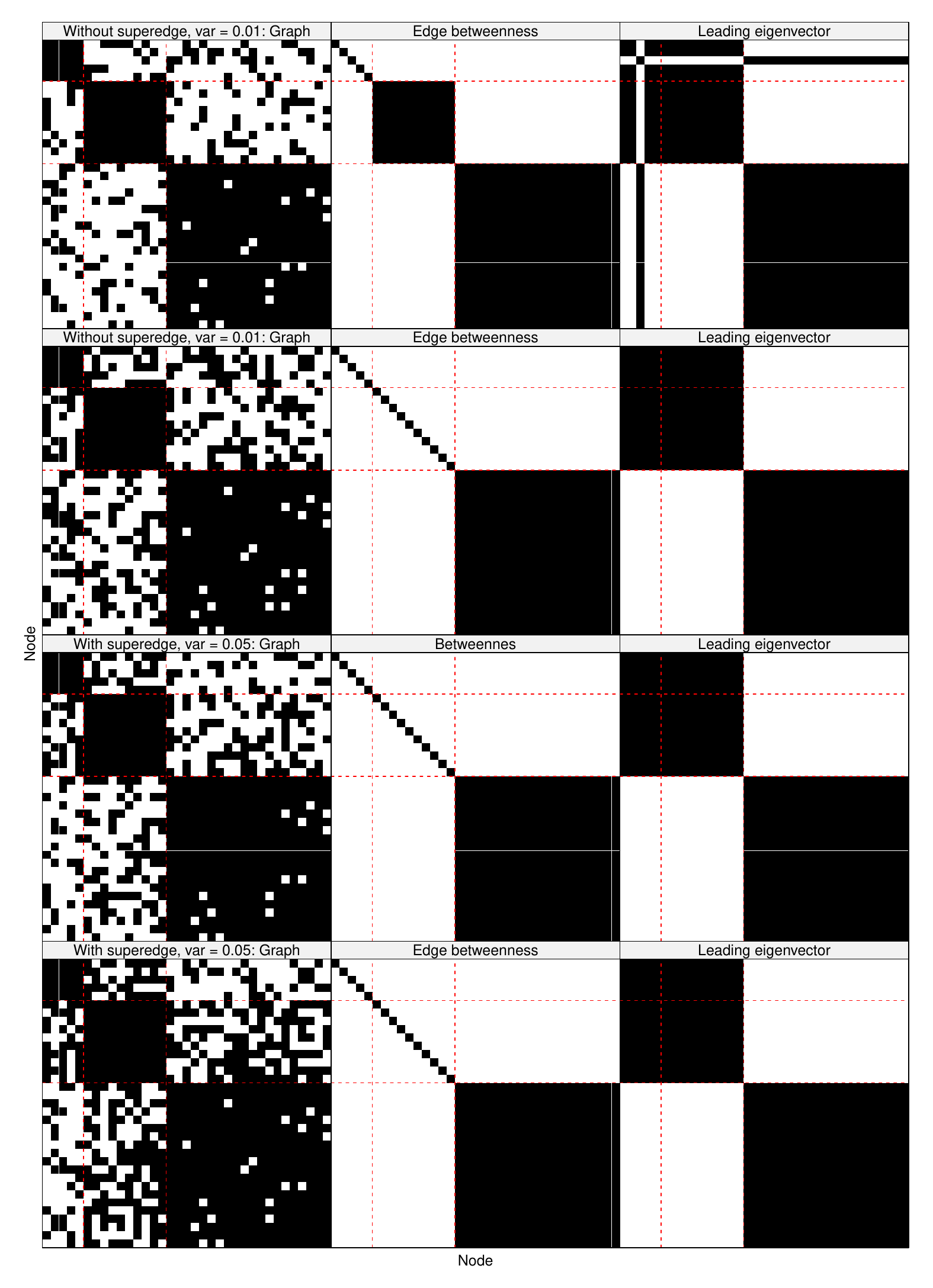}
\caption{
Simulation based on latent factors:
results from two-step approaches.
The left column displays the adjacency matrices of the graphical lasso estimates.
The middle and right columns show the coclustering matrices corresponding to the node partition based on (i) edge betweenness and (ii) the leading eigenvector of the modularity matrix, respectively.
The dashed red lines indicate the latent factor structure used to simulate the data.
\label{fig:simul_lat_fac_2_step}
}
\end{figure}

Figure~\ref{fig:simul_lat_fac_2_step} 
show the results on the simulated data from
Section~\ref{sec:simul_lat_fac}.
The graph estimates from graphical lasso yield partitions of nodes that can be traced back to the true simulation setting.
At the same time, in terms of recovery of these partitions, we see that the two-step approaches perform worse than our methodology
(compare with Figure~\ref{fig:simul_lat_fac_post_sim}).

\clearpage

\subsection{Comparison with a similarity function representing a complete graph}
\label{sec:simul_mvt}

The tree activation $\widetilde{p}(x_k)$
is defined in
Section~\ref{sec:tree_activation}
though a GGM with a tree structure imposed on the graph.
Such graph structure (i) gives rise to the graph of graphs; (ii) results in a $\widetilde{p}(x_k)$ that captures strength of correlations in $x_k$ (see Proposition~\ref{prop:mar_lik} and its discussion in Section~\ref{sec:tree_activation}); (iii) describes the conditional independence structure within a supernode.
To further understand the implications of assuming such structure,
we here compare with results obtained by using a similarity function $\widetilde{p}(x_k)$ that does not involve a distribution over graphs.

We now assume that $T_k$ corresponds to the complete graph. Then, no sparsity is imposed on the precision matrix $\Delta_k$.
Furthermore, $\widetilde{p}(x_k)$ is now defined by
\begin{align*}
    \widetilde{p}(x_k\mid \Delta_k ) &= \prod_i \mathcal{N}(x_{ik}\mid 0_{p_k\times 1},\, \Delta_k^{-1}) \\
     \widetilde{p}(x_k) &=  
     \int \widetilde{p}(x_k\mid \Delta_k )\, \widetilde{p}(\Delta_k)\, d\Delta_k
\end{align*}
where
$\widetilde{p}(\Delta_k)$
is a Wishart distribution with degrees of freedom\footnote{The parameterization is as in \citet{Roverato2002}. More commonly, the Wishart distribution is parameterized with degrees of freedom $\nu = \delta + p_k - 1$.}
$\delta>0$ and positive-definite rate matrix $D$ as in Section~\ref{sec:tree_activation}:
\[
    \widetilde{p}(\Delta_k)
    =
    \frac{
    |D|^{(\delta + p_k - 1) / 2}\,
    |\Delta_k|^{\delta/2 - 1}
    }{
    2^{(\delta + p_k - 1)p_k / 2}\,
    \Gamma_{p_k}\!\left(\frac{\delta + p_k - 1}{2}\right)
    }
    \exp\left\{-\frac{1}{2}\mathrm{tr}(\Delta_k D)\right\}
\]
where $\Gamma_{p_k}(\cdot)$ is the multivariate gamma function.
Then,
$\widetilde{p}(x_k)$
is a Matrix $t$-distribution \citep[Theorem~3.1]{Dickey1967}.
Specifically,
due to conjugacy,
we obtain
\[
    \widetilde{p}(x_k)
    = \frac{
    \Gamma_{p_k}\!\left(\frac{\delta^\star + p_k - 1}{2}\right) |D|^{(\delta + p_k - 1)/2}
    }{
    \pi^{np_k/2}\,
    \Gamma_{p_k}\!\left(\frac{\delta + p_k - 1}{2}\right)
    |D^\star|^{(\delta^\star + p_k - 1)/2}
    }
\]
with
$\delta^\star = \delta + n$ and $D^\star = D + x_k^\top x_k$ as in Section~\ref{sec:tree_activation}.
Now,
using the matrix determinant lemma,
\[
    \widetilde{p}(x_k) \propto |D^\star|^{-(\delta^\star + p_k - 1)/2}
    \propto |I_n + x_k D^{-1} x_k^\top|^{-(\delta^\star + p_k - 1)/2}
\]
such that
we recognize
$\widetilde{p}(x_k)$
as a Matrix $t$-distribution
with degrees of freedom $\delta$,
mean zero, ``rows'' scale matrix $I_n$
and ``columns'' scale matrix $D$.

We apply the graph of graphs methodology
to simulated data
as in Sections~\ref{sec:simul_lat_fac},
but now using as similarity function the coarsened Matrix $t$ distribution
$f_\textnormal{sim.}^{(\zeta)}(x_k) = {\widetilde{p}(x_k\mid T_k)^{\zeta}}\, \widetilde{p}(T_k) = \widetilde{p}(x_k)^\zeta$
instead of the
$f_\textnormal{sim.}^{(\zeta)}(x_k)$
involving the tree activation function in Equation~\eqref{eq:coarsened_sim} of the main manuscript.

The posterior places all nodes in a single supernode.
Specifically, $K=1$ in almost all 5000 recorded MCMC iterations, both with coarsening of the likelihood and with nested MCMC (in more than 90.0\% of iterations).
This contrasts with the results of the tree activation function in Figure~\ref{fig:simul_lat_fac_post_sim}.
A possible explanation for the disagreement in the inference on $K$ is a difference in coarsening: here, the coarsening results in $f_\textnormal{sim.}^{(\zeta)}(x_k) = \widetilde{p}(x_k\mid T_k)^{\zeta}\, \widetilde{p}(T_k) = \widetilde{p}(x_k)^\zeta$.
Instead,
when using the tree activation function,
$f_\textnormal{sim.}^{(\zeta)}(x_k) = \sum_{T_k} \widetilde{p}(x_k\mid T_k)^{\zeta}\, \widetilde{p}(T_k) < \widetilde{p}(x_k)^\zeta$ where the latter follows from Jensen's inequality for $\zeta < 1$. As discussed in Section~\ref{sec:prior_coarsening}, coarsening is instrumental in avoiding that the posterior concentrates on $K=1$ or $K=p$.
Moreover, no superedges are estimated (results not shown).

\clearpage

\section{Application to gene expression data}
\label{ap:gene}

\subsection{Details on the MCMC}
\label{sec:trace}

\begin{figure}
\centering
\includegraphics[width=\textwidth]{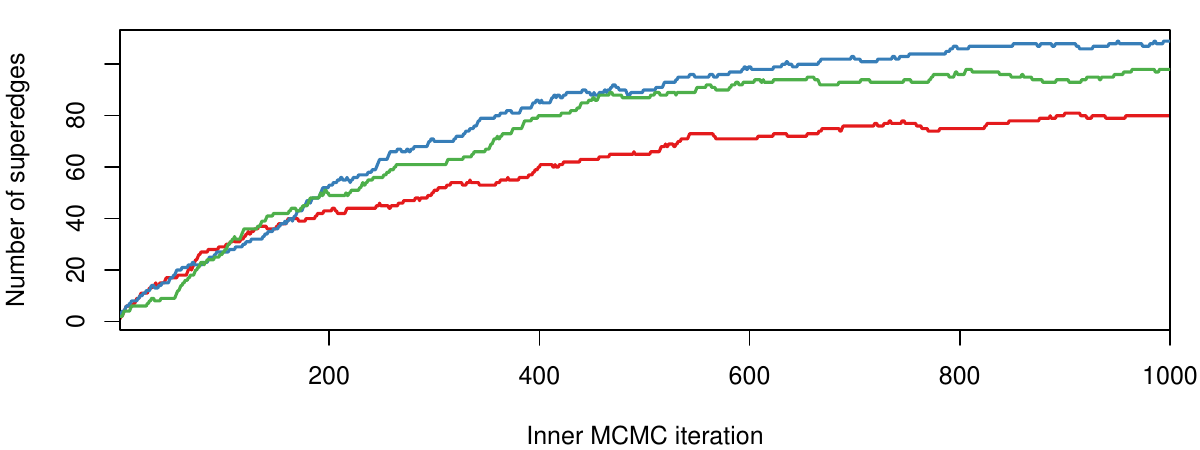}
\caption{
Gene expression data:
trace plots of the number of superedges from the inner MCMC for three iterations of the outer MCMC from the nested MCMC. The inner MCMC chain is initialized to an empty graph.
}
\label{fig:trace_inner}
\end{figure}

For Step~2 of the nested MCMC in Algorithm~\ref{alg:nested_mcmc},
we thin the outer MCMC every 10 iterations resulting in $N_\textnormal{outer} = 1000$
iterations after burn-in and thinning.
Then, we run the MCMC from \citet{vandenBoom2022a} for $N_\textnormal{inner} = 1000$ iterations.
The trace plots in Figure~\ref{fig:trace_inner} suggest that this choice of $N_\textnormal{inner}$ is large enough in the sense that convergence of the inner MCMC is satisfactory.

\begin{figure}
\centering
\includegraphics[width=\textwidth]{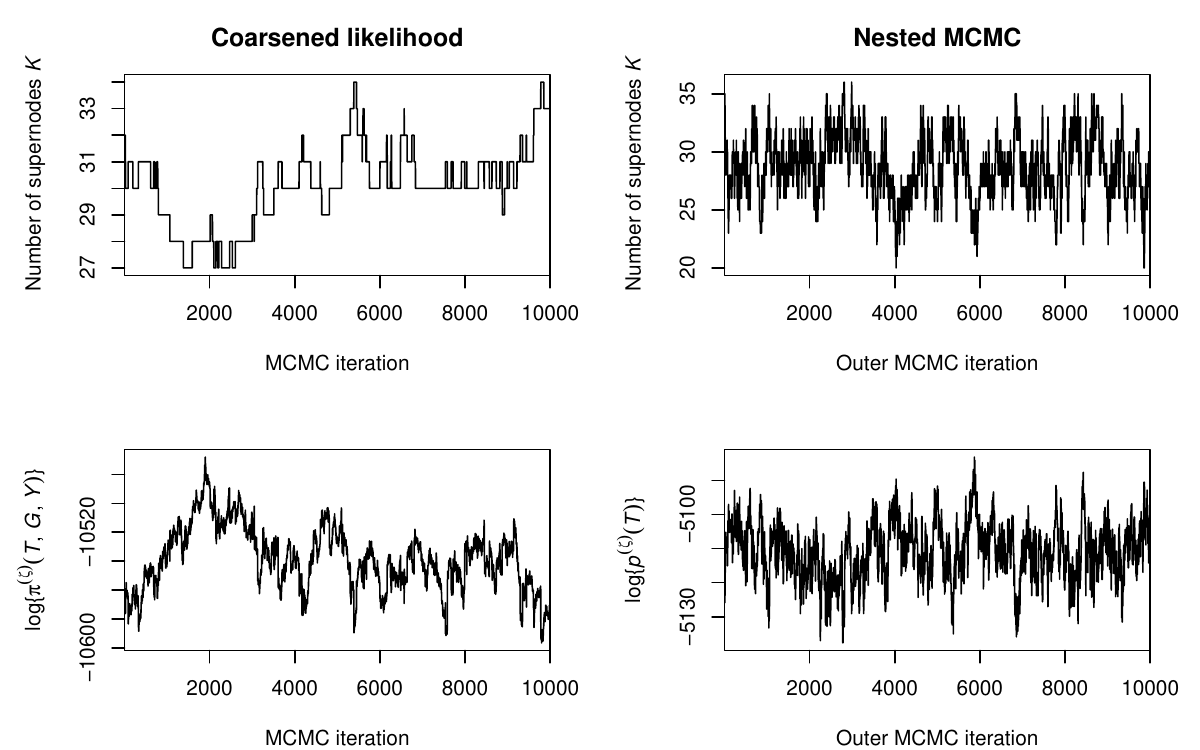}
\caption{
Gene expression data:
trace plots of the number of supernodes and the log of the target distribution for the MCMC iterations after burn-in with coarsening of the likelihood and nested MCMC.
}
\label{fig:trace}
\end{figure}

Figure~\ref{fig:trace} shows trace plots for the MCMC with coarsening of the likelihood and nested MCMC. They suggest that the number of burn-in iterations and total number of iterations are large enough for, respectively, MCMC convergence and sufficient MCMC mixing.

\subsection{Inference on the graph of graphs}

\begin{figure}
\centering
\includegraphics[width=\textwidth]{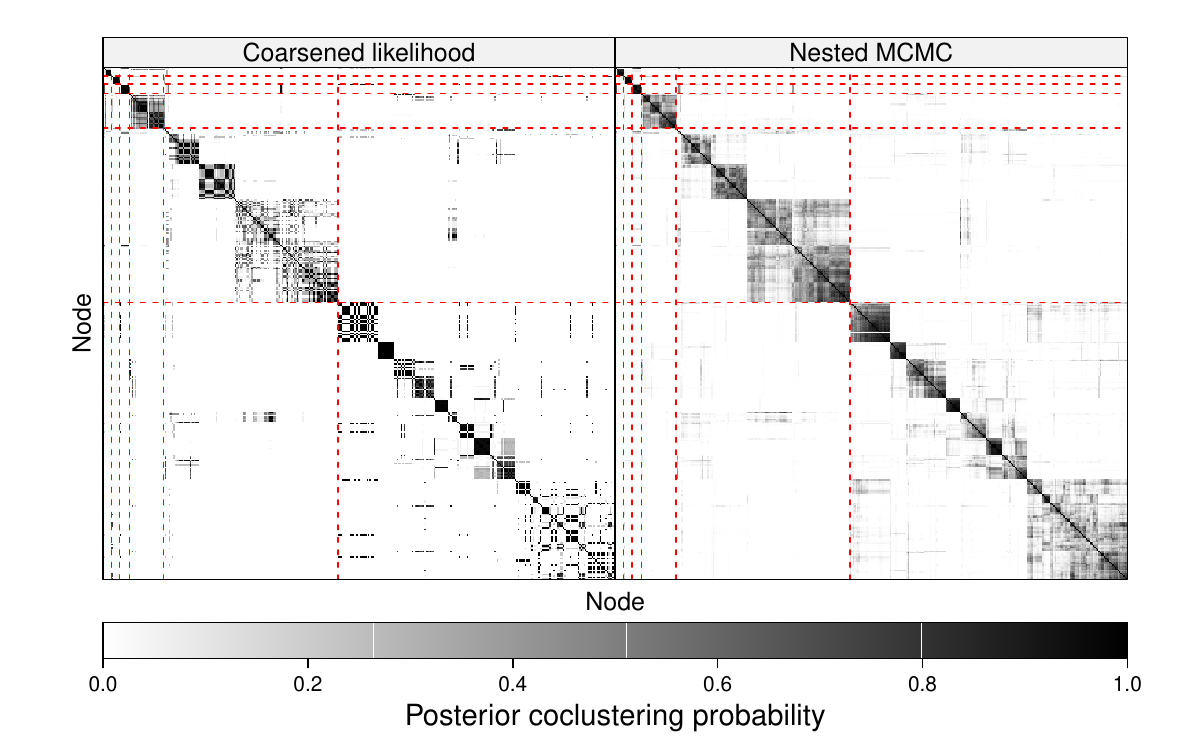}
\caption{
Gene expression data:
posterior co-clustering probabilities.
The panels visualize the posterior probability that a pair of genes (i.e.\ nodes) is allocated to the same supernode.
The dashed red lines demarcate the modules estimated by \citet{Zhang2018}.
}
\label{fig:gene_post_sim}
\end{figure}

We summarize the inference on the tessellation in Figure~\ref{fig:gene_post_sim}.
Nested MCMC results in larger supernodes than coarsening of the likelihood.
Both inference methods result in a finer splitting of genes into supernodes than \citeauthor{Zhang2018}'s modules, but are otherwise consistent with \citet{Zhang2018}.
The two methods also show consistency with each other:
if we consider
the point estimate for the tessellation that minimizes the lower bound to the posterior expectation of the variation of information in \citet{Wade2018}, then the Rand index \citep{Rand1971} between the two estimated tessellations is 0.945.

\begin{figure}
\centering
\includegraphics[width=\textwidth]{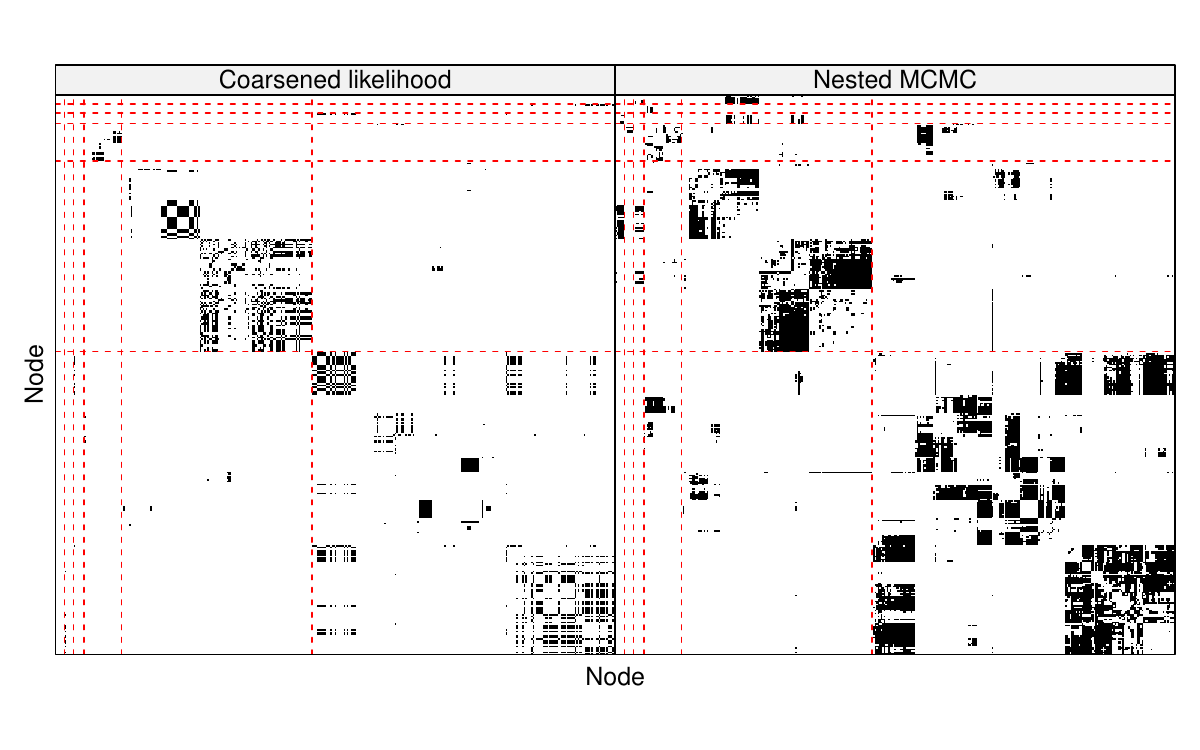}
\caption{
Gene expression data:
posterior estimate of the supergraph.
The panels display the superedges in the supergraph $G^\star$ with posterior inclusion probability greater than $0.5$. The posterior probabilities considered are those between nodes belonging to pairs of supernodes connected in the supergraph.
The dashed red lines demarcate the modules estimated by \citet{Zhang2018}.
}
\label{fig:gene_postedge}
\end{figure}

In the MCMC chains, the number of supernodes and their composition are not fixed.
To summarize
inference on the supergraph,
we first consider, for any pair of nodes,
the posterior probability that they belong to supernodes that are connected by a superedge. Then, we show in Figure~\ref{fig:gene_postedge} the supergraph obtained by considering only those connections whose probability is greater than 0.5.
In general, both methods estimate superedges between supernodes belonging to the same modules, as specified by \cite{Zhang2018}.
With coarsening of the likelihood,
the estimated supergraph connects small supernodes belonging to the two biggest modules in \cite{Zhang2018}.
This behavior is more pronounced in the case of the nested MCMC, where bigger supernodes are grouped together, thanks to the estimated tessellation being composed of fewer and bigger clusters.

\subsection{Supergraph estimate and gene interactions}

To further inspect the supergraph estimate in Figure~\ref{fig:gene_sphere} of the main manuscript,
we compare the prevalence of known gene interactions
(i) within supernodes; (ii) between supernodes that are connected by a superedge and (iii) between supernodes that are not connected by a superedge.
Interactions are taken from the STRING database version 12.0 \citep{Szklarczyk2021}.
It contains gene interactions deriving from a variety of sources including interactions of proteins related to the genes and text mining of scientific literature for co-occurrence of gene names.
Here, we extract gene interactions using the default settings of the Bioconductor R package \texttt{STRINGdb}.

\begin{figure}
\centering
\includegraphics[width=0.5\textwidth]{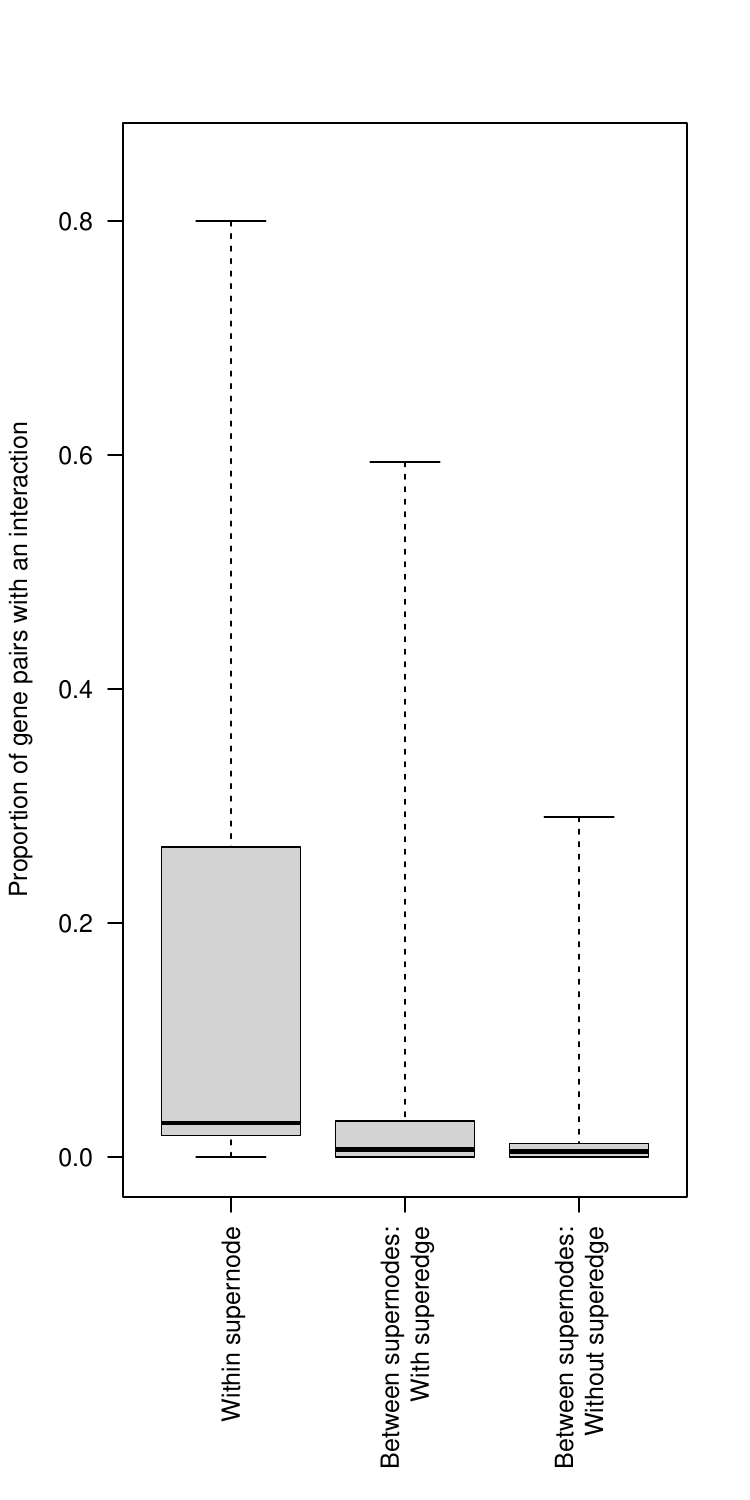}
\caption{
Gene expression data:
box plots of
the proportion of gene pairs that have a gene interaction in the STRING database for all pairs (i) coming from the same supernode, (ii) coming from different supernodes that are connected by a superedge, and (iii) coming from different supernodes that are not connected by a superedge.
The whiskers indicate the range of the proportions.
}
\label{fig:gene_interactions}
\end{figure}

We summarize the prevalence of gene interactions within and between supernodes in Figure~\ref{fig:gene_interactions}.
Gene interactions are most common within a supernode, again suggesting biological validity to the clustering of nodes.
Also, gene interactions exist more often between supernodes that are connected by a superedge than between supernodes that are not connected.
Thus, superedges seem to reflect biological mechanisms between the groups of genes that are defined by the supernodes.

\subsection{Gene Ontology overrepresentation analysis}

\begin{figure}
\centering
\includegraphics[width=\textwidth]{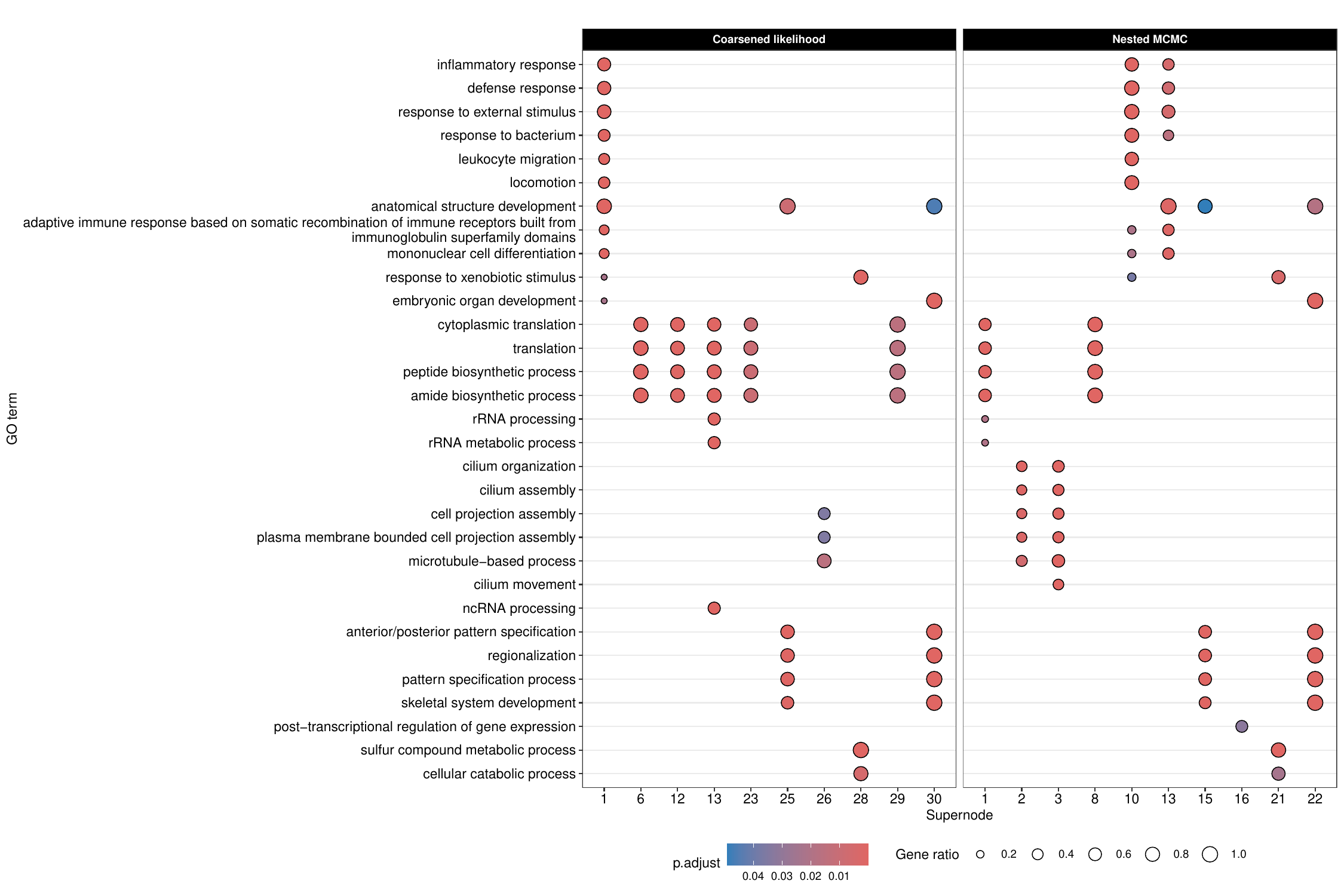}
\caption{
Gene expression data:
GO overrepresentation analysis for the tessellation estimates from coarsening the likelihood and nested MCMC.
The gene ratio is the proportion of genes in the respective supernode that are annotated with the respective GO term.
The $p$-values are adjusted using the Benjamini-Hochberg procedure
to control the false discovery rate.
The supernodes are numbered from large to small (only supernodes with overrepresented GO terms are included).
Note that there is no correspondence between the numberings of supernodes obtained from the two MCMC schemes. 
}
\label{fig:GO}
\end{figure}

This section details the Gene Ontology \citep[GO,][]{Ashburner2000,GO_2023} enrichment analysis mentioned in Section~\ref{sec:application} of the main manuscript.
We summarize
the results of the enrichment analysis in Figure~\ref{fig:GO}.
The plot is generated using version~4.8.3 of the
R package \texttt{clusterProfiler} \citep{Wu2021}. In the analysis, we use the default settings of \texttt{clusterProfiler}.
GO contains annotations of genes with certain terms. GO has three types of terms: molecular function, cellular component and biological process.
We only consider GO terms for biological processes, as in \citet{Zhang2018}.
Furthermore, the annotations considered are for the species \textit{Homo sapiens} as listed in GO on January 1, 2023.
Finally, we exclude GO terms with which fewer than 10 of the background genes are annotated
to avoid the detection of an exceedingly large number of rare GO terms.

As background genes, we use all $p=373$ genes. The gene sets that we consider are the supernodes.
Thus,
this analysis checks for a GO term
whether its relative frequency among genes in a supernode is statistically significantly higher than among all 373 genes.
It does so using a Hypergeometric test.
The resulting
$p$-values are adjusted using the Benjamini-Hochberg procedure
to control the false discovery rate \citep{Yekutieli1999}.
Then, for ease of visualization, for each supernode, only the three GO terms with the smallest adjusted $p$-values, provided $p<0.05$,
are considered. In this way, we build a set of GO terms, and in
Figure~\ref{fig:GO} we show how these GO terms distribute across supernodes. 
Since different supernodes can be enriched with the same GO terms, the total number of statistically significant GO terms shown for one supernode can exceed three.

\subsection{Comparison with a GGM}

We compare posterior inference results obtained using the proposed methodology with the results from a standard GGM.
Specifically, we fit the graphical lasso \citep{Friedman2007} to the data which is a popular method for GGMs.
It requires choosing a regularization parameter.
For a reasonable visual comparison with the graph of graphs methodology,
we tune this parameter such that the number of edges is equal to the total number of within-supernode edges in Figure~\ref{fig:gene_sphere} in the main manuscript, which is 351.
For reference,
the graphical lasso estimates 30610 edges
when choosing
the regularization parameter using cross-validation
via the R package \texttt{CVglasso} \citep{Galloway2018}
with its default settings.

\begin{figure}
\centering
\includegraphics[width=\textwidth]{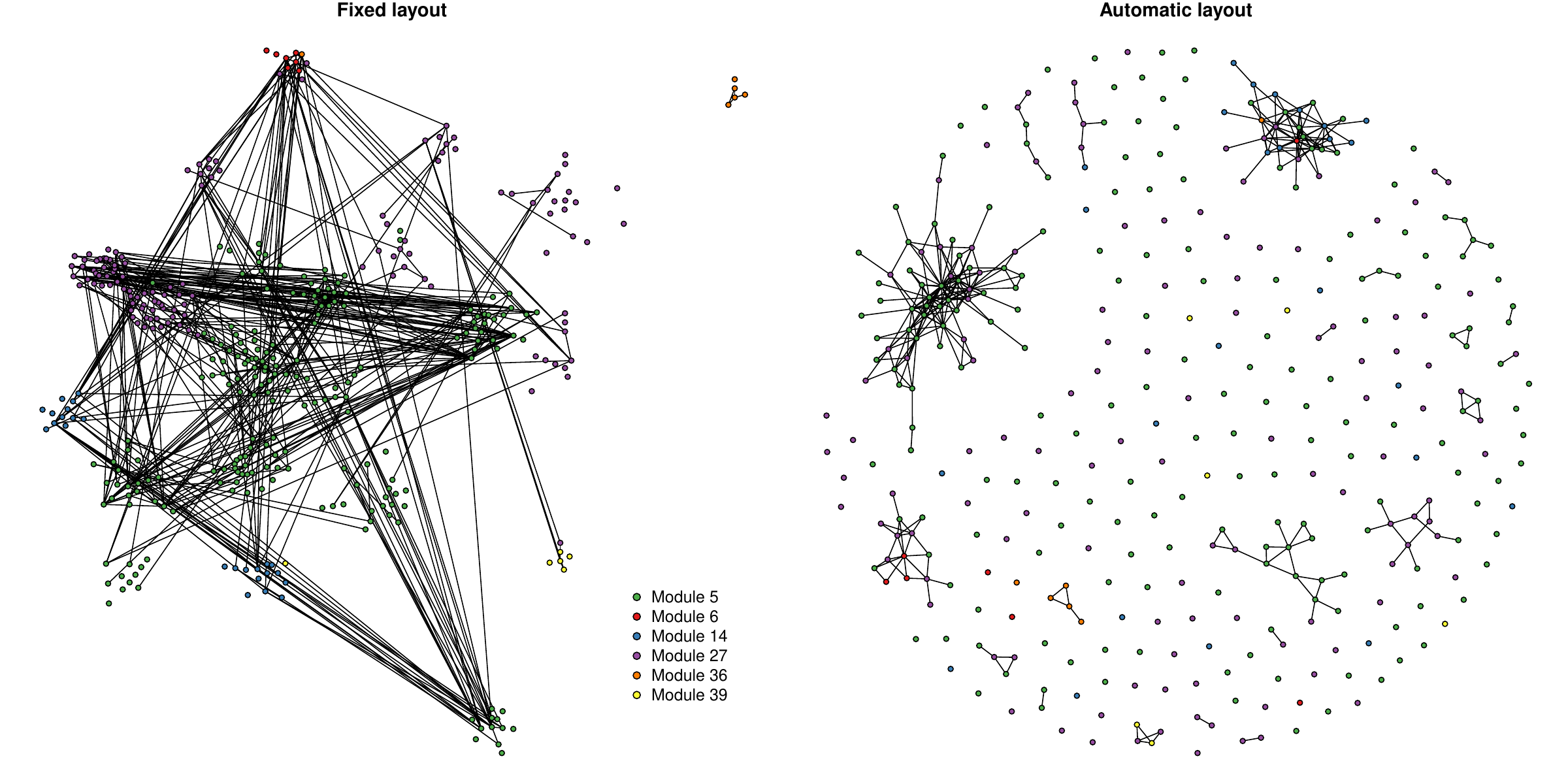}
\caption{
Gene expression data:
graph estimated using the graphical lasso.
The circles represent nodes (i.e.\ genes) which are connected by the edges in black.
The graph is visualized (i) on the left with the same layout as in Figure~\ref{fig:gene_sphere} of the main manuscript and (ii) on the right with a graph layout based on the graphical lasso estimate.
The nodes are colored according to the modules estimated by \citet{Zhang2018}.
}
\label{fig:gene_glasso}
\end{figure}

We visualize the graphical lasso estimate in Figure~\ref{fig:gene_glasso}.
Some large-scale structures arise in terms of connected components in the graph estimate. Nonetheless,
less substructure is highlighted than with the graph of graphs in Figure~\ref{fig:gene_sphere} of the main manuscript.
Furthermore, interpretation of results based on such single-edge inference is a more challenging task.

\clearpage
\bibliographystyle{apalike_modified.bst}
\bibliography{graph}